\documentclass[journal]{IEEEtran}
 \usepackage{multirow}
 \usepackage{enumerate}
\usepackage{setspace}
\usepackage{amstext}
\usepackage{cite}
\usepackage{amsmath}
\usepackage{amsbsy}
\usepackage{amssymb}
\usepackage{mathtools}
\usepackage{graphicx}
\usepackage{caption}
\usepackage{subcaption}
\usepackage{multirow}
\usepackage{booktabs}

\begin{document}
\bstctlcite{IEEEexample:BSTcontrol}
\pagenumbering{arabic}
\title{A Survey of Wideband Spectrum Sensing Algorithms for Cognitive Radio Networks and Sub-Nyquist Approaches}
\author{Bashar I. Ahmad 
\thanks{B. I. Ahmad is with Signal Processing and Communications (SigProC) Laboratory, Department of Engineering, University of Cambridge, Trumpington Street, Cambridge, UK, CB2 1PZ. Email:bia23@cam.ac.uk.}
}
\maketitle

\begin{abstract}
Cognitive Radio (CR) networks presents a paradigm shift aiming to alleviate the spectrum scarcity problem exasperated by the increasing demand on this limited resource. It promotes dynamic spectrum access, cooperation among heterogeneous devices, and spectrum sharing. Spectrum sensing is a key cognitive radio functionality, which entails scanning the RF spectrum to unveil underutilised spectral bands for opportunistic use. To achieve higher data rates while maintaining high quality of service QoS, effective wideband spectrum sensing routines are crucial  due to their capability of achieving spectral awareness over wide frequency range(s)\ and efficiently harnessing the available opportunities. However, implementing wideband sensing under stringent size, weight, power and cost requirements (e.g., for portable devices) brings formidable design challenges such as addressing  potential prohibitively high complexity and data acquisition rates.  This article gives a survey of various wideband spectrum sensing approaches outlining their advantages and limitations; special attention is paid to approaches that utilise sub-Nyquist sampling techniques. Other aspects of CR such as cooperative sensing and performance requirements are briefly addressed. Comparison between sub-Nyquist sensing approaches is also presented.
\end{abstract}
\section{Introduction} \label{sec:Background}
\IEEEPARstart{W}{ith} conventional static spectrum allocation policies, a licensee, i.e. Primary User (PU), is  permitted to use a particular spectrum band over relatively long periods of time. Such inflexible allocation regimes have resulted in a remarkable spectrum under-utilisation in space or time as reported in several empirical studies conducted in densely populated urban environments \cite{2002FCC_Report,2005FCC_Report}. By enabling an unlicensed transmitter, i.e. Secondary User (SU), to opportunistically access or share these fully or partially unused licensed spectrum gap(s), the cognitive radio paradigm (a term attributed to Mitola \cite{mitola1993software}) has emerged as a prominent solution to the persistent spectrum scarcity problem \cite{haykin2005cognitive}. It fundamentally  relies on a dependable  spectrum awareness regime to identify vacant spectrum band(s) and limit any introduced interference, possibly below a level agreed \textit{a priori} with the network PUs. Therefore, spectrum sensing, which involves scanning  the RF spectrum in search of a spectrum opportunity, is considered to be one of the most critical components of a CR network \cite{mitola1993software,haykin2005cognitive,quan2008collaborative,yucek2009survey,haykin2009spectrum,axell2012spectrum}. 

There are several approaches to spectrum sharing and PU/SU coexistence in cognitive radio networks (see \cite{goldsmith2009breaking,yucek2009survey,axell2012spectrum,haykin2009spectrum,Hattab2014} for an overview). Here, we predominantly focus on the interweaving systems, where a secondary user is not permitted to access a spectrum band when a primary user transmission is present, i.e. maintaining minimal interference. With alternative methods, namely underlay and overlay systems, the PU and SU(s) can simultaneously  access a spectral subband. They utilise multi-access techniques such as spread spectrum and/or assume the availability of  a considerable amount of information on the network PUs (e.g., codebooks, operation patterns, propagation channel information, etc.) to constrains  the resulting coexistence interference. These two paradigms still however rely to an extent on the spectrum awareness to establish the spectrum status.

In this article, we present a survey of key wideband  spectrum sensing algorithms for cognitive radio networks. This follows introducing various related background topics  such as sampling rate requirements, standard performance metrics, cooperative and  narrowband sensing techniques. First, wideband sensing methods that fulfill the Nyquist sampling criterion are presented and their  requirements are compared. Subsequently, algorithms that can operate at significantly low, sub-Nyquist, sampling rates are introduced. They aim to circumvent the sampling rate bottleneck of a wideband  sensing capability and deliver a low size, weight, power and cost solution. Sub-Nyquist algorithms are divided into two classes: 1) compressed sensing, and 2) alias-free sampling. The differences and synergies between those two methodologies  are addressed and numerical examples are presented to demonstrate their  performance on example scenarios.

The remainder of this article is organised as follows. In Section \ref{sec:ProblemStatement}, we formulate the wideband sensing problem employing a simplified system model, outline the sampling rates requirements and highlight performance metrics. Basic narrowband and cooperative  sensing approaches are discussed in Section \ref{sec:Background}. Wideband spectrum sensing methods are then categorised according to their data acquisition approach. Algorithms that abide by the Nyquist sampling criterion are succinctly addressed in Section \ref{sec:NyquistWSS}. Sub-Nyquist wideband spectrum sensing methods that are based on the compressed sensing  and alias-free sampling methodologies are discussed in Sections \ref{sec:CSWSS} and \ref{sec:NCSWSS}, respectively. In Section \ref{sec:SubNyquistComparison}, the various sub-Nyquist approaches are compared and numerical examples are presented to illustrate their effectiveness. Finally, conclusions are drawn in Section \ref{sec:Conclusions}.
\section{Problem Statement and Performance Metrics}\label{sec:ProblemStatement}
\subsection{Problem Formulation and System Model} \label{sec:SystemModel}
In CR networks the objective is typically to maximise the \textit{opportunistic} throughput (see Section \ref{sec:PerformanceandOT}), for instance to support the required QoS  and/or high data rate communications via opportunistic dynamic spectrum access. This necessitates achieving spectrum awareness over wide frequency range(s) typically consisting of a number of spectrum bands with different licensed users. If a PU reappears, the availability of several other possible vacant subbands facilitates the seamless hand-off from one spectral channel to another. This can minimise interruptions to the data transmission/exchange between the secondary user and the targeted receiver.

Hence, the wideband spectrum sensing problem entails the SU(s) scanning the frequency range $\mathfrak{B}=\bigcup_{l=1}^{L}\mathcal{B}_{l}$, consisting of $L$ spectral subbands, and determining  which of them are vacant. This is also known as Multiband Spectrum Sensing (MSS). Assuming that the network transmissions are uncorrelated, the wideband spectrum sensing task reduces to the following binary hypothesis testing for  the $l^\text{th}$ subband
\begin{equation}\label{eq:WSS_TestStatistics}
\mathfrak{T}(\mathbf{y}_{l}) \overset{\mathcal{H}_{0,l}}{\underset{\mathcal{H}_{1,l}}{\lesseqgtr}}\gamma_{l},~~l=1,2,...,L, \end{equation}
and the aim is to discriminate between
\begin{align} \label{eq:HT_WSS}
\mathcal{H}_{0,l}&:\mathbf{y}_{l}=\mathbf{w}_{l} ~~~~\text{and} \nonumber \\
\mathcal{H}_{1,l}&:\mathbf{y}_{l}=\mathbf{x}_{l}+\mathbf{w}_{l},
\end{align}
such that $\mathcal{H}_{0,l}$ and $\mathcal{H}_{1,l}$ signify the absence and presence of a transmission in  $\mathcal{B}_{l}$, respectively. Vector $\mathbf{y}_l=\left[y_l(t_{1}),y_l(t_{2}),...,y_l(t_{M_l})\right]'$  encompasses the $M_l$ collected samples of the received  signal over the $l^\text{th}$ subband at the secondary user, i.e. $\mathbf{x}_{l}=\left[x_l(t_{1}),x_l(t_{2}),...,x_l(t_{M_l})\right]'$, and noise. For simplicity, $\mathbf{w_l}\sim\mathcal{N}\left( 0,\boldsymbol{\sigma}_{w,l}^{2} \right)$ is zero mean Additive White Gaussian Noise (AWGN) with covariance $\boldsymbol{\sigma}_{w,l}^{2} $; more general noise models can be considered.    

\begin{figure*}[t] 
\centering
\includegraphics[width=0.7\linewidth]{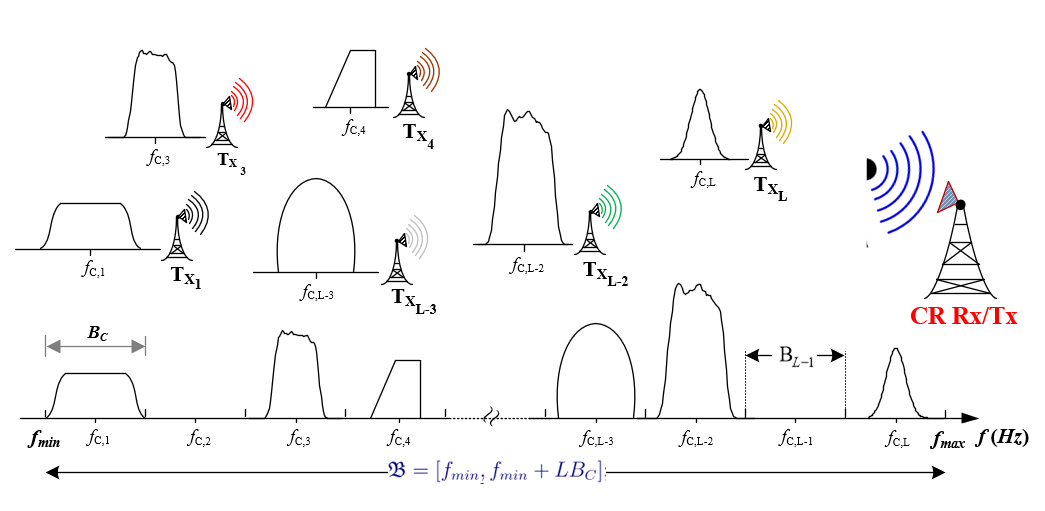}
\caption{Wide monitored bandwidth divided into $L$ non-overlapping contiguous subbands.  used by various transmitters. The cognitive radio scans $\mathfrak{B}$ in search of a spectral opportunity.}
\label{fig:SystemModel}
\end{figure*}

For simplicity and without the loss of generality,  we assume henceforth that the monitored spectral subbands, constituting $\mathfrak{B}$, are contiguous and of equal width denoted by $B_{C}$. Thereby, the overseen wide bandwidth has a total width of $LB_{C}$ and is given by $\mathfrak{B}=\left[f_{min},f_{min}+LB_{C} \right]$; for example, the initial frequency point is $f_{min}=0 $. This  model, which support heterogeneous wireless devices that may adopt different wireless technologies for their transmissions, is depicted in Figure \ref{fig:SystemModel}. At any point in time or geographic location, the maximum number of concurrently active channels is expected to be  $L_{A}$ and the received signal at the secondary user is given by: $y(t)=\sum_{k=1}^{K}{y_{k}(t)}$ such that $K\leq L_{A}$. In accordance with the low spectrum utilisation premise that motivated the cognitive radio paradigm from the outset, we reasonably assume $L_{A}\ll L$ and the single-sided joint bandwidth of the active channels does not exceed $B_{A}=L_{A}B_{A}$.  It is emphasised that the multiband spectrum sensing algorithms discussed in this article are not restricted to the above model; they apply to scenarios where spectrum awareness across disjoint spectral subbands of varying widths is sought.
\subsection{Sampling Rate Requirements}\label{sec:SamplingRate}
In classical Digital Signal Processing (DSP), the sampling instants $\left\{t_{m}\right\}_{m=1}^{M_l}$ are uniformly distributed where $t_{m}=mT_{US,l}$ and $f_{US,l}=1/T_{US,l}$ is the data acquisition rate. To shorten the notation, let $x_l[m]=x_l(mT_{US})$ and $y_l[m]=y(mT_{US,l})$. This explicitly assumes that the data for each monitored subband is digitised/sampled separately. However, digitally performing the simultaneous  sensing of the $L$ spectral channels requires a wide RF\ front-end and sampling the incoming signal $y(t)$ at rates exceeding the Nyquist rate $f_{Nyq}=2B$. No prior knowledge of the activity of the system subbands is presumed and in this case $M$ is the total number of samples for all subbands in the scanned range $\mathfrak{B}$. A sampling rate $f_{US}\geqslant f_{Nyq}$ prevents the adverse effects of the aliasing phenomenon that can hinder accomplishing numerous  processing task (including spectrum sensing or signal reconstruction) as per the Nyquist sampling criterion \cite{vaughan1991theory}. 

For considerably wide bandwidths (e.g., several GHz), $f_{Nyq}$ can be prohibitively high (e.g., tens of GHz) demanding specialised data acquisition hardware and high speed processing modules with high  memory and power consumption requirements. Such solutions can be very challenging and inviable for certain scenarios, e.g. portable devices. It is noted that whilst wideband compact selective antennas are continuously emerging, e.g. \cite{hao2011highly}, the development of Analogue to Digital Converters (ADCs) with high resolution and reasonable power consumption is relatively behind \cite{de2011sigma,balasubramanian2014current}. Therefore, the sampling rate can be the bottleneck in realising efficient wideband spectrum sensing routines. This triggered an immense interest in novel  sampling techniques that can mitigate the Nyquist criterion and permit sampling at remarkably low rates without compromising the sensing quality, i.e. sub-Nyquist data acquisition. Instead of concurrently processing the $L$ subbands, the SU\  can sweep across $\mathfrak{B}$  and filter out data relating to each individual  system subband. A narrowband detector is then utilised to affirm the status of the filtered spectral channel. In this case, $f_{Nyq}=2B_{C}\ll2B$ and this approach is dubbed sequential wideband Nyquist spectrum sensing discussed in Section \ref{sec:NyquistWSS}.        
\subsection{ Sensing Performance and Trade-offs}\label{sec:PerformanceandOT}
Before introducing several wideband spectrum sensing algorithms, here we highlight  the performance measures frequently adopted to assess the sensing quality along  a few of the associated trade-offs. As  will be apparent below,  there is no unified performance metric for wideband spectrum sensing  and the selected measure is dependent on the tackled scenario and its parameters.  
\subsubsection{Probabilities of Detection and False Alarm}
The Receiver Operating Characteristics (ROC)\ is one of the most commonly used detection performance metrics. For a particular spectral subband, e.g. $\mathcal{B}_{l}$, it captures the relation between the probability of false alarm and detection given by
\begin{align}
&P_{FA,l}=Pr\left\{\mathfrak{T}(\mathbf{y}_{l}) \geqslant\gamma_{l}|\mathcal{H}_{0,l}\right\} ~~\text{and}\nonumber \\&~~P_{D,l}=Pr\left\{\mathfrak{T}(\mathbf{y}_{l}) \geqslant\gamma_{l}|\mathcal{H}_{1,l}\right\},
\end{align} 
respectively. They are interrelated via the
detection threshold $\gamma_{l}$ whose value trades $P_{D,l}$ for the probabilities of unveiling a spectrum opportunity $1-P_{FA,l}$ and vice versa. In certain instances, the probability of missed detection $P_{M,l}=1-P_{D,l}$ is examined in lieu of $P_{D,l}$ and the ROC probabilities are plotted against the $\text{SNR}$. The
reliability of a spectrum sensing routine can be  reflected in its ability to  fulfill certain probabilities of detection, i.e.  $P_{D,l}\geq \eta_{l}$,  and false alarm, i.e. $P_{FA,l}\leqslant\ \rho_{l}$. To illustrate the relationship between $P_{FA,l}$, $P_{D,l}$ and the number of collected transmission measurements $M$, next we  consider the narrowband match filter and energy detectors.

From the coherent detector in (\ref{eq:NSS_CoherentDetector}), we have $\mathfrak{T}(\mathbf{y}_{l})\sim\mathcal{N}\left(0,MP_{S,l}\sigma_{w,l}^{2} |\mathcal{H}_{0,l}\right)$ and $\mathfrak{T}(\mathbf{y}_{l})\sim\mathcal{N}\left(MP_{S,l},MP_{S,l}\sigma_{w,l}^{2} |\mathcal{H}_{1,l}\right)$ where the \textit{a priori} known $\mathbf{x}_{l}$ is  deterministic  \cite{quan2008collaborative}. We recall that $P_{S,l}$ and $\sigma_{w,l}^2$ are the signal   power  and AWGN\ variance respectively. It follows that  
\begin{align}\label{eq:CoherentDetector_Nlimit}
P_{D,l}&=Q\left( Q^{-1}\left( P_{FA,l} \right)+MP_{s,l}/\sigma_{w,l}P_{S,l}  \right) \nonumber \\
 \hat M&=\left[ Q^{-1}\left( P_{FA,l} \right)-Q^{-1}\left( P_{D,l} \right) \right]^{2}\text{SNR}^{-1}.
 \end{align}
The number of data samples required to achieve a desired operating point $\left( P_{FA,l},P_{D,l}  \right )$ is denoted by $\hat M$; $Q(x)$ is the tail probability of a zero-mean Gaussian random variable. Whereas, for the energy detector in (\ref{eq:NSS_EnergyDetector}), we have $\mathfrak{T}(\mathbf{y}_{l})\sim\mathcal{N}\left(M\sigma_{w,l}^{2} ,2M\sigma_{w,l}^{2} |\mathcal{H}_{0,l}\right)$ and $\mathfrak{T}(\mathbf{y}_{l})\sim\mathcal{N}\left(M\sigma_{w,l}^{2} +MP_{S,l},2M\sigma_{w,l}^{2} +4M\sigma_{w,l}^{2}P_{S,l}|\mathcal{H}_{1,l}\right)$. It is noted that the Central Limit Theorem (CLT) is employed to approximate the chi-squared distribution of the energy detector by a normal distribution $\mathcal{N}\left( \bar m,\sigma^{2} \right)$ of mean and variance equal to $\bar m$ and $\sigma ^{2}$\ respectively. Subsequently, we can write  
\begin{align}\label{eq:EnergyDetector_Nlimit}
P_{D,l}=\frac{Q\left( \sigma_{w,l}\sqrt{2M}Q^{-1}\left( P_{FA,l} \right)-M\sigma_{w,l}P_{s,l}\right)}{\sigma_{w,l}\sqrt{2M\sigma_{w,l}^{2}+4MP_{s,l}}} \nonumber \\ \
\hat M=2\left[ Q^{-1}\left( P_{FA,l} \right)-\sqrt{1+2\text{SNR}}Q^{-1}\left( P_{D,l} \right) \right]^{2}\text{SNR}^{-2}.
\end{align} 
It can be noticed from (\ref{eq:CoherentDetector_Nlimit}) and (\ref{eq:EnergyDetector_Nlimit}) that the number of data samples $\hat M$ is a design parameter that can be manipulated to achieve  the sought $P_{D,l}\geq \eta_{l}$ and $P_{FA,l}\leqslant\ \rho_{l}$ at the expense of increasing the sensing time since classically $T_{ST}=M/f_{US}$ and  $f_{US}\geqslant f_{Nyq}$ is the uniform sampling rate. This has implication on the delivered opportunistic throughput discussed below. Deciding a subband's status  intrinsically relies  on the test statistics threshold, i.e. $\gamma_{l}$ in (\ref{eq:WSS_TestStatistics}). It dictates the detector  operational point/region and  the complete ROC plot is generated from testing all the possible threshold values. The explicit dependency of $P_{FA,l}$ and $P_{D,l}$ on $\gamma_{l}$ is discarded to simplify the notation.      

For the studied wideband spectrum sensing problem, the ROC can be defined by the two vectors $\mathbf{P}_{FA}=\left[P_{FA,1},P_{FA,2},...,P_{FA,L}\right]^{T}$ and $\mathbf{P}_{D}=\left[P_{D,1},P_{D,2},...,P_{D,L}\right]^{T}$ encompassing  the probabilities of the $L$ system subbands. The aforestated reliability measure can be extended to the multiband environment via
\begin{equation}\label{eq:MSS_Reliability}
\mathbf{P}_{FA}\preceq \boldsymbol{\rho}~~~\text{and} ~~~ \mathbf{P}_{D}\succeq\ \boldsymbol{\eta}
\end{equation}
 where $\boldsymbol{\rho}=\left[\rho_{1},\rho_{2},...,\rho_{L}\right]^{T}$ and $\boldsymbol{\eta}=\left[\eta_{1},\eta_{2},...,\eta_{L}\right]^{T}$. Each of  $\succeq$ and $\preceq$ refers to an element by element vectors comparison. The monitored channels can have different sensing requirements and the available design parameters, e.g. $M$, is selected such that (\ref{eq:MSS_Reliability}) is met for all the system subbands. Alternatively, the SU can combine the probabilities of the spectrum bands via
\begin{equation}
\bar P_{FA}=\sum_{l=1}^{L}{a_{l}P_{FA,l}}~~~\text{and}~~~\bar P_{D}=\sum_{l=1}^{L}{b_{l}P_{D,l}}
\end{equation}
where $\left\{{a_{l}}\right\}_{l=1}^{L}$ and $\left\{{b_{l}}\right\}_{l=1}^{L}$ are the weighting parameters. They reflect the importance, interference provisions and the confidence level of the test statistics per channel. The sensing reliability  can take the form of $\bar{P}_{FA}\leqslant\ \bar \rho_{l}$ and $\bar{P}_{D}\geq \bar \eta_{l}$. Using $b_{l}=1$ leads to a simple averaging approach $\bar P_{D}=\frac{1}{L}\sum_{l=1}^{L}{P_{D,l}}$. This creates  the risk of a low detection rate for  one particular subband (e.g. due to  PU low transmission power)    drastically affecting the SU overall multiband detection across  $\mathfrak{B}$. On the other hand, if a particular subband, e.g. $\mathcal{B}_{l}$, has low interference constraints, a marginal $b_{l}$ value can be assigned.

Recalling that the objective of the wideband spectrum sensing is to unveil a sufficient amount of spectrum opportunities at a  SU without causing harmful interference to the PUs, the probability of missing a spectral opportunity can be  defined as 
\begin{equation} \label{eq:Performance_FA_AllSubbands}
\tilde P_{MSO}=Pr\left\{ \cap_{l=1}^{L} \mathcal{H}_{1,l}\left\vert \cup_{l=1}^{L} \mathcal{H}_{0,l} \right. \right\}.
\end{equation}
Entries  $\cap_{l=1}^{L} \mathcal{H}_{1,l}$ and $\cup_{l=1}^{L} \mathcal{H}_{0,l}$ stipulate that the hypothesis testing outcome for all the surveyed subbands are "1" and at least one channel hypothesis testing results in "0", respectively.  This  implies that a missed spectrum opportunity occurred, since at least one of the monitored spectral channels  was vacant. Whilst in (\ref{eq:Performance_FA_AllSubbands}) one unoccupied subband is sought by the CR,  a more generic formulation  can incorporate multiple opportunities \cite{xin2010efficient}. In \cite{sun2013performance}, multiband detection performance measures that are independent of the sensing algorithm are proposed. Let $N_{SO}$ be the number of correctly identified inactive channels, $N_{DSO}$ is the pursued number of vacant subbands (i.e. spectral opportunities), $N_{I}$ is the number of occupied channels declared vacant and $N_{d}$ is the maximum permitted number of falsely identified vacant subbands (i.e. interference limit).   The sensing quality can be empirically examined in terms  of the \textit{probability of insufficient spectrum opportunities}  $P_{ISO}(N_{SO})=Pr\left\{ N_{SO}<N_{DSO}\right\}$ and \textit{probability of excessive interference} $P_{EI}( N_{I})=Pr\left\{ N_{I}>N_{PI}\right\}$.  
\subsubsection{Opportunistic Throughput and Sensing Time}
\begin{figure}[t] 
\centering
\includegraphics[width=1\linewidth]{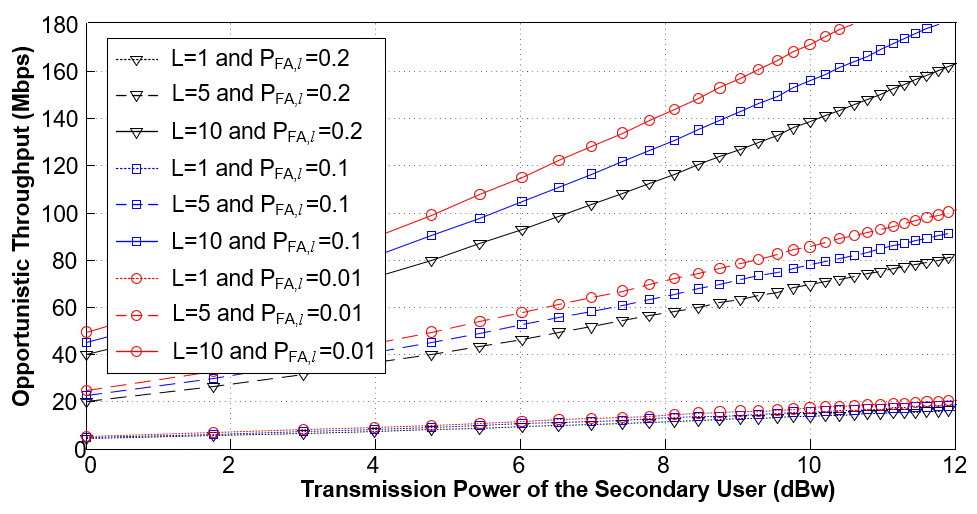}
\caption{Opportunistic throughput for a varying number of surveyed subbands ($5~\text{MHz}$ each) and SU transmitted power such that $P_{D,l}=1$ and $\text{SNR}\geqslant0~\text{dB}$.}
\label{fig:OpportunisticThroughput}
\end{figure}

 The main advantage of wideband spectrum sensing is its ability to provide superior  opportunistic throughput $R_{O}$ to meet onerous QoS requirements for the network secondary users. The $R_{O}$ values is the sum of the possible achieved data transmission rates leveraged by exploiting the network vacant spectral subbands as per
\begin{equation}\label{eq:OpportunisticThroughput}
R_{O}\triangleq\sum_{l=1}^{L}{r_{O,l}\left( 1-P_{FA,l}\right)+r_{I,l}\left( 1-P_{D,l} \right)}.
\end{equation}  
We have  $r_{O,l}=B_{C}\log_{2}\left( 1-P_{S,l}^{SU}/\sigma_{w,l}^{2} \right)$ is furnished correctly unveiling  a spectral opportunity, i.e. $\{\mathcal{H}_{0,l}|\mathcal{H}_{0,l}\}$, and  $r_{I,l}=B_{C}\log_{2}\left( 1-P_{s,l}^{SU}/\left\{P_{s,l}^{PU}+\sigma_{w,l}^{2} \right\}\right)$ is obtained when inadvertently interfering with an active PU transmission, i.e. $\{\mathcal{H}_{0,l}|\mathcal{H}_{1,l}\}$. $P_{S,l}^{SU}$ and $P_{S,l}^{PU}$ are the transmissions power over $\mathcal{B}_{l}$ pertaining to a SU and PU respectively. For an interference free network,  $r_{I,l}=0$. It is clear from (\ref{eq:OpportunisticThroughput}) that  effective  wideband spectrum sensing routine can substantially enhance $R_{O}$. This is depicted in Figure \ref{fig:OpportunisticThroughput}, which displays the opportunistic throughput $R_{O}$ for a varying number of subbands, $P_{FA,l}$ and SU transmission power.

The  opportunistic spectrum access operation at a CR involves spectrum sensing followed by transmitting over the identified vacant   system subband(s). Let $T_{Total}=T_{ST}+T_{OT}$ be the total access time consisting of the sensing functionality slot $T_{ST}$ and the opportunistic transmission time slot $T_{OT}$. It is noted that throughout this article, the sensing time $T_{ST}$ is assumed to incorporate the associated processing time affected by the detectors complexity, computational cost and the available processing resources at the SU. Thereby, the total leveraged throughput according to (\ref{eq:OpportunisticThroughput}) is $R_T\left(T_{ST}\right)=\frac{T_{Total}-T_{ST}}{T_{Total}}R_{O}$ and optimising $T_{ST}$ can be formulated as
\begin{align} \label{eq:OptimisingThroughputVsTime}
&\hat T_{ST}=\underset{0<T_{ST}\leqslant T_{Total}}{\arg\max}R_T\left(T_{ST}\right) \nonumber \\
&\text{s.t.}~~\mathbf{P}_{FA}\preceq \boldsymbol{\rho}~~~\text{and} ~~~ \mathbf{P}_{D}\succeq\ \boldsymbol{\eta}. \end{align}   
Variations of (\ref{eq:OptimisingThroughputVsTime}) with different constraint(s) can be adopted, e.g. optimising the number of captured samples $M$ that maximises $R_{T}$ in lieu of $T_{ST}$  as in \cite{kim2011multi}. Moreover, network Medium Access Control (MAC) and frame structuring techniques other than the  sequential sensing and transmission regime can be applied. This includes a range of  $T_{ST}$ or  $T_{Total}$ partitioning strategies into sub-slots   \cite{fan2010optimal} and administering concurrent sensing-transmission during $T_{Total}$ via parallel decoding and sensing \cite{stotas2010overcoming}. A number of alternative opportunistic throughput boosting techniques are proposed in the literature, e.g. utilising a dual radio architecture with parallel sensing and transmission modules  \cite{ghassemi2010cognitive},  detection  with  adaptive sensing time   algorithms that take the $\text{SNR}$ value into account \cite{paysarvi2011optimal} and many others. 

Finally, there are several practical network considerations that should be weighed when performing  wideband spectrum sensing. For examples, transmission power control and interference trade-offs, vacant single or multiple subbands allocation, number of collaborating SUs simultaneously overseeing $\mathfrak{B}$, etc. They are typically decided based on an sought detection quality and the gained opportunistic throughput; see \cite{haykin2005cognitive,yucek2009survey,haykin2009spectrum,axell2012spectrum} for further details.
\section{Background: Narrowband and Cooperative Spectrum Sensing} \label{sec:Background}
\subsection{Narrowband Sensing Techniques}
\label{sec:NarrowbandSS}
Unlike wideband sensing methods for revealing opportunities across $\mathfrak{B}=\bigcup_{l=1}^{L}\mathcal{B}_{l}$, Narrowband Spectrum Sensing (NBSS) algorithms are applied to a single spectral band, e.g. $\mathcal{B}_{l}$. As background, the coherent, energy and cyclostationary detectors, which are among the most commonly used NBSS methods in cognitive radio networks \cite{haykin2005cognitive,yucek2009survey,quan2008collaborative,axell2012spectrum}, are discussed here and their requirements are compared. 
\begin{itemize}
\item 
\textit{Coherent Detector}: Otherwise  known as matched filter and it utilises the optimal test statistic that maximises the Signal to Noise Ratio (SNR) in the presence of additive noise \cite{kay1998fundamentals,poor1994introduction}. Its test statistic simply involves correlating the  suspected   PU\ transmission $\mathbf{x}$ with the received signal according to
\begin{equation} \label{eq:NSS_CoherentDetector}
\mathfrak{T}(\mathbf{y})\triangleq\mathbf{x}^H\mathbf{y} \overset{\mathcal{H}_{0}}{\underset{\mathcal{H}_{1}}{\lesseqgtr}}\gamma
\end{equation}
where $(.)^{H}$ is the conjugate transpose operate. The  PU\ signal structure is assumed to be perfectly known at the receiver, e.g. signal variance, modulation type, packet format, channel coefficients, etc.  One of the main advantages of the coherent detector is that it requires a small number of data measurements to achieve predefined probabilities of detection $P_{D}$ and false alarm $P_{FA}$ where $\mathcal{O}\left( 1/\text{SNR}\right)$ samples suffices even in low $\text{SNR}$ regions (i.e. when $\text{SNR}\ll1$). The signal to noise ratio is defined as $\text{SNR}=P_{s,l}/\sigma_{w,l}^{2}$ such that $P_{s,l}$ and $\sigma_{w,l}^2$ are the   powers of a present transmission in $\mathcal{B}_{l}$ and AWGN\ variance, respectively. However, in low $\text{SNR}$ conditions,  the detector's performance drastically degrades due to the difficulty in maintaining synchronisation between the transmitter and the receiver. Accurate synchronisation is a fundamental requirement the coherent detector. Additionally, the complexity of match filter grows with the diversity of potential primary users since a distinct detector per signal structure is imperative. The coherent detector is inflexible and can  be unsuitable for CR networks, which often include several PUs using different transmission technologies and dynamically adapting their transmission characteristics.              
\item 
\textit{Energy Detector}:   The energy detector, also known as  radiometer, is a non-coherent detector  widely regarded as one of the simplest approaches for deciding between  $\mathcal{H}_{0}$ and $\mathcal{H}_{1}$. Its test statistic  is given by
  \begin{equation} \label{eq:NSS_EnergyDetector}
\mathfrak{T}(\mathbf{y})\triangleq\sum_{m=1}^{M}\left\vert y[m] \right\vert^{2} \overset{\mathcal{H}_{0}}{\underset{\mathcal{H}_{1}}{\lesseqgtr}}\gamma.
\end{equation}
 This detector does not assume any knowledge of the PU\ signal structure or synchronisation with the transmitter.  It demands   $\mathcal{O}\left( 1/\text{SNR}\right)$ signal measurements in high SNR cases (i.e. when $\text{SNR}\gg1$) and $\mathcal{O}\left( 1/\text{SNR}^{2} \right)$ samples in low SNR regions to deliver the desired $P_{FA}$ and $P_{D}$ \cite{quan2008collaborative}.  It is noted that if the present AWGN noise power/variance is known \textit{a priori,} the energy detector is the optimal detector according to the Neyman-Pearson criterion \cite{kay1998fundamentals}. Whilst in (\ref{eq:NSS_EnergyDetector}) the time domain samples can be used to determine the energy level in the monitored frequency band $\mathcal{B}_{l}$, the energy detector can be implemented in the frequency domain by taking the Fast Fourier Transform (FFT) of  $\left\{y[m]\right\}_{m=1}^{M}$ and summing the squared magnitude of the resultant FFT\ bins that belong to $\mathcal{B}_{l}$ , i.e. $\mathfrak{T}(\mathbf{y})\triangleq\sum_{f_{k}\in \mathcal{B}_{l}}\left\vert Y(f_{k}) \right\vert^{2}$. The FFT is an optimised version of the   Discrete Fourier Transform (DFT) given by:   $Y(f_{k})=\sum_{m=1}^{M}{y[m]e^{-i2\pi km/M}}$ such that $k=0,1,...,M-1$, $f_{k}=kf_{US}/M$ and $f_{US}$ is the uniform sampling rate.    
\item 
\textit{Feature Detection}: Communication signals inherently  incorporate distinct features such as symbol periods, training sequences and  cyclic prefixes  to facilitate their detection at the intended receiver. Feature detectors exploit such unique structures to unveil the presence of a transmission in $\mathcal{B}_{l}$ by formulating its test statistics as a function of the incoming signal second order statistics
\begin{equation} \label{eq:NSS_FeatureDetector}
\mathfrak{T}(\mathbf{y})\triangleq\mathfrak{F}\left\{\mathbb{E}\left[ \mathbf{y}\mathbf{y}^H \right]\right\} \overset{\mathcal{H}_{0}}{\underset{\mathcal{H}_{1}}{\lesseqgtr}}\gamma
\end{equation}  
where $\mathbb{E}\left[. \right]$ is the  expectation operator and $\mathfrak{F}(.)$ is a generic function. A detailed discussion of such detectors is presented in \cite{axell2012spectrum} with the relevant references. Here, we focus on a particular  feature detector known as the cyclostationary feature detector.

Most transmissions are modulated sinusoidal carriers with particular symbol periods. Their means and autocorrelation functions exhibit periodicity, i.e. they are Wide Sense Cyclostationary  (WSCS) signals.   The cyclostationary detector capitalises on these built-in periodicities and uses the  Cyclic Spectral Density (CSD) function of the incoming WSCS signal \cite{gardner1991exploitation,gardner2006cyclostationarity}. Let $T_{p}$ be the underlying cyclostationarity period, the sampled transmission $\left\{x[m]\right\}_{m=1}^{M}$ Cyclic Autocorrelation Function (CAF)  is defined by: $R_{x}^{\tilde c}[m]\triangleq\mathbb{E}\left[ x[n]x^{*}[n+m]e^{-2\pi \tilde c n} \right]$ where $R_{x}^{\tilde c}[m]\neq0$  if $\tilde c=i/T_{p}$ ($i$ is  a non-zero integer) and $R_{x}^{\tilde c}[m]=0$ if $\tilde c \neq i/T_{p}$. The cyclic frequency is $\tilde c\neq 0$. The  CSD is the discrete-time   Fourier Transform (DTFT) of the CAF, i.e. $S_{x}(\tilde c,f_{k})=\sum_{m=-\infty}^{+\infty}R_{x}^{\tilde c}[m]e^{-j2\pi f_{k}m}$ and an FFT-type implementations can be used.  Unlike a PU transmission,  the present  noise is not WSCS. It is   typically assumed to be Wide Sense Stationary (WSS) and the DTFT of $R_{w}^{\tilde c}[m]$ is $S_{w}(\tilde c,f_{k})=0$ for $\tilde c \neq0$. The   cyclic detector test statistics can be expressed by
\begin{displaymath}
\mathfrak{T}(\mathbf{y})=\sum_{\tilde c}\sum_{f_{k}} \hat S_{y}(\tilde c,f_{k})\left[S_{x}(\tilde c,f_{k})\right]^{*}\overset{\mathcal{H}_{0}}{\underset{\mathcal{H}_{1}}{\lesseqgtr}}\gamma
\end{displaymath}
assuming a known transmission $S_{x}(\tilde c,f_{k})$ with multiple periods; $x^{*}$ is the conjugate of $x$ \cite{quan2008collaborative}. The estimated CSD  of the received signal is denoted by $\hat S_{y}(\tilde c,f_{k})$ and it is attained from  $\mathbf{y}$. Whilst the cyclostationary detector can reliably differentiate between various PU modulated signals  and the present noise, its complexity and computationally cost is relatively high as it involves calculating the 2D\ cyclic spectral density function. Its performance and sampling requirements in terms of  delivering desired $P_{FA}$ and ${P_{D}}$  values are  generally intractable \cite{quan2008collaborative}. 

Alternative  feature detection techniques use the properties of the covariance matrix in (\ref{eq:NSS_FeatureDetector}) to identify the presence a PU signal; namely the fact that the signal and noise covariance matrices are distinguishable. An example is the covariance detectors in  \cite{zeng2009spectrum} where the test statistic is expressed in terms of a sample covariance matrix maximum and minimum eigenvalues, i.e. $\mathfrak{T}(\mathbf{y})=\upsilon_{max}/\upsilon_{min}$, and no information on the transmitter signal is required. Other methods  promote particular structures of the PU\ signal  covariance matrix    and usually demand knowledge of certain signal characteristics, e.g.\cite{axell2011unified}.
\end{itemize}
 
  The above narrowband spectrum sensing algorithms are compared in Table \ref{tab:NBSS_Comparison} outlining their advantages and disadvantages. Main features such as the amount of prior information the CR needs to unveil the presence of the PU(s)\ are outlined. If a single  PU is utilising $\mathcal{B}_{l}$ and its transmissions structure   is fully known by the SU, then the coherent detector is the best  candidate   for the spectrum sensing task with the highest performance, lowest complexity and shortest sensing time. It presumes accurate synchronisation between the PU and the SU. Since this scenario is rarely faced in CR\ networks where other transmitters can opportunistically access a vacant $\mathcal{B}_{l}$, other detectors become more viable candidates. For example, feature detectors could be deployed when partial knowledge of the PU\ transmissions is available, e.g. cyclic prefixes, modulation scheme, preambles, etc. They are robust against noise uncertainties, interference and can distinguish between different types of signals. Nevertheless, the feature detectors, e.g. cyclostationary detector, require more complex processing, sensing time and power resources. On the other hand, the energy detector is a simple and low complexity  option, which does not levy any prior knowledge of the PU signal or synchronisation. Its sensing time is also notably low for relatively high SNR\ regions. However, the radiometer does not differentiate between  PU(s), SU(s) and potential interferers. Its performance is highly dependent on accurate estimation of the present noise power/variance to decide on the threshold values to restrain $P_{FA}$. Inaccurate estimation of the noise power can cap the attained detection  quality when the signal to noise ratio is lower than a particular level, i.e.  the $\text{SNR}$ wall reported in \cite{tandra2008snr}.
\begin{table*}[t]
\caption{Comparison between common narrowband spectrum sensing techniques.}
\centering
\small
\begin{tabular}{llll}
\hline \hline
\textbf{Detector}&\textbf{Prior Knowledge}&\textbf{Advantages}&\textbf{Limitations} \\ \hline
\multirow{2}{1cm}{\textbf{Coherent}}& \multirow{2}{3cm}{PU full signal structure} & Optimal performance & PU signal dependent \\ && Low  computational complexity & Demands synchronisation \\ \hline
\multirow{2}{1cm}{\textbf{Energy}}& \multirow{2}{3cm}{Noise power } & No signal knowledge & Does not distinguish between users \\ && Low computational complexity & Limited by noise power estimation \\ \hline
\multirow{2}{1cm}{\textbf{Feature}}& \multirow{2}{3cm}{Partial knowledge  of  the PU structure} & Distinguishes PUs and SUs  & High computational complexity\\ && Robust to noise and interference & Long sensing times \\ \hline\hline
\end{tabular}
\label{tab:NBSS_Comparison}
\end{table*}
\subsection{Cooperative Sensing}
One of the key challenges of realising a spectrum sensing routine is the well known hidden terminal problem  faced in wireless communications. It pertains to the scenario where the PU transmission  is undermined by channel shadowing or multi-path fading and a SU is located in the PU deep fading region. This can lead to the SU reaching a decision that the sensed  spectrum band is vacant; any subsequent utilising can cause severe interference to  the  PU.  To enhance the CR network sensitivity, the  network can fuse the sensing results of a few of its spatially distributed  CRs to exploit their inherent spatial diversity. Each of these CRs experience different channel conditions and their cooperation can alleviate the hidden terminal problem \cite{haykin2005cognitive,quan2008collaborative,yucek2009survey,letaief2009cooperative,axell2012spectrum,Ibnkahla2014book}.

A key aspect of collaborative sensing is efficient cooperation schemes that substantially improve the network reliability. They should minimise the bandwidth and power requirements  that are associated with the control channel over which information is exchanged among the network SUs. Below, we briefly address the three common information fusing schemes to combine the sensing results of $I$ collaborating SUs in a CR network. 
\begin{itemize}                                                  
\item 
\textit{Hard Decision Fusion}: With hard combining each of the SUs makes a decision on the presence of the PU and shares a single bit to represent its binary decision $d_{i}=\left\{ 0,1 \right\}$; i.e. "0" and "1" signify $\mathcal{H}_{0}$ and $\mathcal{H}_{1}$ respectively. The final decision on the spectrum band status is based on a voting metric that can be expressed by
\begin{equation} \label{eq:CooperativeSensing_HardCombining}
\mathcal{V}_{HF}=\sum_{i\in I}{d_{i}}\overset{\mathcal{H}_{0}}{\underset{\mathcal{H}_{1}}{\lesseqgtr}}\nu
\end{equation}
where $\nu$ is the voting threshold. The final decision based on (\ref{eq:CooperativeSensing_HardCombining}) is simply a combining logic  that takes the following forms: 1) AND logic where  $\nu=I$ and all the collaborating SUs should decide that the subband is not in use to deem it vacant , 2) OR logic where $\nu=1$ and a PU transmission is considered to be present if one of the SUs reached the decision $d_{i}=1$ and 3) majority vote where $\nu= \left\lceil I/2\right\rceil $ ensuring that at least half of the SUs detected an active PU before deciding $\mathcal{H}_{1}$. The ceiling function $\left\lceil x\right\rceil$ yields the smallest integer greater than or equal to $x$. Each of the above voting strategies reflect  a different view on  opportunistic access, e.g. the OR logic  guarantees   minimum  network interference at the excess of missed spectral opportunities. On the contrary, the AND logic prioritises increasing the opportunistic throughput without restraining the possible interference. It is noted that the overheads of the hard combining in terms of the information exchange is minimal as 1-bit is shared by each SU.                                                          
\item 
\textit{Soft Decision Fusion}: In this approach the $I$  SUs share their sensing statistics , i.e. $\mathfrak{T}_{i}(\mathbf{y}), i=1,2,...,I$. A weighted sum of the sensing statistics is used as the decision metric according to
\begin{equation} \label{eq:CooperativeSensing_SoftCombining}
\mathcal{V}_{SF}=\sum_{i\in I}{\varpi_{i}\mathfrak{T}_{i}(\mathbf{y})}
\end{equation}
 where  $\varpi_{i}$ is the weight allocated to the $i^{th}$ SU. A simple choice of the weights is a uniform prior without considering the quality of the channel between the SU and the PU, i.e.  $\varpi_{i}=1$. The  weights $\left\{\varpi_{i}\right\}_{i=1}^{I}$ can be proportional to the $i^{th}$ link quality, e.g. SNR of the channel between the PU and the $i^{th}$ participating SU \cite{mishra2006cooperative}. Although  soft combining in (\ref{eq:CooperativeSensing_SoftCombining}) necessitates the exchange of large quantities of data\ compared with the hard fusion, it can lead to optimal cooperative spectrum sensing \cite{axell2012spectrum}.          
\item 
\textit{Hybrid Decision Fusion}: This approach combines both soft and hard combining techniques seeking to harness the hard fusion low transmission overhead and soft fusion superior performance. Generally, sharing more statistical information among the SUs  results in a better  fusion outcome and vice versa.  An example is the hybrid technique proposed in \cite{ma2008soft} where each SU sends two bits of information related to the monitored subband, i.e. softened hard combining, to enhance the network sensing dependability.   
\end{itemize}
In practice, several other  cooperative sensing design challenges should be taken into account, such as feasibility issues of the control channel, optimising the overheads associated with information exchange, collaborative network implementation or clustering (e.g. centralised or  distributed or ad-hoc fusion centre), cooperative sequential detection, censoring or sleeping,  etc. Whilst the objective of this subsection is to briefly introduce cooperative spectrum sensing in CR\ networks, several comprehensive overviews on this topic exist and the reader is referred to  \cite{haykin2005cognitive,quan2008collaborative,yucek2009survey,haykin2009spectrum,letaief2009cooperative,axell2012spectrum,Ibnkahla2014book} with extensive references lists  therein.Most importantly, cooperative sensing is typically implemented at a network level higher than the considered physical layer. We are predominantly  interested in  determining the status of  a given monitored subband (or a number of them) at a single CR  and the cooperative sensing concept can leverage the obtained statistic at each of the these CRs. Nonetheless, in Sections \ref{sec:NyquistWSS} and \ref{sec:CSWSS} we address certain  wideband spectrum sensing techniques that are particularly amenable to collaborative multiband detection.

\section{Nyquist Multiband Spectrum Sensing} \label{sec:NyquistWSS}
In this section, we described two wide spectrum sensing approaches that use the classical Nyquist data acquisition paradigm where $y[m]=y(t_{m})=y(m/f_{US})$ and $f_{US}\geqslant f_{Nyq}$ is the uniform sampling rate. They are: 1) Sequential Multiband Nyquist Spectrum Sensing (SMNSS) and 2) Parallel Multiband Nyquist Spectrum Sensing (PMNSS).   
\subsection{Sequential Multiband Spectrum Sensing Using Narrowband Techniques}\label{sec:SWSS}
\begin{figure}[t]
\centering
\begin{subfigure}[t]{1\linewidth}
\centering
\includegraphics[width=0.95\linewidth]{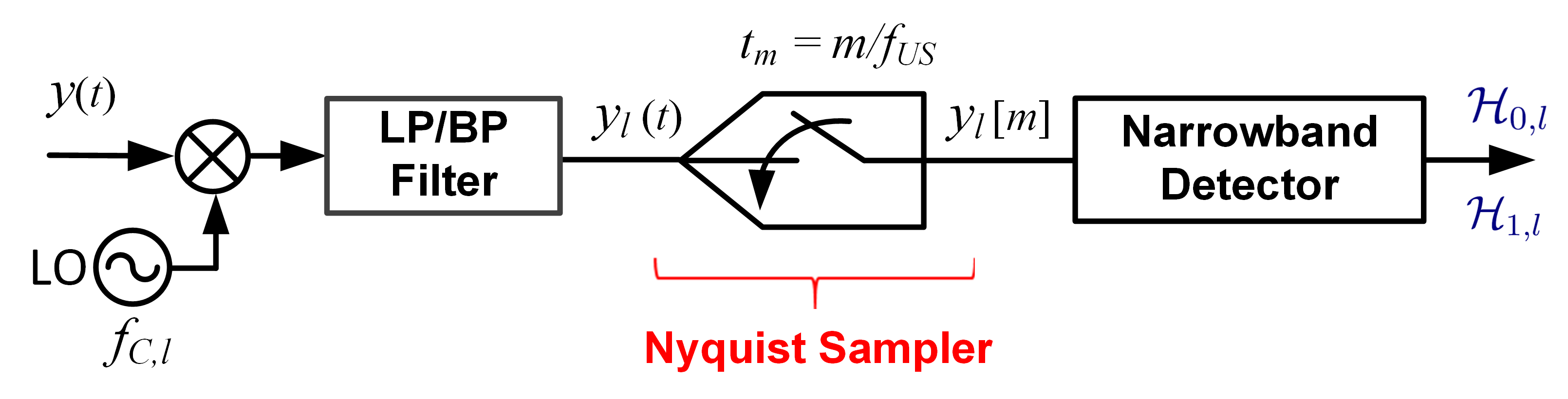}
\caption{Demodulation using a local oscillator.}
\label{fig:SMSS_LO}
\end{subfigure}\
\begin{subfigure}[t]{1\linewidth}
\centering
\includegraphics[width=0.9\linewidth]{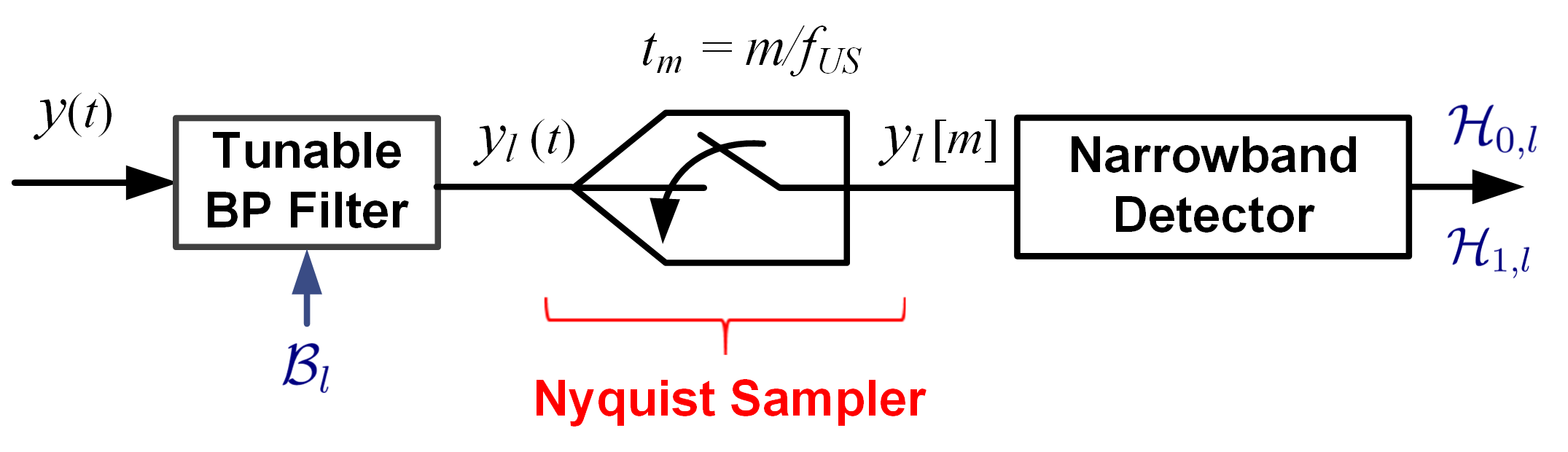}
\caption{Utilising a tunable bandpass filter.}
\label{fig:SMSS_TBBF}
\end{subfigure}%
\caption{Two common approaches to sequential Nyquist multiband spectrum sensing.}
\label{fig:SMNSS}
\end{figure}

In SMNSS, a narrowband detector (see Section \ref{sec:NarrowbandSS}) is applied to one of the system subbands at a time. This circumvents the need to digitise the wide monitored   frequency  range $\mathfrak{B,}$ and instead processes each channel $\mathcal{B}_{l}$ separately where  $f_{Nyq}=2B_{C}$. Below, two common SMNSS methods are outlined and their block diagrams are   depicted in Figure \ref{fig:SMNSS}.
\begin{enumerate}
\item 
Demodulation: a Local Oscillator (LO)  is used at the SU. It down-converts the signal in each subband to the origin (or any intermediate frequency) by multiplying $y(t)$ by the channel's carrier frequency followed by filtering and  low rate sampling. This is a widely used technique in wireless communications (i.e. superheterodyne receiver architecture) and intrinsically relies on prior knowledge of the location/carrier-frequency of the present transmission. When the positions of the active subbands and their carrier frequencies are not known as in multiband spectrum sensing, the standard demodulation technique cannot be implemented efficiently. Furthermore, the accurate generation of the  carrier frequencies can demand bulky energy hungry  phased locked loop circuit(s).
\item 
Tunable bandpass filter: a tunable analogue bandpass filter is used to filter out the data belonging to each of the monitored subbands prior to sampling. Implementing a tunable power-efficient analogue bandpass filter with a sharp cut-off  frequency and high out-of-band attenuation poses serious design challenges, especially for portable devices.     
\end{enumerate}
A critical limitation of the demodulation and tunable bandpass filtering methods is the delay introduced by sweeping the spectrum where one subband is inspected at a time. This severely increases the aggregate sensing time $T_{ST}$ necessary to scan the system $L$ channels, hinders fast processing and degrades the network opportunistic throughput. Sequential  techniques are also inflexible, requiring fine tuning of analogue components  to a  particular channels' layout. The concept of initially performing coarse wideband  sensing, i.e. low quality detection, to minimise the number of subbands  ought to be searched is proposed in \cite{luo2009two}. It is a two stage sequential spectrum sensing where a robust narrowband detector, e.g. a cyclostationary detector, can be employed in the second stage for high quality results. Other  sequential methods, e.g. sequential probability ratio tests, exist and a good overview is given in \cite{axell2012spectrum}.  
\subsection{Parallel Multiband Detection at Nyquist Rates}\label{sec:PWSS}
Few detection algorithms that simultaneously sense all the monitored subbands are discussed here. A trivial solution to the parallel sensing problem is to  use a bank of $L$ sequential multiband sensing modules, e.g. those in Figure \ref{fig:SMNSS}. Each uses a  sampling rate of $f_{US}\geqslant2B_{C}$ and is dedicated to a particular system channel. Different narrowband detectors can be assigned   to scan $\mathfrak{B}$ based on the subbands' requirements, i.e. heterogeneous architectures. This analogue-based solution requires a bulky inflexible power-hungry analogue front-end filter bank with high complexity, especially if different detectors are used. On the contrary, we are predominantly interested in digitally performing the wideband  sensing task, i.e. "software-based" solution with minimal analogue front-end infrastructure and acclaimed flexibility. This can be realised by estimating the spectrum of the incoming multiband signal from its  samples collected at sufficiently high rates, i.e. $f_{US}\geqslant f_{Nyq}$ and $f_{Nyq}=2LB_{C}$.  Next, three spectrum-estimation-based PMNSS techniques are addressed. \subsubsection{Multiband Energy  Detector}
\begin{figure}[t] 
\centering
\includegraphics[width=1\linewidth]{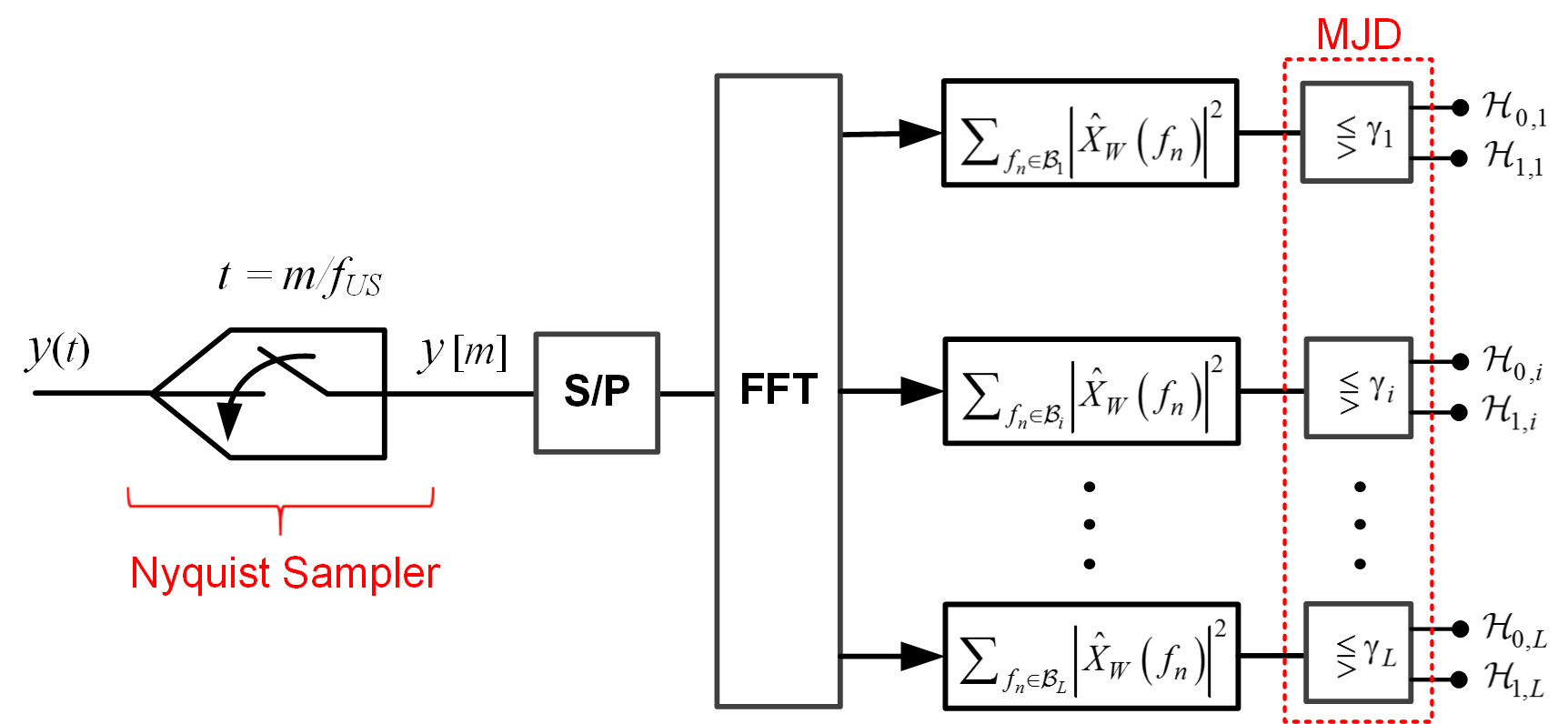}
\caption{Block diagram of the multiband energy detector.}
\label{fig:EnergyDetector_MSS}
\end{figure}

The Multiband Energy Detector (MBED) is an extension of the classical narrowband energy detector; its block diagram is shown in Figure \ref{fig:EnergyDetector_MSS}. It one of the most widely used multiband detection methods and relies on estimating the energy in each subband using the simplest Power Spectral Density (PSD) estimator, i.e. periodogram. The periodogram can be viewed as a simple estimate of the PSD\ formed using a digital filter bank of bandpass filters and it involves the scaled squared magnitude of the signal's DFT/FFT \cite{hayes1996statistical}. The test statistic is given by
\begin{equation} \label{eq:EnergyDetector_MSS}
\mathfrak{T}(\mathbf{y})\triangleq\sum_{f_{n}\in\mathcal{B}_{l}}\vert \hat X_{W}\left( f_{n} \right) \vert^{2} \overset{\mathcal{H}_{0,l}}{\underset{\mathcal{H}_{1,l}}{\lesseqgtr}}\gamma_{l},~~~l=1,2,...,L
\end{equation}
where  $\hat X_{W}\left( f_{n} \right)$ is the windowed DFT/FFT of the received signal and only the frequency-bins  that fall in $\mathcal{B}_{l}$ are considered. A  windowing function $w_{i}(t)$ can be  introduced,   $\hat X(f_{n})=\sum_{m=1}^{M}{y[m]w_{i}[m]e^{-j2\pi mn/M}}$, to minimise the experienced  spectral leakage. The windowing function is defined within a signal time analysis window $\mathcal{T}_{i}=\left[\tau_{i},\tau_{i}+T_{W}\right]$  starting at the initial time instant $\tau_{i}$ and is of width $T_{W}=M/f_{US}$ such that $w_{i}(t)=w(t)$  if $t\in\mathcal{T}_{i}$ and  $w_{i}(t)=0$ if $t\notin\mathcal{T}_{i}$. The fixed tapering template $w(t)$ is  chosen from a wide variety of available windowing functions, each with distinct characteristics.   An extensive  seminal overview of windowing/tapering functions is given in \cite{harris1978use}. Clearly, if no tapering is applied then   $w_{i}(t)=1$ if $t\in\mathcal{T}_{i}$ and zero otherwise. A number of estimates over overlapping or non-overlapping time windows  are often averaged to improve the periodogram PSD\ estimation accuracy, e.g. Bartlett and Welch periodograms \cite{hayes1996statistical}. This results in adding an averaging block to Figure \ref{fig:EnergyDetector_MSS} preceding the thresholding operation where    $T_{ST}=\left\vert\cup_{j=1}^{J}\mathcal{T}_{j}\right\vert$ and $J$ is the number of averaged spectrum estimates.   

It can be noticed from (\ref{eq:EnergyDetector_MSS}) and Figure \ref{fig:EnergyDetector_MSS} that the test statistic for each subband  has its own threshold value. Quan \textit{et. al }proposed jointly choosing the threshold values  across all the system subbands, i.e. $\boldsymbol{\gamma}=\left[\gamma_{1},\gamma_{2},...,\gamma_{L} \right]^{T}$, to optimise the network opportunistic throughput  constrained by   satisfying (\ref{eq:MSS_Reliability}) and keeping the overall network interference below a certain level \cite{quan2008collaborative}. This is known by Multiple Joint Detection (MJD), which is a benchmark multiband spectrum sensing algorithm.
Several extensions of MJD emerged, e.g. MJD\ with dynamically changing sensing time is proposed in \cite{paysarvi2011optimal}. A weighted version of   (\ref{eq:EnergyDetector_MSS}), i.e. $\mathfrak{T}(\mathbf{y}_{l})\triangleq\sum_{f_{n}\in\mathcal{B}_{l}}a_{n}\vert \hat X_{W}\left( f_{n} \right) \vert^{2}$, is investigated in \cite{hossain2011wideband} to reflect correlated concurrent transmissions over the system subbands. It brings notable improvements over the original MJD when the level of correlation among the present transmissions is known in advance. Earlier joint detection  in \cite{da2007distributed} utilised a bank of feature detectors.  
\subsubsection{Multitaper Spectrum Estimation}
The Multitaper  Power Spectral Density Estimator (MT-PSDE)  achieves superior  estimation results by using carefully designed tapering functions, unlike the periodogram where a fixed windowing function $w(t)$ is deployed   \cite{zhang2011theoretical,thomson1982spectrum,haykin2009spectrum}. It utilises multiple orthogonal prototype filters with Slepian sequences or discrete prolate spherical wave function as coefficients to improve the variance of estimated spectrum without compromising  the level of incurred spectral leakage. The general  discrete-time multitaper PSD estimator is   expressed by
\begin{equation}
\hat X_{MT}(f_{n})=\frac{1}{\bar\lambda}\sum_{k=1}^{N_{MT}}\lambda _{k}\left\vert X_{k}^{ES}(f_{n})\right\vert^{2}
\end{equation}
such that the eigenspectrum $X_{k}^{ES}(f_{n})$, which deploys  the $k^{th}$ discrete orthogonal taper function $\boldsymbol{\nu}_{k}=\left[\nu_{k}[1],\nu_{k}[2],...,\nu_{k}[M]\right]^{T}$, is defined by
\begin{equation}
X_{k}^{ES}(f_{n})= \sum_{m=1}^{M}y[m]\nu_{k}[m]e^{-j2\pi m f_{n}},~~~k=1,2,...,N_{T}.
\end{equation}   
The average of the eigenspectra scaling factors $\left\{ \lambda_{k}\right\}_{k=1}^{N_{MT}}$ is given by    $\bar\lambda=\sum_{k=1}^{N_{MT}}\lambda_{k}$ where  $\lambda_{k}$ is the eigenvalue of the  eigenvector $\boldsymbol{\nu}_{k}$ such that $\mathbf{M}\boldsymbol{\nu}_{k}=\lambda_{k}\boldsymbol{\nu}_{k}$ and $(k,m)^{th}$ entry of the $M\times M$ matrix  $\mathbf{M}$ is $\sin\left(2\pi B_{C}(k-m)\right)/\pi(k-m)$. Frequently, $\lambda_{k}=1$ is assumed since the dominant eigenvalues are typically  close to unity and/or   $\bar\lambda$ is  discarded since it does not  influence the detector  success rate. In \cite{thomson1982spectrum}, an adaptive multitaper  estimator is introduced where $\left\{\lambda_{k} \right\}_{k=1}^{N_{MT}}$ scaling entries are replaced with weights that are optimised for a particular processed signal and its PSD\ characteristics. It is noted that the number of employed tapers can be bound by  $N_{MT}\leqslant\left\lfloor MB_{C} \right \rfloor$ representing the number of degrees of freedom available to control the estimation variance, e.g. $N_{MT} \in \left\{1,2,...,16 \right\}$ is recommended in \cite{haykin2009spectrum}. After obtaining the PSD\ estimate $\hat X_{MT}(f_{n})$, the energy of the signal in each of the system channels is measured similar to the classical multiband  energy detector. The  resultant values per spectral channel are  compared to predetermined threshold values to decide between $\mathcal{H}_{0,l}$ and $\mathcal{H}_{1,l}$. Haykin  in \cite{haykin2005cognitive,haykin2009spectrum} proposed a multitaper-singular-value-decomposition based cooperative scheme among CRs to measure the level of present interference and improve the quality of spectrum sensing; it is a soft-combining approach. It was shown in \cite{zhang2011theoretical} that the multitaper technique delivers premium sensing quality compared with the classical energy detector for a single CR and multiple collaborating SUs. Interestingly, a number of   digital filter-bank sensing approaches are addressed in \cite{farhang2008filter} where it was shown that the theory of multitaper spectrum estimation can be formulated within the  filterbank framework. Filterbank techniques have a long history in the DSP\ field with established solutions.

Although the MT-PSDE has nearly optimal performance and is robust against noise  inaccuracies, it has notably high computational and implementation complexities compared to the periodogram. It is noted that filterbank-based multicarrier communication
techniques, e.g.  OFDM, is widely viewed as the modulation schemes of choice for CRs \cite{axell2012spectrum,yucek2009survey}. This is due to their ability to flexibly
adapt the spectrum shape of the transmitted signal based on the available spectral opportunities.
They have built-in FFT/IFFT\ processors that can be
utilised to estimate the spectrum over wide frequency ranges making periodogram-type estimators a prime candidate. It was shown in \cite{farhang2008filter} that such filterbank  estimators  adequately adapted to a  multicarrier communication
technique  can produce accurate PSD\ estimates  similar to the MT-PSDE. The use of filter-bank-type detector has the added advantage that the filterbank can be used
for the CR sensing and transmission functionalities. 
\subsubsection{Wavelet-based Sensing}
\begin{figure*}[t] 
\centering
\includegraphics[width=0.85\linewidth]{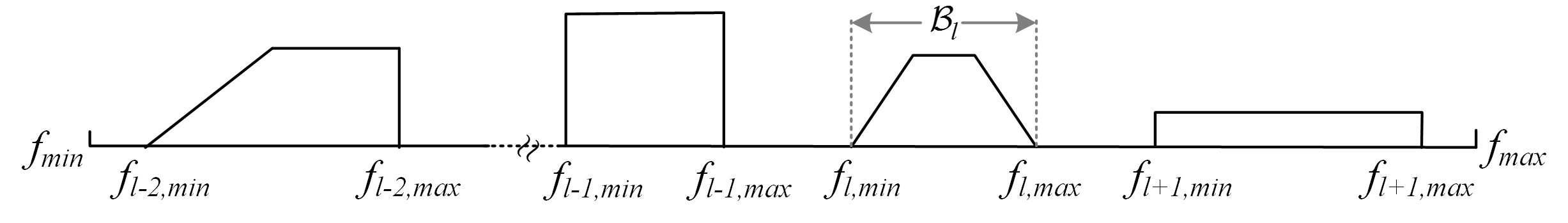}
\caption{The PSD of arbitrarily placed transmissions of varying widths in $\mathfrak{B}=\left[ f_{min},f_{max} \right]$.}
\label{fig:Wavelet_demo}
\end{figure*}

The underlying assumption in the adopted system model is that the secondary users survey spectral subbands with known locations and widths (see Section \ref{sec:SystemModel}). However, the present heterogeneous transmissions  can have different bandwidths   and occupy part(s) of the predefined subbands. In \cite{tian2006wavelet} a wavelet-based MSS approach was proposed where the present transmissions have arbitrary  positions and boundaries within the wide overseen frequency range $\mathfrak{B}$.  The wavelet-based detector models the incoming signal as a train of  transmissions each bandlimited/confined to a spectral subband with unknown position or width as demonstrated in Figure \ref{fig:Wavelet_demo}. The signal PSD within each of the active channels is assumed to be smooth, but exhibits discontinuities or singularities at the subbands' boundaries or edges.  The Continuous Wavelet Transform (CWT) of the power spectral density  of the incoming wide sense stationary signal facilitates detecting these singularities, which reveal the location and width of the present transmissions. 

In the frequency domain, the CWT can be expressed by
\begin{equation}
\mathcal{W}_{\vartheta}^{X}(f)=\mathcal{P}_{Y}(f)\ast\varphi_{\vartheta}(f)
\end{equation}
where $\mathcal{P}_{Y}(f)$ is the PSD of  the received wideband signal $y(t)$, $\ast$ is the convolution operator, 
\begin{equation}
\varphi_{\vartheta}(f)=\frac{1}{\vartheta} \varphi\left(\frac{f}{\vartheta}\right),
\end{equation}
$\varphi(f)$ is the frequency response of the wavelet smoothing function and  $\vartheta$ is the dilation factor. The latter dyadic scale can take values that are power of 2. To make the pursued discontinuities more pronounced and their characterisation easier, derivatives of the $\mathcal{W}_{\vartheta}^{X}(f)$ are used \cite{tian2006wavelet}. Such approaches are known as wavelet-modules-maxima. For example,  the local maxima of the first order derivative of $\mathcal{W}_{\vartheta}^{X}(f)$  and the zero-crossings of the second order derivative  are employed to find the active subbands boundaries and locations, respectively. Controlling $\varphi_{\vartheta}(f)$ provides additional flexibility  rendering wideband wavelet-based detection   suitable  to  dynamic spectrum structures. They also possess the  requisite properties to adaptively tune the time and frequency resolution where a high frequency resolution aids locating the subband edges. 

To  improve the edges detection procedure at the  expense of higher complexity, the wavelet multi-scale product $\mathcal{\tilde W}_{N_{W}}^{X}(f)$ or sum $\mathcal{\bar W}_{N_{W}}^{X}(f)$ can be applied where   
 \begin{equation} \label{eq:wavelet_SumAdd}
\mathcal{\tilde W}_{N_{W}}^{X}(f)=\prod_{n=1}^{N_W}{\frac{d\mathcal{W}_{2^{n}}^{X}(f)}{df} } ~~~ \text{and}~~~~ \mathcal{\bar W}_{N_{W}}^{X}(f)=\sum_{n=1}^{N_{W}}{\frac{d\mathcal{W}_{2^{n}}^{X}(f)}{df} }.
\end{equation}
The $N_W$ summands/multiplicands in (\ref{eq:wavelet_SumAdd}) are the first order derivatives of the CWT with the dyadic scales $2^{n},~n=1,2...,N_{W}$. Higher order derivatives and/or $N_{W}$ values can be exploited to enhance the multiband detection sensitivity.  

The wavelet-based  multiband detector is not robust against interferers and noise. Their impact can be minimised by appropriately setting the detection threshold  or/and increase $N_{W}$ in (\ref{eq:wavelet_SumAdd}) as suggested in \cite{zeng2011edge}.      It is noted that a digital implementation of the wavelet-based multiband detector consist of the following four blocks: 1)\ Uniform sampler $f_{US}\geqslant f_{Nyq} $, 2) discrete-time PSD estimator (e.g. periodogram that involves the scaled squared magnitude of the FFT\ of the received signal), 3)\ wavelet transform of the estimated PSD and 4)\ local maximum detector to extract the edges of the active subbands. Other advanced PSD estimation techniques, e.g. multitaper estimator, can be used and the wavelet-based detector generally has a higher complexity compared with the multiband energy detector.
\subsection{Comparing Nyquist Multiband Detection Methods}
\begin{table*}[t]
\caption{Comparison between Nyquist MSS techniques where $f_{US}\geqslant f_{Nyq}$.}
\centering
\small
\begin{tabular}{llll}
\hline \hline
\textbf{Category}&\textbf{Detector}&\textbf{Advantages}&\textbf{Limitations} \\ \hline
\multirow{3}{1.8cm}{\textbf{Sequential}}& Demodulation & Widely used and simple & Slow and complex inefficient hardware \\  &Tunable BPF& Simple and effective& Complex hardware and requires tuning\\ &Two-stage sensing &Faster high quality sensing&Complex hardware and expensive \\ \hline
\multirow{4}{1cm}{\textbf{Parallel}}& Bank of SMNSS  & High quality sensing & Complex, bulky, inefficient and expensive \\  &Multiband Energy& Simple\ and low complexity& Not\ robust against noise or interference\\ & Multitaper &Accurate and robust&Relatively high complexity  \\
& Wavelet-based &Used for unknown subbands&Not robust against noise or interference    \\
 \hline\hline
\end{tabular}
 \label{tab:NWSS_Comparison}
\end{table*}
All the above Nyquist wideband spectrum sensing techniques are compared in Table \ref{tab:NWSS_Comparison}. Sequential  sensing methods demand an analogue filtering module that permit processing one subband at a time. Whilst this facilitates sampling at relatively low rates $f_{US}\geqslant2B_{C}$, it introduces severe delays and imposes stringent space as well as power consumption requirements. On the other hand, the parallel detectors simultaneously scan all the system subbands by digitally processing the entire overseen frequency range. They use excessively high sampling rates, especially for ultra-wide bandwidths. This inevitable trade-off  motivated researchers to study novel sampling approaches  to overcome the data acquisition bottleneck of digitally accomplishing the sensing task. Such algorithms are dubbed sub-Nyquist detectors and are discussed in the remainder of this article. We divide them into two categories: compressive and non-compressive; their pros and cons are outlined in Section \ref{sec:SubNyquistComparison}.
\section{Compressive Sub-Nyquist Wideband Sensing} \label{sec:CSWSS}
Compressed Sampling (CS) or compressive sensing promote  the  reconstruction of sparse signals from a small number of their  measurements collected at significantly low sub-Nyquist rates. In \cite{duarte2011structured,foucart2013mathematical}   comprehensive overviews of  CS and its various aspects are given with an extensive list of references. Noting the low spectrum utilisation premise in CR networks, the processed wideband signal is inherently sparse in the frequency domain since only a few of the overseen subbands are concurrently active $L_{A}\ll L$. CS enables  accomplishing the wideband spectrum sensing with data acquisition  rates $\alpha\ll f_{Nyq}=2LB_{C}$. These rates are proportional to the joint bandwidth of the active subbands, i.e. $B_{A}$, representing the information rate in lieu of the entire monitored spectrum $\mathfrak{B}$. Due to the current immense interest in CS, new compressive multiband detection algorithms are regularly emerging.  In this section,  a number of widely cited and state-of-the-art CS approaches are addressed, see \cite{choi2017compressed} for a review of applications in wireless communications.

In CS, the the secondary user collects $M$ sub-Nyquist samples  of the signal of interested that encompasses the present transmissions, i.e. $x(t)$,  via 
\begin{equation} \label{eq:CS_First_Equation}
\mathbf{y}=\boldsymbol{\Phi}\mathbf{x}
\end{equation} 
where $\mathbf{y}\in\mathbb{C}^{M}$ is the samples vector and  $\mathbf{x}\in \mathbb{C}^{N}$ is the discrete-time representation of the transmissions captured at/above the Nyquist rate. The measurement matrix is $\mathbf{\Phi}\in\mathbb{C}^{M\times N}$ such that $M<N$ and noiseless observations are assumed in ($\ref{eq:CS_First_Equation}$). The signal is analysed within the time window $\mathcal{T}_{j}=\left[ \tau_{j},\tau_{j}+T_{W}\ \right]$ and the sensing time is $T_{ST}=\left\vert\cup_{j=1}^{J} \mathcal{T}_{j} \right\vert$; usually $J=1$ and $T_{ST}=T_{W}$. According to the Nyquist criterion, the  number of  Nyquist samples in $\mathcal{T}_{j}$  is given by $N=\left\lfloor T_{W}f_{US} \right\rfloor$ where $f_{US}\geqslant f_{Nyq}$. Since the CS average sampling frequency is defined by $\alpha=M/T_{W}$, the achieved reduction in the data acquisition rate is reflected in the compression ratio  $\mathcal{C}=f_{Nyq} /\alpha\approx N/M$. 

For the DFT transform basis matrix $\mathbf{D}\in \mathbb{C}^{N\times N}$, we have $\mathbf{x}=\mathbf{\Psi}^{-1}\mathbf{f}$ such that $\mathbf{\Psi}=\mathbf{D}$ and $\mathbf{\Psi}^{-1}$ is the inverse DFT matrix. The sparse vector $\mathbf{f}\in \mathbb{C}^{N}$, which is the frequency representation of the present transmissions, is characterised by   $\left\Vert \mathbf{f}\right\Vert_{0}\leq K_{S}$ where  $K_{S}$ is the sparsity level and $K_{S}\ll N$. The $\ell_{0}$ "norm"  $\left\Vert \mathbf{f}\right\Vert_{0} $ is defined as the number of nonzero entries in $\mathbf{f}$. The relationship between the compressed samples and the signal spectrum can be expressed by
\begin{equation} \label{eq:CS_basic_equation}
\mathbf{y}=\mathbf{\Upsilon}\mathbf{f}
\end{equation}
and $\mathbf{\Upsilon}=\mathbf{\Phi}^{-1}\mathbf{\Psi}$ is the sensing matrix. With CS, we can  exactly recover  $\mathbf{f}$ from the $M<N$ noise-free linear measurements, e.g. $M=\mathcal{O}\left( K_{S}\log\left( N/K_{S} \right)\right)$ suffices, furnishing substantial reductions in the sampling rate. This is facilitated by the sparsity constraint on $\mathbf{f}$, which makes solving the underdetermined system of linear equations in (\ref{eq:CS_basic_equation}) feasible with close form performance guarantees. Such guarantees impose certain conditions on the measurement matrix $\boldsymbol{\Phi}$ or more generally on the sensing matrix $\mathbf{\Upsilon}$ \cite{duarte2011structured,foucart2013mathematical}. Recovering $\mathbf{f}$ from $\mathbf{y}$ entails solving an optimisation whose basic statement is given by
\begin{equation} \label{eq:CS_Optimisation}
\mathbf{\hat f}=\underset{\mathbf{f}\in\Sigma_{K_{S}}}{\arg\min}\left\Vert \mathbf{f}\right\Vert_{0}~~\text{s.t.}~~\mathbf{y}=\mathbf{\Upsilon}\mathbf{f}
\end{equation}
and $\Sigma_{K_{S}}=\left\{ \mathbf{f}\in\mathbb{C}^{N}:\left\Vert \mathbf{f}\right\Vert_{0}\leqslant K_{S} \right\}$. Whilst tackling (\ref{eq:CS_Optimisation}) directly has combinatorial computational complexity, a plethora of  effective and efficient sparse recovery techniques were developed, e.g.  convex-relaxation, greedy, Bayesian, non-convex and brute-force algorithms, see \cite{tropp2010computational,duarte2011structured,eldar2012compressed} for an overview. It is noted that noise can be added to (\ref{eq:CS_basic_equation}) to represent noisy signal observations, i.e. $\mathbf{y}=\boldsymbol{\Phi}\mathbf{x}+\boldsymbol{\varepsilon}$ and $\boldsymbol{\varepsilon}$ is the additive measurements noise vector. Sparsifying\ basis/frames other than the DFT\ can be adopted to promote the sparsity property. Motivated by the wavelet-based detection to identify the edges of the active spectral channels, the earliest papers on CS-based MSS utilise  $\boldsymbol{\Psi}=\mathbf{\Gamma}\mathbf{D}\mathbf{W}$ where  $\mathbf{\Gamma}$ is an $N\times N$ differentiation matrix, $\mathbf{W}$ applies the wavelet-smoothing operation and $\mathbf{x}$ is a realisation/estimate of the discrete-time signal autocorrelation function \cite{tian2007compressed,polo2009compressive}.      

\begin{figure}[t] 
\centering
\includegraphics[width=0.8\linewidth]{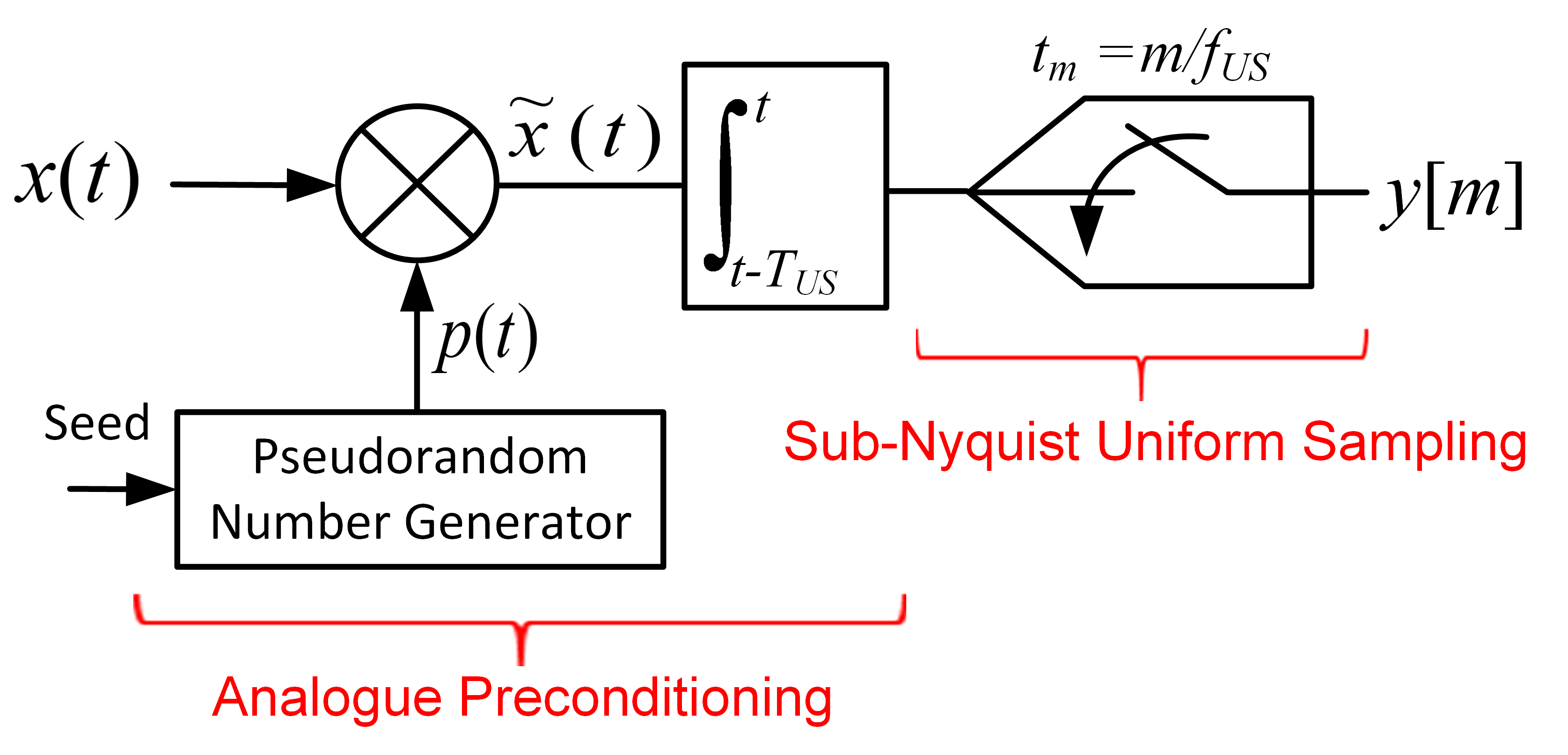}
\caption{Block diagram of the random demodulator sub-Nyquist CS sampler \cite{tropp2010beyond}.}
\label{fig:RandomDemodulator}
\end{figure}

In\ the majority of the theoretical treatments of the CS problem in (\ref{eq:CS_basic_equation}),  the measurement matrix  $\mathbf{\Phi}$ is assumed to be random and drawn from a sub-Gaussian distribution \cite{duarte2011structured}. This implies the availability of signal measurements collected at/above the Nyquist rate  as in \cite{tian2007compressed}, which defies the objective of sub-Nyquist sampling. On the contrary, the compressed samples  in $\mathbf{y}$  should be collected directly from the received wideband analogue signal $y(t)$ without the need to capture the Nyquist samples first. This can be achieved by using the Analogue to Information Converter (AIC) shown in Figure \ref{fig:RandomDemodulator};  it is known as the Random Demodulator (RD). The incoming signal is multiplied by a pseudorandom chipping sequence switching at a rate of $f_{p}=1/T_{p}\geqslant f_{Nyq}$ followed by an integrator and a low rate (sub-Nyquist) uniform sampler of period $T_{US}$. Generating a fast chipping sequence can be easily achieved in practice unlike sampling at excessively high  rates with specialised ADCs. RD is devised to process multitone signals made up of pure sinusoids, e.g. located at multiples of an underlying resolution frequency $\Delta_{f}$. Whilst the integrator can be implemented using a Low Pass Filter (LPF),  $f_{US}$ is proportional to $\mathcal{O}\left( K_{S}log\left(2B/K_{S}\Delta_{f} \right) \right)$  and $K_{S}$ is the number of present pure tones in the double-sided processed bandwidth $2B$. Several other lower bounds on $f_{US}\ll f_{Nyq}$ for $K_{s}\ll2B/\Delta_{f}$ are derived in \cite{tropp2010beyond}. Although compressed model-based spectrum estimation algorithms are proposed in \cite{duarte2013spectral} to improve the random demodulator performance, it remains unsuitable for multiband signals with  each transmission occupying a particular system subband. The RD\ is also very sensitive to modeling mismatches levying fine hardware-software calibration procedures \cite{pankiewicz2013model,duarte2011structured}. An alternative  AIC approach employs a slow uniform sampler that randomly skips samples. It is equivalent to using $\mathbf{\Phi}$ that randomly selects rows out of $N\times N$ unity matrix \cite{duarte2011structured}. This  results in a nonuniform sampling scheme , i.e. random sampling on grid, discussed  in Section \ref{sec:NCSWSS}.

\begin{figure}[b] 
\centering
\includegraphics[width=1\linewidth]{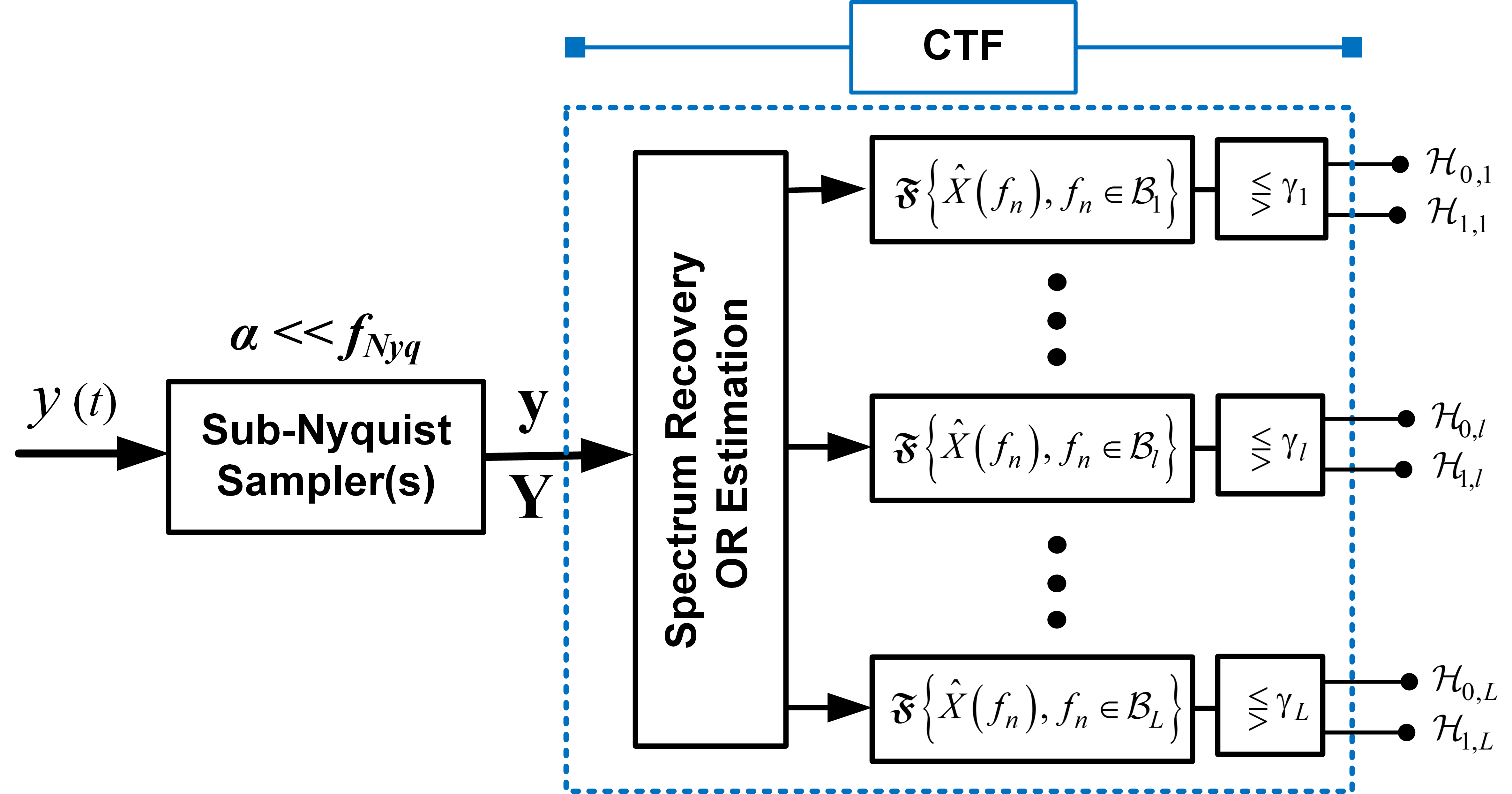}
\caption{Block diagram of several  sub-Nyquist MSS techniques; $\hat X(f)$ is the recovered/estimated spectrum and $\alpha$ is the data acquisition rate such that $\alpha \ll f_{Nyq}$.}
\label{fig:Sub_Nyquist_MSS}
\end{figure}

\begin{figure}[t] 
\centering
\begin{subfigure}[t]{1\linewidth}
\centering
\includegraphics[width=0.8\linewidth]{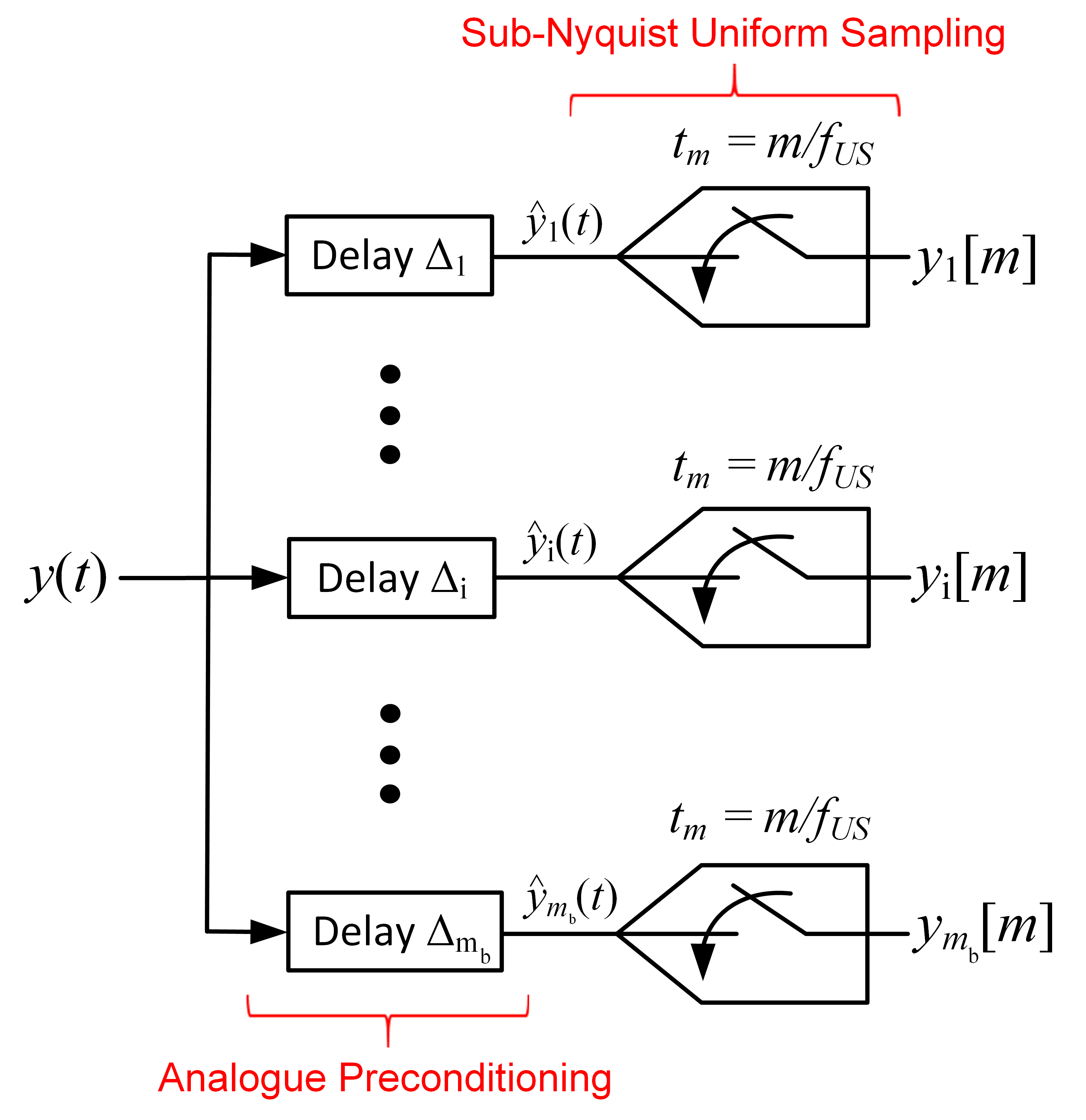}
\caption{}
\label{fig:PNS}
\end{subfigure}
\begin{subfigure}[t]{1\linewidth}
\centering
\includegraphics[width=0.85\linewidth]{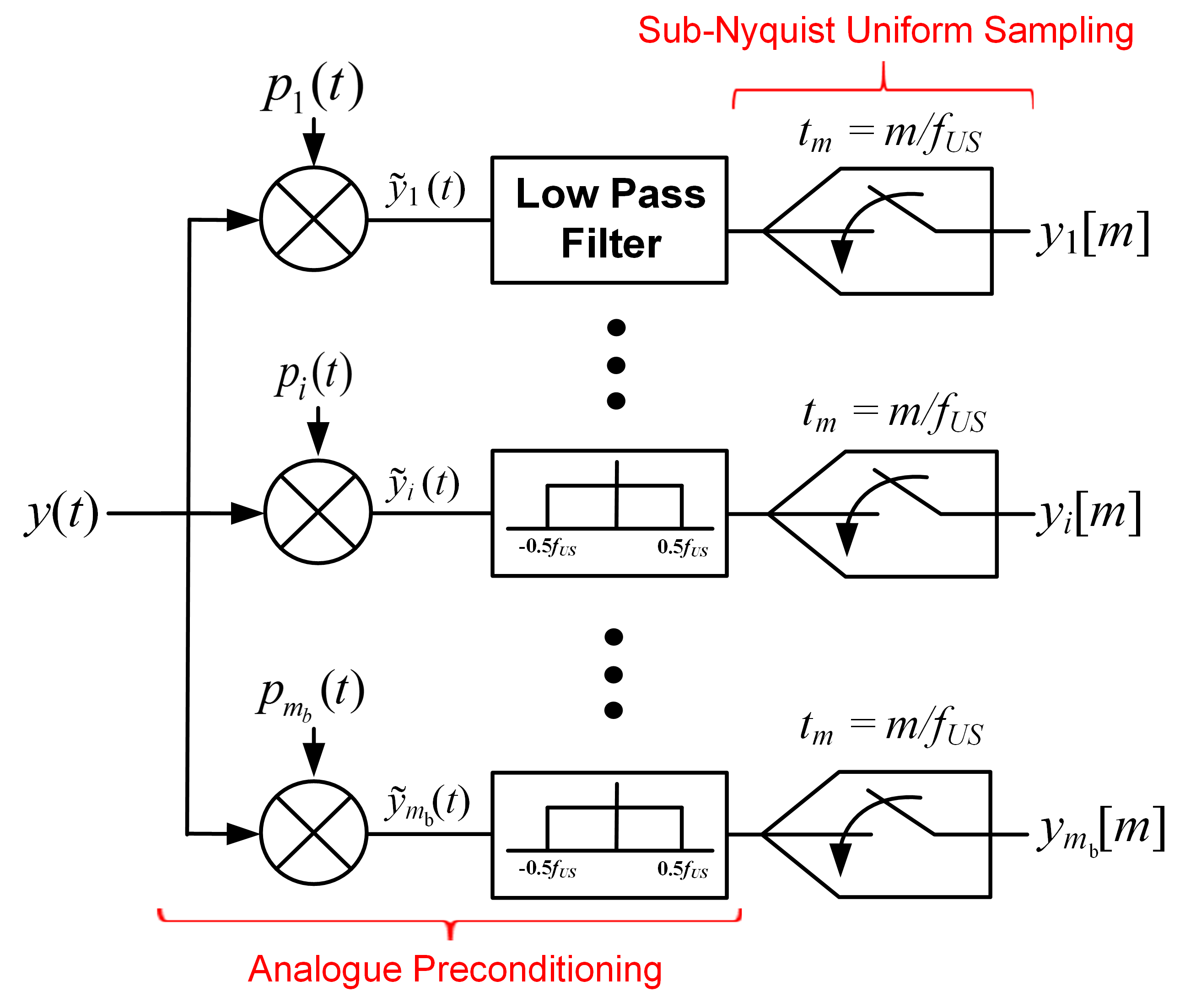}
\caption{}
\label{fig:MWC}
\end{subfigure}
\caption{Block diagrams of two sub-Nyquist CS\ samplers.(a) Mulitcoset sampling \cite{mishali2009blind}. (b) Modulated wideband converter \cite{mishali2010theory}. }
\label{fig:PNS_MWC}
\end{figure}

Assuming the availability of $\mathbf{y}$ and that DFT is the sparsifying basis, multiband detection can be performed via 
\begin{equation} \label{eq:EnergyDetector_CS}
\mathfrak{T}(\mathbf{y})\triangleq\sum_{f_{n}\in\mathcal{B}_{l}}\vert \hat X( f_{n} ) \vert^{2} \overset{\mathcal{H}_{0,l}}{\underset{\mathcal{H}_{1,l}}{\lesseqgtr}}\gamma_{l}, ~~~~l=1,2,...,L.
\end{equation}
The estimated spectrum $\hat X\left( f_{n} \right)$ from the compressed samples in  $\mathbf{y}$ refers to the entry in  vector $\mathbf{\hat f}$ representing the frequency point $f_{n}$. We recall that  $\mathbf{\hat f}$ is obtained from solving  (\ref{eq:CS_basic_equation})
using a sparse approximation algorithm. Therefore, the detector in (\ref{eq:EnergyDetector_CS}) is a sub-Nyquist multiband energy detector whose sampling rate is $\alpha\ll f_{Nyq}$. The number of frequency points per subband depends on spectral resolution dictated by the number of  Nyquist samples  $N$ in the signal analysis time window where $N=\left\lfloor T_{W}f_{Nyq}\right\rfloor$. Additionally, $\boldsymbol{\Psi}$ can be a frame to increase the spectral resolution where $\mathbf{D}\in \mathbb{C}^{N\times \hat N}$ and $\hat N>N$. Figure \ref{fig:Sub_Nyquist_MSS} displays the block diagram of several  sub-Nyquist wideband spectrum sensing techniques. For the test statistic $\mathfrak{T}(\mathbf{y})$ in (\ref{eq:EnergyDetector_CS}), we have $\mathfrak{F}\{ \hat X\left( f_{n} \right), f_{n}\in \mathcal{B}_{l}\}=\sum_{f_{n}\in \mathcal{B}_{l}}\vert \hat X\left( f_{n} \right) \vert^{2}$. Two simply CS-based sensing techniques for a generic $\mathbf{\Phi}$ can be expressed as follows:
\begin{enumerate}
\item 
CS Method 1 (CS-1) \cite{axell2011unified,wang2012sparsity}: the estimated spectral points that belong to each channel are grouped to calculate $\mathfrak{T}\left( \mathbf{y} \right)$  and $N=\left\lfloor T_{ST}f_{Nyq}\right\rfloor$. 
\item
CS Method 2 (CS-2)\cite{fanzi2011distributed}: Unlike CS-1, $L$ DFT points are calculated, i.e. $\mathbf{f}\in\mathbb{C}^{L}$, and one DFT point is recovered per monitored subband. The sensing time $T_{ST}$ is divided into  $\left\lfloor T_{ST}/T_{W}\right\rfloor$ sub-windows each of width $T_{W}=L/f_{Nyq}$. Within each of these partitions an $\mathbf{\hat f}$ solution is determined where  $\mathbf{\Phi}\in\mathbb{C}^{M\times L}$. To improve the estimation accuracy, $J$ of the CS   estimates are averaged over $T_{ST}=JT_{W}$. This emulates the scenario of $J$ spatially distributed CRs collaboratively overseeing $\mathfrak{B}$.
\end{enumerate}
Compressive  samplers that are particularly suitable for  multiband signals include Mulitcoset Sampling (MCS) and Modulated Wideband Converter (MWC) shown in Figure \ref{fig:PNS_MWC}. They consist of a bank of  $m_{b}$  samplers collecting uniformly distributed measurements  at sub-Nyquist rates. The resulting measurements vectors for all the data acquisition branches can be combined in the $M\times m_{b}$ matrix $\mathbf{Y}=\left[\mathbf {y}_{1}, \mathbf{y}_{2},...\mathbf{y}_{m_{b}} \right]$, dubbed Multiple Measurements Vector  (MMV). Similarly, the targeted vector from each bank is stacked in  $\mathbf{F}=\left[\mathbf {f}_{1}, \mathbf{f}_{2},...\mathbf{f}_{m_{b}} \right]$ and 
\begin{equation} \label{eq:MMV_CS_basic_equation}
\mathbf{Y}=\mathbf{\Upsilon}\mathbf{F}
\end{equation}
where all columns $\mathbf{F}$ have the same sparsity pattern. Several sparse recovery algorithms can be applied  to solve the MMV case, e.g. minimum variance distortionless response and range of extended greedy techniques \cite{duarte2011structured}. Next, we briefly  describe MCS, MWC and   the Multirate Asynchronous   sub-Nyquist Sampling (MASS) systems highlighting their main features. 
\subsection {Multicoset Sampling and Blind Spectrum Sensing}
\begin{figure}[b] 
\centering
\includegraphics[width=1\linewidth]{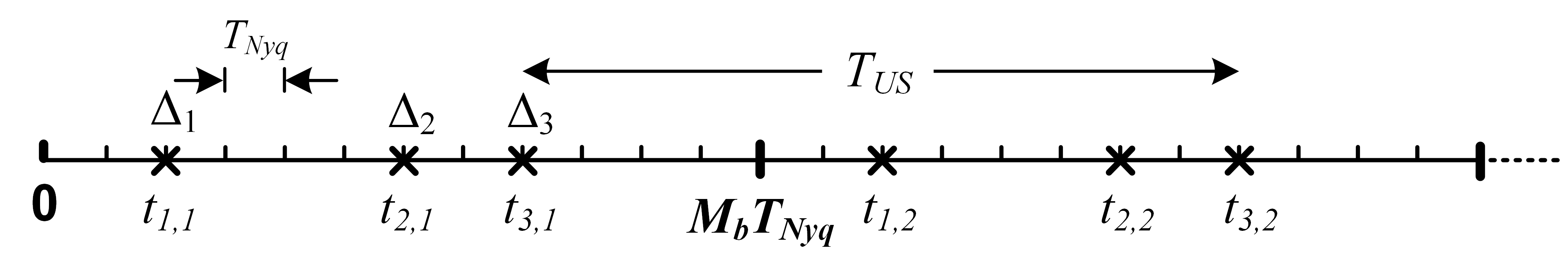}
\caption{MCS with $M_{b}=12$ and $m_{b}=3$. Uniform samplers 1, 2 and 3  capture the measurements  sets $\left\{y(t_{1,1}),y(t_{1,2}),...\right\}$, $\left\{y(t_{2,1}),y(t_{2,2}),...\right\}$ and $\left\{y(t_{3,1}),y(t_{3,2}),...\right\}$,respectively.}
\label{fig:PNS_Seq}
\end{figure}

The MCS, otherwise known as periodic nonuniform sampling, was  proposed by Feng \cite{Feng1997PNS}  as a sub-Nyquist data acquisition approach that promotes the accurate reconstruction of deterministic multiband signals. MCS selects a number of measurements out of an underlying grid whose equidistant points are separated by a period less or equal to   $T_{Nyq}=1/f_{Nyq}=1/2B$. The uniform grid is divided into $M_{b}$ blocks of uniformly distributed samples. In each block, the fixed set $\mathcal{D}$ of length $\left\vert \mathcal{D} \right\vert= m_{b}<M_{b}$   denotes the indices of the retained samples in the   block; the remaining $M_{b}-m_{b}$ samples are discarded. The set  
\begin{equation}
\mathcal{D}=\left\{ \varrho_{1},\varrho_{2},...,\varrho_{m_{b}} \right\},~~~0\leqslant \varrho_{1}<\varrho_{2}<...<\varrho_{m_{b}}\leqslant M_{b}-1
\end{equation}
is referred to as the sampling pattern. The samples of the received wideband signal $y(t)$   in the $i^{th}$ branch/coset are given by 
\begin{equation} \label{eq:PNS_Sequence}
y_{i}[m]=y\left(\left(mM_{b}+\varrho_{i}\right)T_{Nyq}\right),~~~m \in \mathbb{Z}
\end{equation}
  and they are all shifted by the delay $\Delta_{i}=\varrho_{i}T_{Nyq}$ with respect to the origin. An example of  a MCS\ sequences is shown in Figure \ref{fig:PNS_Seq} with  $M_{b}=12$   and  $m_{b}=3$  instants are selected.  

The multicoset sampling scheme  can be implemented by a bank of $m_{b}$  uniform samplers each running at an acquisition rate of $f_{US}=f_{Nyq}/M_{b}$  and preceded by a delay as depicted in Figure \ref{fig:PNS}. The  rate of the uniform samplers $f_{US}$ in each of the system branches should fulfill \begin{equation}\label{eq:PNS_resolution}
f_{US}^{MCS}\geqslant\ B_{C}.
\end{equation}
and typically $f_{US}= B_{C}$. The MCS  average sampling rate is  $\alpha_{MCS}=m_{b}f_{Nyq}/M_{b}$ , which  is lower than Nyquist for $m_{b}<M_{b}$. For signals with unknown spectral support (i.e  locations of the active subbands in $\mathfrak{B}$ are unknown), the MCS minimum permissible  rate  is  
\begin{equation}\label{eq:PNS_SamplingRateLimit}
\alpha_{MCS}\geqslant 4L_{A}B_{C},
\end{equation} 
recalling that $L_{A}$ is the maximum number of concurrently active spectral channels \cite{Feng1997PNS,mishali2009blind}. Accordingly, the minimum number of required multicoset sampling channels  is $m_{b}^{MCS}\geq4L_{A}$. The rate in (\ref{eq:PNS_SamplingRateLimit})
is twice the Landau rate $f_{Landau}=2L_{A}B_{C}$, which is the theoretical minimum sampling rate that permits the exact recovery of the continuous-time signal from its measurements as per Landau theorem \cite{landau1967necessary}; maximum spectrum occupancy is assumed \cite{duarte2011structured}.

From (\ref{eq:PNS_Sequence}), it can be  shown that the spectrum of the incoming signal, i.e. $Y(f)=\int_{-\infty}^{+\infty}y(t)e^{-j2\pi ft}dt$, and that attained from the samples in the $i^{th}$ branch are related via
\begin{align} \label{eq:PNS_DTFT}
Y^{d}_{i}(f)=\sum_{m=-\infty}^{+\infty}{y_{i}[m]e^{-j2 \pi f \left( (mM_{b}+\varrho_{i})T_{Nyq}\right)}}\nonumber \\=B_{C}\sum_{n=-L}^{L-1}{Y\left(f+nB_{C}\right)e^{j2\pi \varrho_{i}n/M_{b}}}
\end{align}   
such that $f\in \mathcal{\hat B}$, $\mathcal{\hat B}=\left[0,B_{C}\right]$, $y(t)$ is real and $M_{b}=2L$. It is noted that (\ref{eq:PNS_DTFT}) incorporates a frequency point per monitored spectral subband. Subsequently, we can write 
\begin{equation} \label{eq:PNS_IntialCSFormulation}
\mathbf{y}(f)=\mathbf{A}\mathbf{x}(f),~~~ f\in\mathcal{\hat B}
\end{equation}
where $\mathbf{y}(f)$ is a vector of length $m_{b}$ combining all $\left\{Y^{d}_{i}(f)\right\}_{i=1}^{m_{b}}$. The $(i,k)^{th}$ entry of the  $m_{b}\times M_{b}$ matrix $\mathbf{A}$ is  $B_{C}e^{j2\pi \varrho_{i}k/M_{b}}$ and $\mathbf{x}(f)$ contains the $M_{b}$ sought unknowns for each frequency $f$, i.e. $\left\{Y\left(f+nB_{C}\right)\right\}_{n=-L}^{L-1}$. To fully recover the signal's infinite resolution spectrum, equation (\ref{eq:PNS_IntialCSFormulation}) has to be solved for  $f\in\mathcal{\hat B}$  where $\mathbf{x}(f)$ is sparse since only few of the subbands are simultaneously active. The resulting infinite number of equations in  (\ref{eq:PNS_IntialCSFormulation}) for  $\{\mathbf{y}(f), f\in\mathcal{\hat B}\}$ is referred to as an Infinite Measurement Vector (IMV).  It is reasonable to assume that the  set of all vectors $\{\mathbf{x}(f), f\in\mathcal{\hat B}\}$   have common support since the non-zero values for each $f$\ pertain to the same active subbands \cite{duarte2011structured}. Sampling patterns that permit the exact recovery of $Y(f)$  and thereby the underlying continuous-time signal from  its  multicoset samples were studied in \cite{venkataramani2000perfect}; searching  all possibilities is a combinatorial problem. 

\begin{figure}[t] 
\centering
\includegraphics[width=1\linewidth]{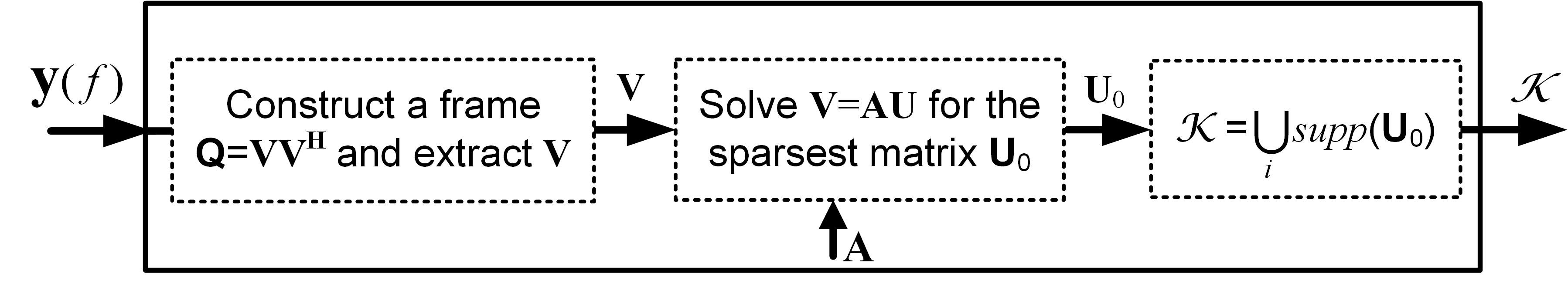}
\caption{Block diagram of the CTF module utilised in the MCS and MWC systems \cite{mishali2010theory}.}
\label{fig:CTF}
\end{figure}

In \cite{mishali2009blind}, signal reconstruction from its multicoset samples was introduced within the compressed sampling framework. Let the support  set defined by $\mathcal{\mathcal{K}}=\{ i:Y(f+iB_{C})\neq0,f\in\mathcal{\hat B},i=-L,-L+1,...,L-1 \}$ be the indices of the $K_{S}\leqslant 2L_{A}$ active subbands and $\left\vert\mathcal{\mathcal{K}}\right\vert=K_{S}$ (positive and negative frequencies  are included). For the wideband spectrum sensing problem, unveiling the unknown $\mathcal{\mathcal{K}}$  suffices. The Continuous To Finite (CTF) algorithm/block robustly  detects $\mathcal{K}$ and reduces the IMV to  a MMV of finite dimensions via  
\begin{equation} \label{eq:CTF_Formulation}
\mathbf{V}=\mathbf{A}\mathbf{U}
\end{equation}
where frame  $\mathbf{Q}=\mathbf{V}\mathbf{V}^{H}$ can be constructed using  $\mathbf{Q}=\sum_{f\in \mathcal{\hat B}}\mathbf{y}(f)\mathbf{y}^{H}(f)$ from roughly $2K_{S}$ snapshots of $\mathbf{y}(f)$ \cite{mishali2009blind,duarte2011structured}.  The decomposition performed to  attain $\mathbf{V}$ from $\mathbf{Q}$ minimises the impact of the present noise in  the received signal where $y[m]=x[n]+w[n]$ and $x[n]$\ represents the active transmissions. The received signal $y(t)$ in (\ref{eq:PNS_Sequence}) is presumed to consist of noiseless transmissions to simplify the notation. Assuming that the conditions in (\ref{eq:PNS_resolution}) and (\ref{eq:PNS_SamplingRateLimit}) are satisfied, it is shown in \cite{mishali2009blind} that the underdetermined system in (\ref{eq:CTF_Formulation}) has a unique solution matrix $\mathbf{U}_{0}$ with the minimal number of non-identically zeros rows. The indices of the latter rows  coincide with the support set $\mathcal{K}$, which in turn reveals the occupied  system subbands. A block diagram of the continuous to finite module is depicted in Figure \ref{fig:CTF}. Its input  $\mathbf{y}(f)$ is the DTFT or FFT  of the data samples produced in all the MCS branches. As well as accomplishing multiband detection, the CTF module plays a critical role in recovering the detected transmissions at the secondary user for postprocessing tasks, if required \cite{mishali2009blind,duarte2011structured}. Other approaches to the IMV problem exist, e.g.   MUSIC-type\ algorithms  \cite{Feng1997PNS}. An interesting discussion on the correlation and relationship between  the initial work on MCS  and CS is given in \cite{bresler2008spectrum}.

Therefore, the MCS facilitates wideband spectrum sensing with substantially low sub-Nyquist sampling rates and $\alpha_{MCS}\ll f_{Nyq}$ in  the low spectrum utilisation regime, i.e. $L_{A}\ll L$ . The CTF algorithm can also significantly reduce the sub-Nyquist MSS\ computational complexity as the sensing matrix $\mathbf{A}$  in (\ref{eq:CTF_Formulation}) is of fixed dimension $m_{b}\times M_{b}$  for an infinite spectral resolution.  On the contrary, directly applying  CS with DFT sparsifying basis/frame to the detection problem, e.g. CS-1, yields sensing matrices whose dimensions grow proportional to the  sought resolution. Nevertheless, implementing  multicoset sampling involves accurate time interleaving among the $m_{b}$ relatively slow uniform samplers that directly process the incoming wideband signal. This necessitates a high bandwidth track and hold sampling device, which is difficult  to build and might require specialised fine-tuned ADCs \cite{mishali2010theory}. Maintaining accurate time shifts in the order of $1/f_{Nyq}$ as per (\ref{fig:PNS_Seq}) is challenging to realise in hardware, especially for $f_{Nyq}$ in excess of several GHz. Inaccurate-shifts can notably degrade the quality of the spectrum recovery.  

\subsection {Modulated Wideband Converter}
The MWC data acquisition system depicted in Figure \ref{fig:MWC} aims to exploit advances in the CS field and circumvents the MCS drawbacks \cite{mishali2010theory}. It is comprised of $m_b$ bank of modulators and low-pass filters. In the $i^{th}$  branch, $i=1,2,..,m_b$, the received signal $y(t)$ is multiplied by a periodic chipping waveform $p_{i} (t)$ of period $T_p$. The modulated output $\tilde y_{i}(t)=y(t)p_{i}(t)$ is low-pass filtered and subsequently sampled at a sub-Nyquist sampling rate equal to $f_{US}=1/T_{US}$; the analogue filter cut-off frequency is $0.5f_{US}$. To be able to recover the spectrum of the sampled signal or identify the active subbands, a typical MWC configuration imposes
\begin{equation}\label{eq:MWC_Req1}
f_{b}=\frac{1}{T_{p}}\geqslant B_{C},~f_{US}^{MWC}\geqslant f_{p},~\alpha_{MWC}\geqslant 4L_{A} B_{C} 
\end{equation}
as the   frequency of the periodic waveform, uniform sampling rate per branch and the modulated wideband converter overall average sampling rate, respectively. This implies that
\begin{equation}\label{eq:MWC_branches}
m_{b}^{MWC}\geqslant4L_{A}
\end{equation}
sampling channels are required for $f_{US}=B_{C}$. With  MWC, the number of deployed modulators $m_{b}$ can be reduced at the expense of increasing the uniform sampling rate per branch \cite{mishali2010theory}. 
It can be shown that the DTFT\ of the samples in the $i^{th}$ branch is given by
\begin{align}\label{eq:MWC_DTFT}
Y^{d}_{i}(f)&=\sum_{m=-\infty}^{+\infty}{y_{i}[m]e^{-j2 \pi fmT_{US}}}\nonumber \\
&=\sum_{n=-L_{0}}^{L_{0}}{c_{i,n}Y\left(f-nf_{p}\right)},~~ f\in \mathcal{\tilde B}
\end{align}
where $\mathcal{\tilde B}=\left[-0.5f_{US},0.5f_{US}\right]$ is the range dictated by the low pass filter. The coefficients $c_{i,n}$ in (\ref{eq:MWC_DTFT}) are the Fourier expansion coefficients of  $p_{i}(t)$ such that $c_{i,n}=f_{p}\int_{0}^{T_{p}}p_{i}(t)e^{-j2\pi nt/T_{p}}dt$. To ensure that   the $2L$ overseen subbands (including negative frequencies) are present in $Y^{d}_{i}(f)$, we have $L_{0}=\left\lceil 0.5T_{p}(f_{Nyq}+f_{US}) \right\rceil-1=L$  for $f_{p}=B_{C}$. The mixing periodic function $p_{i}(t)$ should have a transition speed $f_{Tran}\gtrsim f_{Nyq}$ within $T_{p}$, i.e. $T_{p}$ is divided into $M_{PS}\geqslant2L+1$ slots within which the chipping sequence can alter its  values. Equation (\ref{eq:MWC_DTFT}) leads to 
\begin{equation}\label{eq:MWC_CSFOrmulation}
\mathbf{y}(f)=\mathbf{B}\mathbf{x}(f),~~~  f\in\mathcal{\tilde B}
\end{equation} 
where $\mathbf{y}(f)$ is a vector of length $m_{b}$ with the $i^{th}$ element being $Y^{d}_{i}(f)$. The $(i,n)^{th}$ entry of  $m_{b}\times 2L_{0}+1$ matrix $\mathbf{B}$ is the $c_{i,n}$ Fourier coefficient and the entries of vector $\mathbf{x}(f)$ are the sought $\{ Y\left(f-nf_{p}\right)\}_{n=-L_{0}}^{L_{0}}$ for $f\in\mathcal{\tilde B}$. It is noticed that the multicoset sampling formulation in (\ref{eq:PNS_IntialCSFormulation}) and that of the MWC in  (\ref{eq:MWC_CSFOrmulation}) are very similar. Thus, the CTF\ algorithm is also utilised in the MWC system and similar performance guarantees are derived in \cite{mishali2010theory}. We note  that  the CTF\ in MWC is less computationally demanding compared with  MCS. Constructing frame $\mathbf{Q}$ in MWC does not involve interpolating the slow sub-Nyquist data streams where $\mathbf{Q}=\int_{f \in \mathcal{\tilde B}} \mathbf{y}(f)\mathbf{y}^{H}(f)df=\sum_{m=-\infty}^{+\infty}\mathbf{\bar y}[m]\mathbf{\bar y}^{T}[m]$ and $\mathbf{\bar y}[m]=[y_{1}[m],y_{2}[m],...,y_{m_{b}}[m]]^{T}$.        

In principle,  any periodic function, i.e.  $p_{i}(t)=p_{i}(t+T_{p})$, with low-mutual correlation and high-speed transitions exceeding $f_{Nyq}$ is admissible. A popular choice is the sign altering function with $M_{PS}$ sign intervals within $T{p}$; other sign patterns can be used \cite{duarte2011structured}. This flexibility is crucial and the high speed chipping signals can be easily generated using a standard shift-register. Synchronising   the $m_{b}$ uniform samplers can be enforced by driving all samplers from a single master-clock.  The low-pass  filters in MWC 
do not have to be ideal since mismatches, e.g. rugged filter response(s),  can be compensated for in the digital domain \cite{duarte2011structured}.
 They also limit the bandwidth of the digitised signal in  each of the system branches  to approximately $\pm B_{C}$, i.e. off-the-shelf ADCs can be employed. In general, the MWC is  robust
against noise and model mismatches compared with RD and MCS. Similar to the latter, the lower bound on the modulated wideband converter  sub-Nyquist sampling rate  implies $\alpha_{MWC}\ll f_{Nyq}$ for low spectrum occupancy. Since  the objective here is MSS, recovering the signal's spectral support using the CTF block is sufficient.   Finally, it can be noticed that the MWC  consists of a bank of   random demodulators. The relationship between the RD and MWC is thoroughly treated in \cite{lexa2012reconciling}.  
\subsection {Multirate Asynchronous Sub-Nyquist   Sampler}
\begin{figure}[t]  
\centering
\includegraphics[width=1\linewidth]{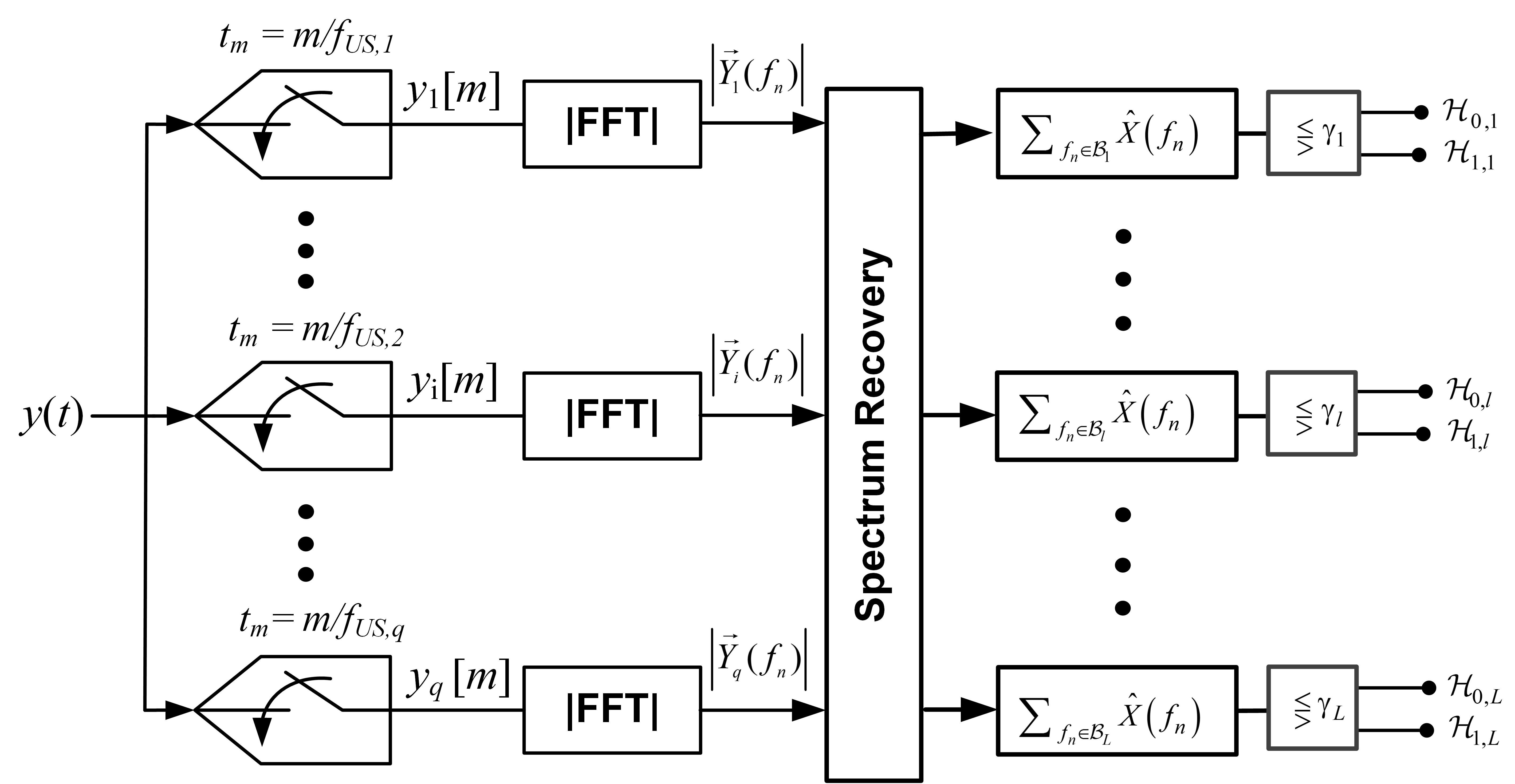}
\caption{Block diagram of the MASS-based sub-Nyquist multiband detection system \cite{sun2013wideband}.}
\label{fig:MASS}
\end{figure}

The MASS system in Figure \ref{fig:MASS}  was proposed in \cite{sun2013wideband} as a CS-based multiband detection approach. It  utilises a bank of $q$  uniform samplers each running at a distinct sub-Nyquist sampling rate  $f_{US,i}$, $i=1,2,...,q$. Most notably, the  MASS does not impose synchronisation among its $q$ channels, i.e. asynchronous. 
Let $T_{ST}$\ be the width of the signal observation window (in seconds) and $\mathcal{M}=\left\{ M_{i} \right\}_{i=1}^{q}$ is the set encompassing the number of captured uniform samples in all the system branches. Thus, the multirate approach average sampling rate is: $\alpha_{MASS}=\sum_{i=1}^{q}f_{US,i}=M/T_{ST}$ and $M=\sum_{i=1}^{q}M_{i}$ is the  total number of collected measurements. The aim is to achieve $M\ll N$  such that $N=\left\lfloor T_{ST}f_{Nyq}\right \rfloor$ denotes the number of Nyquist samples and $f_{Nyq}=2B$. In a typical MASS configuration, we have 
\begin{equation}\label{eq:MASS_OperatingCoditions}
M_{i}=\tilde \rho_{i}\sqrt{N},~M_{i}\in\mathcal{M},~ \tilde\rho_{i}\in\mathbb{P},~M_{1}<M_{2}\cdots<M_{q}
\end{equation} 
where $\mathbb{P}$ is the set of prime numbers and the chosen measurements numbers in  $\mathcal{M}$    usually have consecutive values. Both  $\mathcal{M}$ and $q$ dictate the achieved compression ratio   $N/M$. 

According to Figure \ref{fig:MASS}, discrete Fourier transform or FFT\ is applied  in each sampling branch and the absolute value of the resultant  is taken.  Let $Y^{d}_{Nyq}(f)=\sum_{n=0}^{N-1}y(nT_{Nyq})e^{-j2\pi f nT_{Nyq}}$ be the discrete-time Fourier transform of the incoming wideband signal sampled at the Nyquist rate and $T_{Nyq}=1/f_{Nyq}$. It can be shown that the relationship between the DFT\ from  the $M_{i}$  measurements  in the $i^{th}$ branch  and $Y^{d}_{Nyq}(f)$  can be expressed by
\begin{equation} \label{eq:MASS_DFT}
Y^{d}_{i}[m]=\frac{M_{i}}{N}\sum_{n=-\left\lfloor0.5N\right\rfloor}^{\left\lfloor0.5N\right\rfloor}{Y^{d}_{Nyq}[n]\sum_{l=-\infty}^{+\infty}}\delta[n-(m+lM_{i})],
\end{equation}
such that $m\in \left[-\left\lfloor0.5M_{i}\right\rfloor,\left\lfloor0.5M_{i}\right\rfloor  \right]$ and $\delta[n]$ is a Kronecker delta.    

By expressing (\ref{eq:MASS_DFT}) in a matrix format, we can write
\begin{equation} \label{eq:MASS_CSForm1}
\mathbf{y}_{i}=\mathbf{C}_{i} \mathbf{f}
\end{equation}
where $\mathbf{y}_{i} \in \mathbb{C}^{M_{i}\times1}$ is output of the FFT\ block in the $i^{th}$ sampling channel, $\mathbf{C}_{i}\in\mathbb{R}^{M_{i}\times N}$ is the sensing matrix and   $\mathbf{f}\in\mathbb{C}^{N}$ is the sought signal spectrum. The $(m,n)^{th}$ entry of  $\mathbf{C}_{i}$ is given by $\frac{M_{i}}{N}\sum_{l=-\infty}^{+\infty}{\delta[n-(m+lM_{i})]}$. This implies that each column of the  sensing matrix contains only one non-zero value equal to $M_{i}/N$ and in each row the maximum number of non-zero entries is $\left\lceil N/M_{i}\right\rceil$. Whilst $f_{Nyq}$ guarantees no aliasing is present in  $Y^{d}_{Nyq}(f)$, the sub-Nyquist  rates   in each of the MASS\ branches wrap the wideband signal
spectrum content onto itself in $Y^{d}_{i}[m]$. Nonetheless, the sparsity constraint $\left\Vert \mathbf{f}\right\Vert_{0}\leqslant K_{S}$ and the operating conditions set in (\ref{eq:MASS_OperatingCoditions}) ensure that the probability of a spectral overlap is very small \cite{sun2013performance}. By aggregating the data from all the $q$ system branches, we can write
\begin{equation} \label{eq:MASS_CSFormulationFinal}
\mathbf{\breve y}=\mathbf{\breve C} \mathbf{\breve f}
\end{equation}
where $\mathbf{\breve y}=\left[ \left\vert \mathbf{y}_{i}\right\vert,\left\vert \mathbf{y}_{2}\right\vert ,...,\left\vert \mathbf{y}_{q}\right\vert\right]^{T}$ and $\mathbf{\breve C}=\left[ \mathbf{C}_{1},\mathbf{C}_{2},...,\mathbf{C}_{q}\right]^{T}$  are the concatenated absolute values of the FFT outputs and  the associated  disjoint sensing matrices, respectively. It is noted that $\mathbf{\breve y}\in \mathbb{R}^{M\times 1}$ and $\mathbf{\breve C}\in\mathbb{R}^{M\times N}$ can be of  high dimensions since $M=\sum_{i=1}^{q}M_{i}$. Whereas, the sought sparse real vector is $\mathbf{\breve f}=\left\vert \mathbf{f} \right\vert$ and determining the spectrum magnitude at a resolution of  $1/N$   accomplishes the MSS task. The spectrum recovery block in the MASS\ system entails solving (\ref{eq:MASS_CSFormulationFinal}) using one of the standard sparse recovery algorithms from the CS literature. The energy per subband can be subsequently measured to establish its status. It is shown in \cite{sun2013performance} that  $\mathbf{\breve f}$ can be reliably reconstructed from (\ref{eq:MASS_CSFormulationFinal}) provided that
\begin{equation}
q_{\text{MASS}}>2K_{S}-1.
\end{equation}
sampling branches are employed. For  low spectrum utilisation, i.e. $K_{S}\ll N$, the multirate asynchronous sub-Nyquist sampling detector can  substantially reduce  the  sampling rate where $M\sim\mathcal{O}( K_{S}\sqrt{N} )$.

Whilst MASS is asynchronous, its computational complexity can be high given the sizes of the handled matrices. Nevertheless, it circumvents  the need for specialised analogue preconditioning modules as in the RD, MCS and MWC systems that can render the CS  sampler inflexible and  expensive. Most importantly,  MASS is particularly amenable to be implemented by spatially distributed CRs. Each radio can have one (or a few) sampling channel(s) and transmits its $M_{i}$ measurements to a fusion centre; i.e.  soft combining. Since synchronisation among different channels is not required, SUs do not need to share their  sensing matrices. The latter aspect is a limiting factor for implementing other CS approaches across a network since each row of their sensing matrix is uniquely generated (e.g. chipping sequences, etc.) and synchronisation among the collaborating radio is essential to ensure a reasonable detection/spectrum-recovery quality. Thus, MASS  can effectively leverage spatial diversity in CR networks.             
\subsection {Remarks on Compressed Sampling Detectors}
The original MCS and MWC design objective is to be able to fully recover the processed multiband signal $x(t)$ from its sub-Nyquist samples. For the wideband spectrum sensing task,   their average sampling rates are bounded by $\alpha\geq 4L_{A}B_{C}=2f_{Landau}$  (the maximum expected spectrum occupancy is assumed). Additionally, the other addressed CS\ detection techniques levy similar requirements by  requesting    $M\sim\mathcal{O}(\kappa K_{S})$ measurements in $\mathcal{T}_{j}$.  The spectrum sparsity level denoted by  $K_{S}\sim\mathcal{O}(2L_{A}B_{C})$  depends on the spectral resolution and the constant   $\kappa$ is typically significantly larger than 2 as in MASS. 

If the final goal is multiband detection  for cognitive radio, full signal reconstruction is not necessary. This can  ease the  data acquisition requirements and even alleviate the sparsity constraint on the overseen spectrum  \cite{lexa2011compressive,ariananda2012compressive,cohen2013sub,cohen2017sub}. These two advantages can be acquired by formulating the detection problem in terms of recovering the power spectral density of the received wideband signal, which is  assumed to consist of  $K\leqslant L$ wide sense stationary transmissions. Energy in each system subband can be subsequently measured to determine the channel's status, see Figure  \ref{fig:Sub_Nyquist_MSS}. It is shown in  \cite{cohen2013sub,cohen2017sub} that the MCS and MWC systems permit the exact recovery of the incoming signal PSD (not the underlying signal realisation) when its average sampling rate satisfies
\begin{equation}\label{eq:MWC_MCS_limit2_Sparse}
\alpha_{SC}\geqslant 2L_{A}B_{C}, ~~L_{A}\ll L
\end{equation}
assuming  sparse spectrum. Hence for low spectrum utilisation,  the  sampling rate can be half of that imposed by the original MCS and MWC  systems. In this case, the computationally efficient CTF algorithm  can be used. Most remarkably, a sampling rate exceeding
\begin{equation}\label{eq:MWC_MCS_limit2_NonSparse}
\alpha_{NSC}\geqslant 0.5f_{Nyq}
\end{equation} 
is sufficient for non-sparse signals . This implies that the sampling rates of the multiband detector can  be  as low as half of the Nyquist rate   even for high spectrum occupancy, i.e.   the joint bandwidth of the simultaneously active system channels can be arbitrarily close to the total overseen bandwidth. Since\ majority of the CS-based  detectors  are prone to interference and noise uncertainty, a CS-based  feature detector is proposed  in \cite{tian2012cyclic}. It is robust against such adverse effects and exploits the cyclostationarity feature of the incoming transmissions assuming that their 2-D cyclic spectrum is sparse. Its main ideas are derived from the classical cyclostationary detector described in Section \ref{sec:NarrowbandSS}. 

In summary,  CS facilitates performing parallel wideband spectrum sensing at significantly low sub-Nyquist sampling rates, i.e. mitigates the faced data acquisition rate limitation. This comes at the expense of more complex processing techniques, e.g. sparse recovery algorithms that can be highly nonlinear, and specialised hardware to precondition the digitised signal.  There are several remaining challenges that require further analysis, e.g. devising flexible CS samplers, effect of noise on CS-based detector, model-based recovery techniques that take the communication signal structures into account, efficient implementations, to name a few.
\section{Alias-free-based Sub-Nyquist Wideband spectrum  Sensing} \label{sec:NCSWSS}
\begin{figure}[b]  
\centering
\includegraphics[width=0.95\linewidth]{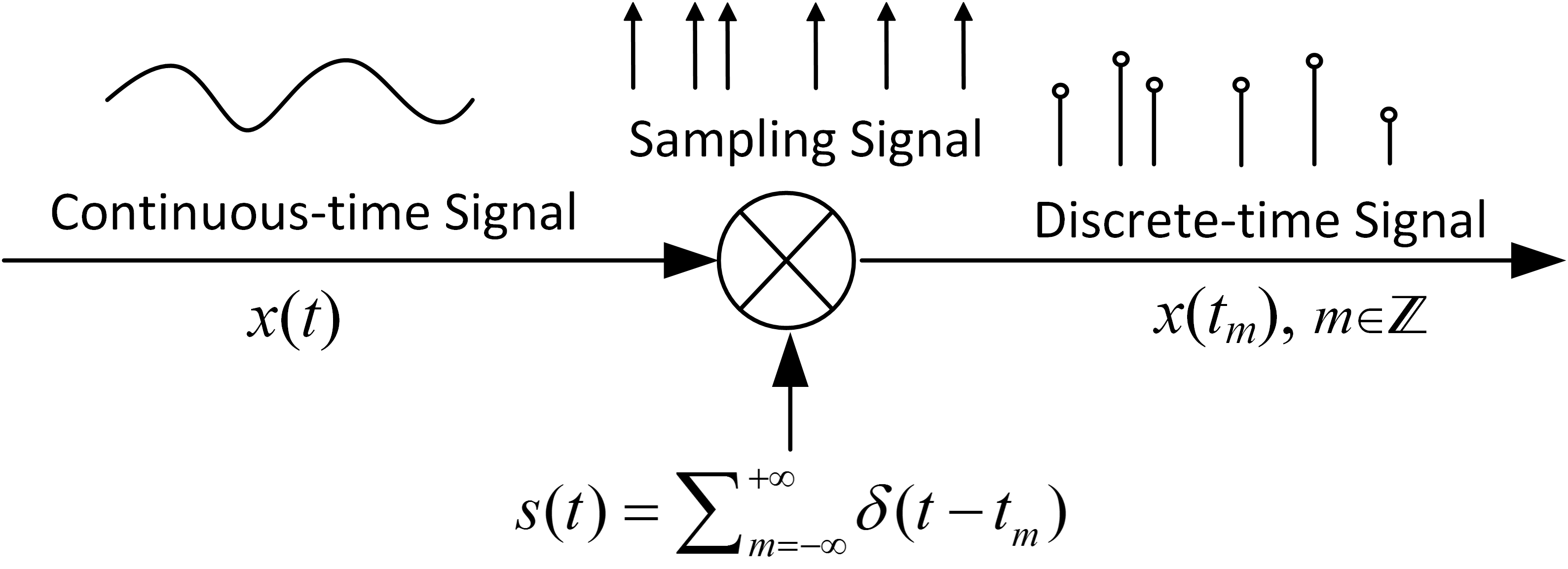}
\caption{An ideal sampler and for uniform sampling $t_{m}=mT_{US}$.}
\label{fig:IdealSampler}
\end{figure}

The sampling process, which converts a continuous-time signal $x(t)$ into its discrete-time
representation $x(t_{m})$, is typically modeled by
\begin{equation}\label{eq:NUS_IdealSampler}
x(t_{m})=x(t)s(t),~m\in\mathbb{Z},
\end{equation}
as depicted in Figure \ref{fig:IdealSampler}. The sampling signal $s(t)=\sum_{m\in \mathbb{Z}}\delta(t-t_{m}) $ comprises an infinite series of Dirac delta pulses positioned at the data acquisition time instants $\left\{t_{m}\right\}_{m\in\mathbb{Z}}$. In classical DSP,  uniform sampling is utilised and the captured  measurements are equidistant, i.e.    $t_{m}=mT_{US}=m/f_{US}$. The multiplication in (\ref{eq:NUS_IdealSampler}) translates into a convolution in the frequency domain and the power spectral density of the sampled waveform is given by
\begin{equation} \label{eq:NUS_FreqResponse}
\mathcal{P}_{X}^{d}(f)=\mathcal{P}_{X}(f)\ast\mathcal{P}_{S}(f)
\end{equation}
where $\mathcal{P}_{X}(f)$ is the PSD of the processed wide sense stationary signal and $\mathcal{P}_{S}(f)$ is the spectrum of the sampling signal. For uniform sampling, it can be shown that $\mathcal{P}_{S}(f)=f_{US}\sum_{n\in \mathbb{Z}}\delta(f-nf_{US})$ and the spectrum of the discrete-time signal can be expressed by
  \begin{equation} \label{eq:US_PSD}
\mathcal{P}_{X}^{d}(f)=f_{US}\sum_{n\in \mathbb{Z}}\mathcal{P}_{X}(f-nf_{US}).
\end{equation}
It is made up of identical  copies (i.e. aliases) of the  continuous-time signal spectrum $\mathcal{P}_{X}(f)$ shifted by multiples of the uniform sampling rate.

Assume that a transmission $x_{l}(t)$ occupies an unknown spectral band $\mathcal{B}_{l}$, $\left\vert \mathcal{B}_{l} \right\vert=B_{C}$, within the overseen frequency range $\mathfrak{B}$ of total width  $B=LB_{C}$. If $f_{US} <f_{Nyq}$ ($f_{Nyq}=2LB$), more than one spectral replica   $\mathcal{P}_{X_{l}}(f-nf_{US}),~n\in\mathbb{Z}$ of the transmission will be  present in $\mathcal{P}_{X}^{d}(f)$. Without prior knowledge of  the $\mathcal{B}_{l}$ position,  we have no means of identifying  which of the overseen system subbands  is truly occupied by examining the spectrum of sampled signal, i.e. $\mathcal{P}_{X}^{d}(f)$. The simultaneous presence of more than one transmission will  lead to overlap among their replicas in $\mathfrak{B}$ rendering  multiband detection  infeasible when $f_{US} <f_{Nyq}$. This lack of ability to unambiguously identify the spectral component(s) of the underlying continuous-time signal from the spectrum of the sampled data is referred to as spectrum aliasing, which can cause irresolvable processing problems. Sampling above the Nyquist rate eliminates aliasing phenomenon and satisfies the Shannon sampling theorem requirements to fully recovers  $x(t)$ from $x(mT_{US}),~m\in\mathbb{Z}$.  If the position of the sole active subband is known in advance, bandpass sampling  can be used and $f_{US}\geqslant2B_{C}$ suffices\cite{vaughan1991theory}; this is not the case in  wideband spectrum sensing. 

Nonuniform Sampling (NUS) poses as an alternative data acquisition approach that offers additional flexibility and new opportunities due to its potential to suppress  spectrum aliasing. It intentionally uses nonuniformly distributed sampling instants unlike scenarios where the irregularity of the collected measurements  is viewed as a deficiency, e.g.  inaccessibility of the  signals in certain periods, hardware imperfections, etc. Here, we consider randomised nonuniform sampling (RNUS) that can be 
regarded as an aliasing repression measure. It promotes  performing wide multiband detection at remarkably low  sub-Nyquist rates as illustrated in \cite{ahmad2010reliable,ahmad2011wideband,MustafaAsilomar2013WSS,ahmad2011sars,ahmad2012spectral}. The utilisation of randomised sampling  in conjunction with appropriate processing algorithms, e.g. adapted spectrum estimators, to eliminate/suppress the effect of aliasing is a methodology referred to as Digital Alias-free Signal Processing (DASP).  Few monographs on the topic exist, e.g. \cite{marvasti2001nonuniform,Wojtiuk2000Thesis,Papenfuss2007Thesis,bilinskis2007digital,babu2010spectral}.

\begin{figure}[t] 
\centering
\begin{subfigure}[t]{1\linewidth}
\centering
\includegraphics[width=1\linewidth]{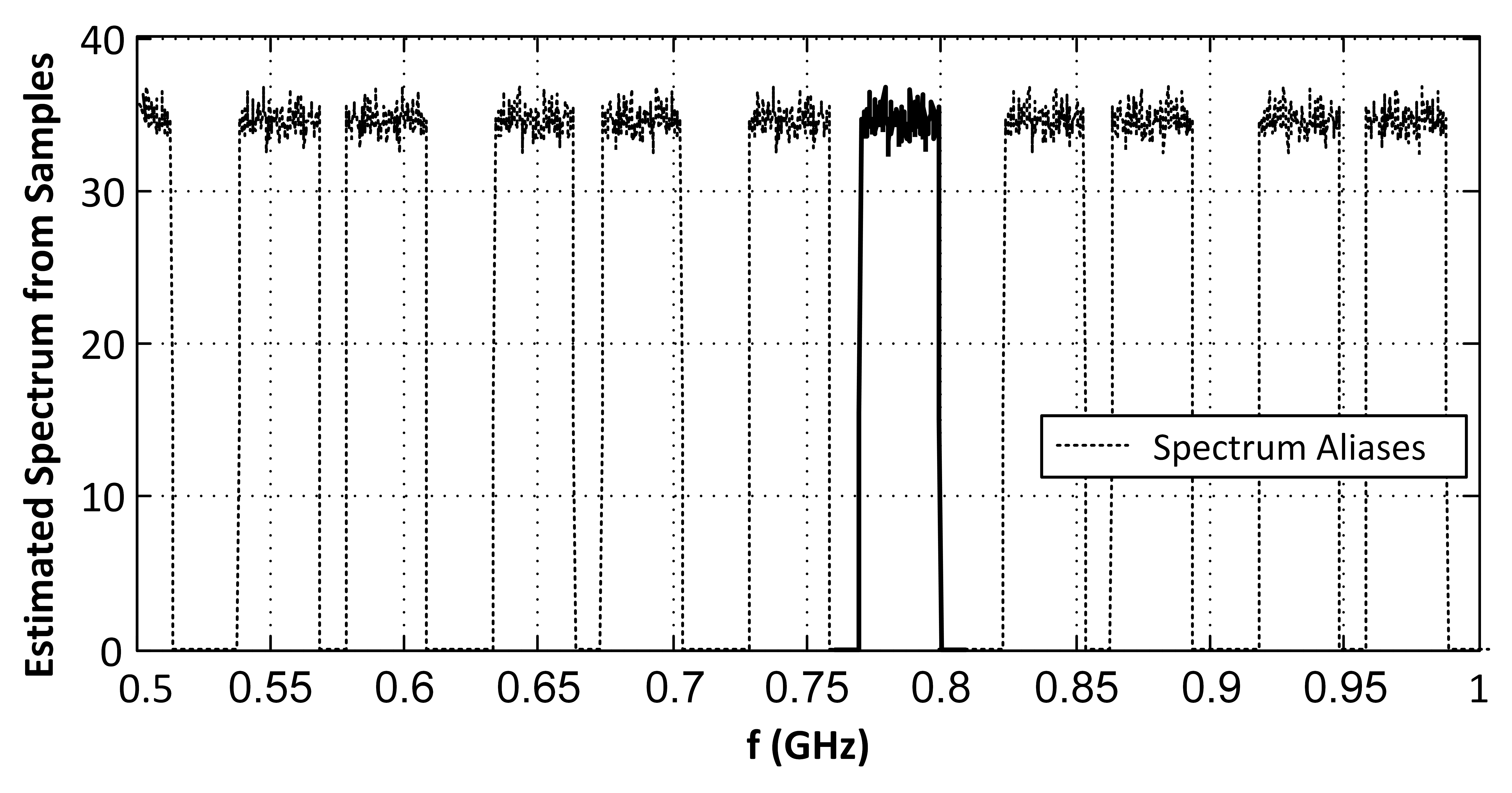}
\caption{Uniform sampling.}
\label{fig:NUS_Examplea}
\end{subfigure}
\begin{subfigure}[t]{1\linewidth}
\centering
\includegraphics[width=1\linewidth]{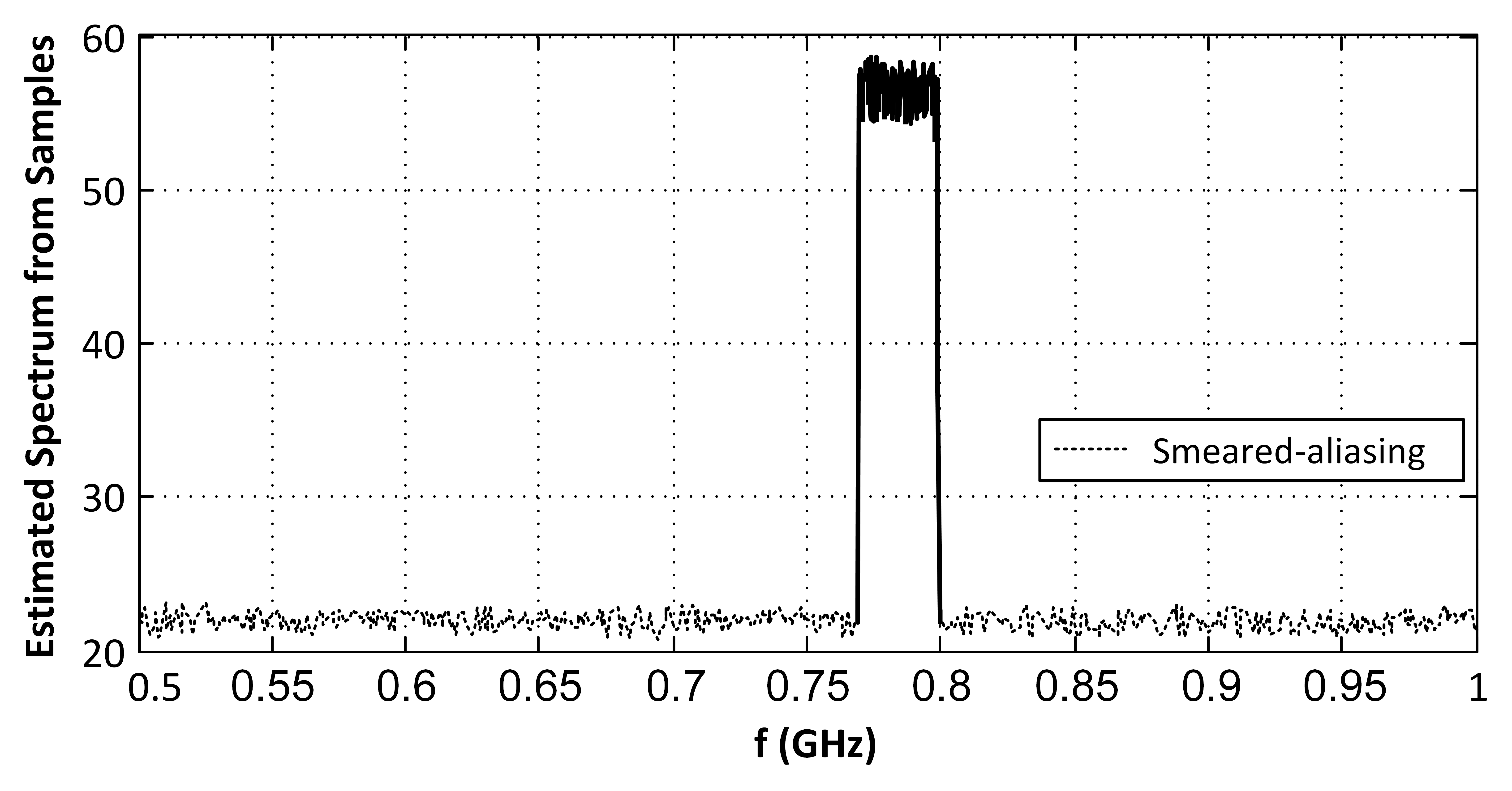}
\caption{Random nonuniform sampling.}
\label{fig:NUS_Exampleb}
\centering
\end{subfigure}
\caption{Estimated spectrum within the overseen wideband  $\left[ 0.5,1 \right]\text{GHz}$ from $\left\{x(t_{m})\right\}_{m=1}^{M}$ captured at  $\alpha=95~\text{MHz}$. One   transmission is present (solid-line) at an unknown location.}
\label{fig:NUS_PSDExample}
\end{figure}

In\ Figure \ref{fig:NUS_PSDExample}, a CR is surveying the frequency range $\left[ 0.5,1 \right]\text{GHz}$ by estimating the spectrum of incoming signal using  the sub-Nyquist sampling rate $\alpha=96~\text{MHz}$. A single PU  transmission residing in $\mathcal{B}_{l}=\left[780, 800\right]\text{MHz}$ is present; its  location is unknown to the CR. With uniform sampling, replicas of the single transmission are spread all over $\mathfrak{B}$ as shown in Figure \ref{fig:NUS_Examplea}. They are indistinguishable from one another and  most of the overseen spectral subbands can be  erroneously regarded  as occupied. In  Figure \ref{fig:NUS_Exampleb} where RNUS is employed, the previously observed  stiff-coherent aliasing  is no longer present. It is significantly suppressed and instead a broadband white-noise-component is added. The latter is  known by smeared or incoherent aliasing and it does not hinder the correct identification of the active subband(s) as evident from Figure \ref{fig:NUS_Exampleb}. This demonstrates the aliasing-suppression capabilities of randomised sampling, which is leveraged here  to devise effective sub-Nyquist multiband detection routine. Next, we briefly discuss the notion of alias-free sampling, list few RNUS schemes and introduce the DASP-based detection. 
\subsection{Alias-free Sampling Notion}
The alias-free behaviour is typically related to the spectral
analysis of a randomly sampled signal, e.g. $x(t)$, e.g. estimating the PSD $\mathcal{P}_{X}(f)$ from $\left\{x(t_{m})\right\}_{m=1}^{M}$,  rather than to reconstructing  \cite{marvasti2001nonuniform,Wojtiuk2000Thesis,bilinskis2007digital,babu2010spectral}.
 As per (\ref{eq:NUS_FreqResponse}), the total elimination of spectral aliasing  is achieved  when 
\begin{equation} \label{eq:DASP_Condition}
\mathcal{P}_{S}(f)=\delta(f).
\end{equation}
Early papers on alias-free sampling, e.g. \cite{shapiro1960alias},  showed that (\ref{eq:DASP_Condition}) can be fulfilled and $\mathcal{P}_{X}(f)$ can be exactly estimated from arbitrarily  slow nonuniformly distributed signal samples collected over infinitely long periods of time. Literal alias-free behavior is only observed in asymptotic regimes, i.e. as  $M$ tends to infinity. In practice, the signal is analysed for a limited duration of time $\left\vert\mathcal{T}_j\right\vert=T_{W}$ and $M$ is finite. Thus,   eradicating spectrum aliasing is unattainable and the benign smeared-aliasing component is typically sustained. As a result, several  criteria were proposed in the literature to affirm  the alias-free nature of a RNUS  scheme  for a finite $T_{W}$ \cite{marvasti2001nonuniform,Wojtiuk2000Thesis,bilinskis2007digital}. For example, a scheme is alias-free if it satisfies the following stationarity condition  \cite{bilinskis2007digital}
\begin{equation}
\sum_{m=1}^{M}p_{m}(t)=\alpha
\end{equation}
such that $p_{m}(t)$ is the Probability Density Function (PDF) of the sampling instant $t_{m}$. The average sampling rate is defined by $\alpha=M/T_{W}$ and $\mathcal{T}_{j}=\left[ \tau_{i},\tau_{i}+T_{W}\ \right]$. Different alias-free criteria can lead to contradicting assessment results for the same scheme. 

In the context of the studied wideband spectrum sensing problem, alias-free sampling and processing simply refers to the ability of the randomised
sampling scheme and the  deployed estimator  to sufficiently attenuate spectrum aliasing within the overseen wide frequency range $\mathfrak{B}$. As long as this suppression permits the reliable identification of the  active system subbands,  the sampling process is deemed to be suitable. We acknowledge that the  term  alias-free can be misleading since $\mathcal{P}_{X}^{d}(f)$  in practice can never be completely free of aliasing.
\subsection{Randomised Sampling Schemes}
Below, we outline a number of
data acquisition strategies that are typically used in DASP\ and are adequate for the MSS task.
Each  scheme has its own spectrum aliasing suppression characteristics necessitating separate analysis as noted in \cite{Wojtiuk2000Thesis,bilinskis2007digital}. 
 \begin{itemize}
\item
\textit{Total Random Sampling (TRS)}:\  its concept is drawn from Monte-Carlo integration over a finite integral.   All the $M$ sampling
instants of a TRS sequence are Independent Identically Distributed (IID) random variables. Their PDFs $\left\{p_{m}(t)\right\}_{m=1}^{M}$ have non-zero values only within the signal time analysis window $\mathcal{T}_{j}$, and for a uniform prior case they are given by    
\begin{equation}
p_{m}(t)=\begin{cases}1/T_{W} & t\in\mathcal{T}_{j} \\
0 & \text{elsewhere},~~~~~~~m=1,2...,M.
\end{cases}
\end{equation}
\item
\textit{Random Sampling on Grid (RSG)}: it randomly selects  $M$ samples out of the total $N_{g}$ possible sample
positions that can be in general arbitrarily distributed within $\mathcal{T}_{j}$. For simplicity, let the nominal time-locations be   $T_{g}=1/f_{g}$  apart, i.e. form an underlying uniform grid. Any of the grid points can be selected only once with equal probability and $\binom{N_{g}}{M}$ possible distinct sampling sequences
of length $M$ exist. Typically, we set  $f_{g}=f_{Nyq}$  and $N_{g}=N=\left\lfloor T_{W}f_{Nqy}\right \rfloor$. The random sampling on grid scheme  accommodates the practical constraint of having   a minimum distance between any two consecutive samples.
\item \textit{Stratified Random Sampling (SRS)}:
it divides $\mathcal{T}_{j}$  into $N_{S}$ disjointed subintervals, i.e. $\left\{ \mathcal{S}_{k} \right\}_{k=1}^{N_{S}}$ where $\mathcal{T}_{j}=\cup_{k=1}^{N_{S}}\mathcal{S}_{k}$. Let $M_{k}$ be the number of collected samples in  stratum $\mathcal{S}_{k}$ and $M=\sum_{k=1}^{N_{S}}M_{k}$ is the total number of  measurements. Various
methods exist for choosing the number of samples per subinterval. For simplicity,  assume $M_{m}=1$  and  the PDF of the $M$ random independent sampling instants is given by
\begin{equation}
p_{m}(t)=\begin{cases}1/\left\vert \mathcal{S}_{m}\right\vert & t\in\mathcal{T}_{j} \\
0 & \text{elsewhere},~~~~~~~m=1,2...,M.
\end{cases}
\end{equation}
Choosing $\left\vert \mathcal{S}_{m}\right\vert$, e.g. to improve the spectrum estimation quality, demands a priori knowledge of the processed signal. A practical approach is to assume equal subintervals $\left\vert \mathcal{S}_{k}\right\vert=T_{W}/M=1/\alpha$, i.e. Stratified Sampling with Equal Partitions (SSEP). Figure \ref{fig:SSEP} depicts a realisation of an SSEP sequence. Another popular SRS scheme is the  antithetical stratified sampling where $M_{m}=2$ and these two samples are equidistant from the stratum centre.
\begin{figure*}[t]  
\centering
\includegraphics[width=0.7\linewidth]{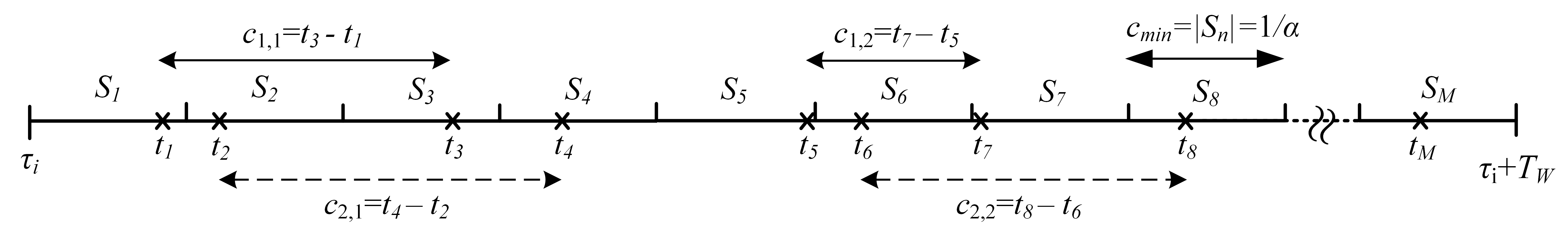}
\caption{An SSEP sequence (crosses are the sampling instants). For two ADCs, samples 
collected by ADC 1  and 2 are $\{c_{1,1},c_{1,2},...\}$ and  $\{c_{2,1},c_{2,2},...\}$  seconds apart, respectively.}
\label{fig:SSEP}
\end{figure*}
\item 
\textit{Jittered Random Sampling (JRS)}: it can be expressed as the intentional departure of the sampling instants from
their nominal uniform sampling grid. It is modeled by: $t_{m}=mT_{US}+\epsilon_{m},~m=1,2,...,M$
where $T_{US}$ is a sub-Nyquist uniform sampling period and $\left\{\epsilon_{m}\right\}_{m=1}^{M}$ are zero mean IID random variables with a PDF $p_{\epsilon}(t)$. The PDF of the $m^{th}$ sample point is $p_{m}(t)=p_{\epsilon}(t-mT_{US})$. Uniform and Gaussian PDFs with varying widths/variances are among the common choices of $p_{\epsilon}(t)$. 
\item 
\textit{Additive Random Sampling (ARS)}: its   sampling instants are described by: $t_{m+1}=t_{m}+\epsilon_{m}, ~m=1,2,...,M$ where $\left\{\epsilon_{m}\right\}_{m=1}^{M}$ are zero mean IID random variables with a PDF $p_{\epsilon}(t)$. ARS is one of the earliest alias-free schemes and was proposed in \cite{shapiro1960alias}. Its $m^{th}$ sampling instant is the sum of $m$  IID random variables and hence its PDF is given by
\begin{equation}
p_{m}=\overset{m}\circledast p_{\epsilon}(t),~~~m=1,2,...,M
\end{equation}
where $\overset{m}\circledast$ is the $m-\text{fold}$ convolution operation. Steps $\left\{\epsilon_{m}\right\}_{m=1}^{M}$ often have a Gaussian or Poisson distribution. In \cite{Wojtiuk2000Thesis}, correlated $\left\{\epsilon_{m}\right\}_{m=1}^{M}$  is suggest as a means to improve the ARS aliasing-suppression impact, i.e. correlated ARS. 
\end{itemize}
There are plenty of other DASP-oriented  randomised and deterministic nonuniform sampling schemes. For example, a data acquisition driven by the level of the processed signal, i.e.  zero crossing and level crossing sampling, and  the previously discussed  multicoset sampling scheme. The MCS aliasing-suppression characteristics is  studied in \cite{tarczynski2005optimal}. 
\subsection{Reliable Alias-free Sampling Based Multiband Spectrum Sensing}
  DASP-based sub-Nyquist wideband spectrum sensing approach  relies on nonparametric spectral
analysis  similar to the majority of parallel sensing methods. It can be represented by the block diagram in Figure \ref{fig:Sub_Nyquist_MSS} and involves the following three steps: 1) randomly sample the incoming signal  at a rate $\alpha\ll f_{Nyq}$,  2)  estimate the spectrum of the multiband signal at selected frequency points and 3) compare the estimation outcome with pre-set thresholds \cite{ahmad2011applications}. Revealing the status of the overseen $L$ system subbands  does not require determining the details spectral shape within $\mathfrak{B}$. This premise is exploited here and estimating a frequency representation  that facilitates the multiband detection task is pursued (i.e. not necessarily the signal's exact power spectral density).

The DASP-based detector adopts the  periodogram-type spectrum estimator given by
\begin{equation} \label{eq:NUS_Estimator}
\hat X_{NUS}\left(f_{n}\right)=\sum_{j=1}^{J}\beta\left\lvert \sum_{m=1}^{M}y(t_{m})w_{j}(t_{m})e^{-i2\pi f_{n}t_{m}} \right\rvert^{2}
\end{equation}
where $\beta$ is a scalar dependent on the sampling scheme and the processed signal is assumed to be wide sense stationary. The $M$ nonuniformly distributed  measurements are contaminated with AWGN, i.e. $\left\{y(t_{m})=x(t_{m})+n(t_{m})\right\}_{m=1}^{M}$. They are collected within a time analysis window $\mathcal{T}_{j}=\left[\tau_{j},\tau_{j}+T_{W}\right]$. The total signal observation window or sensing time  is given by  $T_{ST}=\left\vert\cup_{j=1}^{J} \mathcal{T}_{j}\right\vert$ and the average sampling rate is $\alpha_{NUS}=M/T_{W}$. The windowing function $w_{j} (t)$ is introduced to minimise spectral leakage where
$w_{j}(t)=w(t)$ for $t\in \mathcal{T}_{j}$ and $w_{j}(t)=0$ for $t\notin \mathcal{T}_{j}$. Recalling that spectrum sensing does not require determining the signal exact PSD,  $\hat X_{NUS}\left(f\right)$  is shown to yield a frequency representation that facilitates MSS regardless of the value of $\alpha$  \cite{Wojtiuk2000Thesis,ahmad2010reliable,ahmad2011wideband,ahmad2011sars,ahmad2012spectral}. It is noted that the statistical characteristics of (\ref{eq:NUS_Estimator}) is dependent on the  randomised sampling scheme, i.e. the PDFs of the sampling instants. To demonstrate the suitability of (\ref{eq:NUS_Estimator}) to the detection task, assume that $\left\{ t_{m} \right\}_{m=1}^{M}$ are generated according to the total random sampling scheme and $J=1$. It can be shown that 
\begin{align}\label{eq:NUS_TRS_Mean}
C(f)=\mathbb{E}\left[ \hat X_{NUS}\left(f\right) \right]=\frac{M}{(M-1)\alpha}\left[ P_{S}+\sigma_{w}^{2}\right] \nonumber \\+\frac{1}{E_{W}}\mathcal{P}_{X}(f)\ast \left\vert W(f)\right\vert^{2}
\end{align}
where  $\beta=M/(M-1)E_{W}$, $E_{W}=\int_{\tau_{j}}^{\tau_{j}+T_{W}}w^{2}(t)dt$ is the energy of the employed windowing function $w(t)$, $W(f)=\int_{-\infty}^{+\infty}w(t)e^{-i2\pi ft}dt$ and $\mathcal{P}_{X}(f)$ is power spectral density of the present transmissions. The powers of the processed multiband signal and measurements noise are denoted by $P_{S}$ and $\sigma_{w}^{2}$, respectively. From (\ref{eq:NUS_TRS_Mean}), $C(f)$ consists of a detectable feature given by the signals windowed PSD, i.e. $\mathcal{P}_{X}(f)\ast W(f)/E_{W}$, plus a component that represents the smeared-aliasing phenomenon. Unlike the stiff-coherent spectrum aliasing experienced in uniform sampling, $M\left[ P_{S}+\sigma_{w}^{2}\right]/(M-1)\alpha$ is a frequency-independent component and merely serves as an amplitude offset. It does not hamper the sensing operation, see Figure \ref{fig:NUS_PSDExample}. Therefore, $\hat X_{NUS}\left(f\right)$ is an unbiased estimator of $C(f)$, a detectable frequency representation that allows unveiling any activity within the monitored bandwidth. The width of $\mathcal{T}_{j}$ is chosen such that the  distinguishable spectral  features of the active subband(s) are reserved by  $\mathcal{P}_{X}(f)\ast W(f)/E_{W}$  and  the spectral leakage is kept below a certain level. Similar to the CS-2 method and to minimise the sensing routine computational complexity,  one frequency point per system spectral channel can be inspected to decide between $\mathcal{H}_{0,l}$ and $\mathcal{H}_{1,l}$. To maintain relatively smooth spectrographs, $T_{W}\geq c/B_{C}$, $c>1$ serves as a practical guideline  \cite{ahmad2010reliable}.

The average sampling rate $\alpha$ and the number of averaged estimates $J$   are the available design parameters that can restrain the level of estimation error in  $\hat X_{NUS}\left(f\right)$;  the  variance expressions should be  derived to evaluate the estimation accuracy. To ensure satisfying certain probabilities of detection and false alarm, i.e.$P_{FA,l}\leqslant\  \rho_{l}$ and $P_{D,l}\geq \eta_{l}$, $l=1,2,...,L$, prescriptive guidelines can be attained for the DASP-based detection. They are based on the statistical analysis of the undertaken spectrum estimation. Closed form formulas are presented in \cite{ahmad2010reliable,ahmad2011wideband,ahmad2011sars,ahmad2012spectral} for a number of RNUS\ schemes illustrating that the sensing time is a function of the sampling rate, signal to noise ratio, maximum spectrum occupancy $B_{A}$ and requested system probabilities $\mathbf{P}_{D}$ and $\mathbf{P}_{FA}$, thus 
\begin{equation} \label{eq:NUS_Guidelines}
T_{ST}\geqslant\mathfrak{F}\left( \alpha,\mathbf{P}_{D},\mathbf{P}_{FA},B_{A},\text{SNR} \right).
\end{equation}
Such recommendation clearly depict the trade-off between the sensing time, sub-Nyquist sampling rate and achievable detection performance. Assuming transmissions of equal power levels, system subbands of identical performance requirements ($P_{FA}\leqslant  \rho$ and $P_{D}\geq \eta$) and non-overlapping signal windows, the SSEP randomised scheme has
\begin{align}\label{eq:NUS_SSEP_Guidelines}
T_{ST}\geqslant\{2B_{A}Q^{-1}&(\rho)(1+\text{SNR}^{-1})/\hat D\nonumber \\&-Q^{-1}(\eta)[2B_{A}(0.5+\text{SNR}^{-1})+\alpha]/\hat D \}^{2}
\end{align}  
where $\hat D=(\alpha-B_{A})/T_{W}$.
\begin{figure}[t] 
\centering
\begin{subfigure}[t]{1\linewidth}
\centering
\includegraphics[width=1\linewidth]{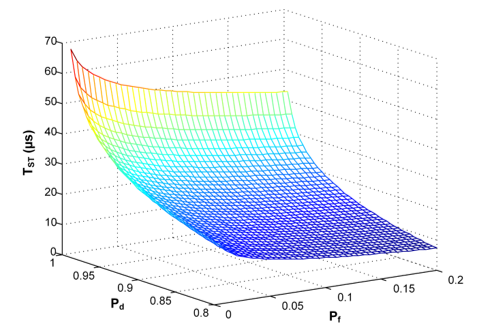}
\caption{Fixed $\alpha=56~\text{MHz}$ and varying detection requirements. }
\label{fig:DASP_Example1}
\end{subfigure}
\begin{subfigure}[t]{1\linewidth}
\centering
\includegraphics[width=1\linewidth]{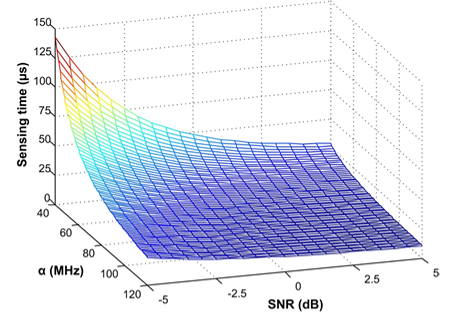}
\caption{$P_{FA,l}\leq 0.08$ and $P_{D,l}\geq0.965.$}
\label{fig:DASP_Example2}
\centering
\end{subfigure}
\caption{Recommended minimum SSEP\ sensing time for $B_{A}/B=0.1$.}
\label{fig:NUS_Tst_Trade-offs}
\end{figure} 

Most\ of all, the provided MSS reliability guidelines, e.g. (\ref{eq:NUS_SSEP_Guidelines}), affirms that the DASP-based detector sampling rate can be arbitrarily low at a predetermined additional sensing time and vice versa. Hence, there is no lower bound on the sampling rate on contrary to the compressive sensing counterpart. Different NUS schemes have different properties and guidelines in form of  (\ref{eq:NUS_Guidelines}) can be procured. In Figure \ref{fig:NUS_Tst_Trade-offs},  we display the impact of  the sub-Nyquist sampling rate $\alpha$, requested ROC probabilities and $\text{SNR}$ on the sensing time $T_{ST}$ when the overseen frequency range is of width $B=100~\text{MHz}$ and $B_{A}=5~\text{MHz}$ ($L=20$ and $B_{C}=5~\text{MHz}$). This figure can give a system designer the necessary tools to assess the requirements and viability of the sub-Nyquist multiband detector. 

Several other randomised sampling schemes were introduced that can further reduce  the sampling and sensing time requirements for alias-free (sub-Nyquist) sampling wideband spectrum sensing, for instance due their ability to acheive higher uniform convergence rates for Fourier transform or biased PSD estimations such as in \cite{MustafaAsilomar2013WSS} and Hybrid Stratified Sampling (HySt) in  \cite{MustafaAsilomaral2018,tarczynski2016estimation,ahmad2016ICASSP}.  
\subsection{Implementation Considerations of DASP, Signal Reconstruction and Final Remarks}\label{sec:CSVsNUS_Remarks}
\begin{figure}[t] 
\centering
\includegraphics[width=0.6\linewidth]{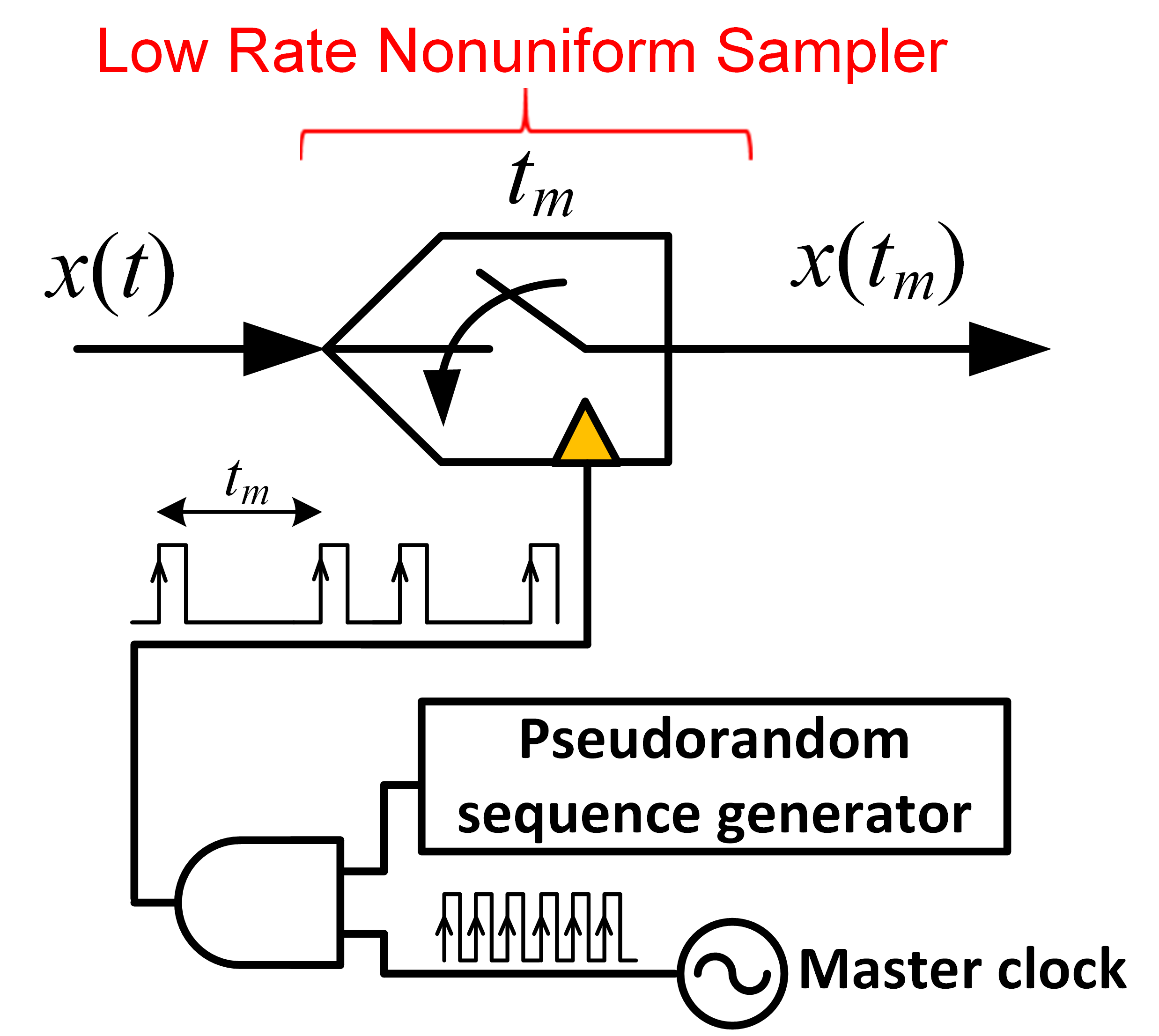}
\caption{A way to realise a nonuniform sampler using conventional ADC(s) driven by a clock with irregularly spaced rising edges.}
\label{fig:NUS_Implementation}
\end{figure}

Each nonuniform sampling scheme exhibits distinct spectrum aliasing suppression capabilities depending on the utilised spectral analysis tool, clearly more advanced estimation methods can used \cite{babu2010spectral}. However, the simplicity of the periodogram and the fact that its main building block is an FFT/DFT makes it particularly appealing for spectrum sensing in CRs. A key limitation of RNUS is its implementation feasibility. Certain schemes are more amenable  to be realised using off-the-shelf components than others. For example, the stratified sampling scheme can be implemented using two or more interleaved conventional ADCs each running at significantly low sub-Nyquist rate as shown in Figure \ref{fig:SSEP}.  Instead of collecting a data sample every $T_{US}$ seconds, the sampler collects measurements at
nonuniformly distributed time instants dictated by a pseudorandom generator that
drives the ADC. For SSEP with two ADCs in Figure \ref{fig:SSEP}, ADC 1 and 2 capture the sampling instants with odd, i.e. $\{t_{1},t_{3},...\}$ and even, i.e. $\{t_{2},t_{4},...\}$, indices, respectively.  Figure \ref{fig:NUS_Implementation} exhibits a sampler architecture that generates irregularly spaced signal measurements. Synchronisation among the interleaved ADCs can have a marginal effect on the detection quality since randomness is an integrated part of the proposed sampler.  ADC architectures that support random sampling are emerging, e.g.\cite{wakin2012nonuniform,Papenfuss2007Thesis}.  

Similar to the classical energy detector, DASP-based sensing is not immune against interference and noise level uncertainties. However, the threshold values that limit its $P_{FA,l}$ is a function of the combined overall signal and noise powers, i.e. $\gamma_{l}=\mathfrak{F}\left( P_{S}+\sigma_{w}^{2} \right)$ in lieu of $\gamma_{l}=\mathfrak{F}\left(\sigma_{w}^{2} \right)$ as in uniform-sampling-based multiband energy detector \cite{ahmad2011spectrum}. This can render the DASP-based detectors more resilient to noise estimation errors as the combined signal plus noise powers can be continuously measured at the receiver using a cheap analogue integrator. It is noted that nonuniform-sampling-based feature detectors that can be robust against  noise and interference effects remain an explored area outside the CS framework.

To preserve the reconstrutability of the detected transmissions, the sampling rates $\alpha$ should exceed at least twice the total bandwidth of the concurrently active subbands, i.e. $\alpha\geq f_{Landau}=2B_{A}$. Despite the fact that the DASP\ methods can operate at sub-Landau rates, in practice $\alpha$ should be proportional to joint bandwidth of the simultaneously active channels regardless of the total width of monitored frequency range bandwidth $\mathfrak{B}$. Prior to the emergence the compressive sensing methodology, a range of signal reconstruction techniques were available to recovery signals from their nonuniformly distributed measurements, e.g. see \cite{aldroubi2001nonuniform}. They are customarily based on minimising the $\ell_{2}$  Euclidean norm of the recovery error. Since the signal spectrum is typically sparse, CS reconstruction algorithms can be applied and they are expected to outperform an $\ell_{2}$-based ones. 

Random nonuniform Sampling can be viewed as a possible CS data acquisition approach and that DASP belongs to the general compressed sensing framework. This can be the case for certain sampling schemes, such as multicoset sampling and random sampling on grid. However, the CS performance guarantees can impose over-conservative sampling requirements and have a limited scope. They are often  applicable to  abstractly constructed sensing matrices. Most importantly, the fundamental difference between DASP\ and CS\ methodologies is in their sought objectives and the utilised processing techniques to extract the pursued signal information. Whilst the former takes advantage of the incoherent spectrum aliasing of a nonuniformly sampled signal and uses a relatively simple spectrum estimators, compressive sensing  focuses on the exact signal reconstruction and uses rather complex recovery techniques. This serves as a impetus to  further research into a unified sub-Nyquist framework for the multiband detection problem.
\section{Comparison Between Sub-Nyquist Spectrum Sensing Algorithms}\label{sec:SubNyquistComparison}
Below, we succinctly compare the considered sub-Nyquist multiband spectrum sensing approaches and  evaluated the detection performance of a number of selected techniques as in \cite{ahmad2013CSVsAliasfree_EUSIPCO2013}.
\subsection{CS Versus Alias-free Sampling for Multiband Detection}\label{sec:CSVsNUS}
\subsubsection{Performance Guarantees and Minimum Average Sampling Rates}
Whilst CS provides performance guarantees in terms of the quality of the reconstructed spectrum (e.g.  Fourier transform or PSD), the achieved  multiband detection quality (e.g. in terms of probabilities of detection and false alarm) is not typically addressed. The time consuming Monte Carlo simulations are commonly used to examine the performance of the CS-based wideband spectrum sensing algorithm. On the contrary, DASP offers clear guidelines on the attainable detection quality and equips the user with perspective recommendations on how to ensure meeting certain sensing specifications, i.e.  reliable MSS routine. Both CS and DASP, sampling rates are affected by the level of spectrum occupancy, i.e.  sparsity level. However, the DASP  minimum admissible sampling rates can be arbitrarily low at a predetermined cost of longer sensing time and vice versa. In CS, $\alpha$ has a lower bound and the impact of the sensing time on the multiband detection\ operation is  unpredictable. Hence, DASP-based detectors\  are more suitable candidates when substantial reductions on the data acquisition rates are pursued.  
\subsubsection{Computational Complexity}
The main attributes of the DASP-based wideband sensing is simplicity and low computational complexity; it only involves DFT or optimised FFT-type operations. Whereas, the CS-based techniques entail solving underdetermined sets of linear equations. This is usually computationally expensive even for state-of-the-art sparse recovery methods. It is noted that multicoset-sampling and MWC techniques adopt a more efficient approach to spectrum sensing compared with other CS methods such as CS-1 and MASS. They utilise the CTF algorithm where the processed matrices are approximately of size $m_{b}\times 2L$ in lieu of $M\times N$ ; $N\gg L\geqslant m_{b}$ is the number of Nyquist samples in the signal time analysis window and it can be very large.
\subsubsection{Postprocessing and Related Cognitive Radio Functionalities}
Alias-free sampling facilitates spectrum sensing at low sub-Nyquist rates and in its current formulation does not offer a means to estimate the signal power spectral density or power level. Nonetheless, it can be argued that the present smeared-aliasing can be easily determined and subsequently removed  to establish the underlying  transmissions PSDs. On the other hand, compressed-sampling-based multiband detectors can exactly recover the signal spectrum. Estimating the signal power level in a particular active system subband  can be important in CR\ networks to characterise the primary users, SU transmission power control  and avoid detrimental interferences. Additionally, the received multiband signal at the CR\ incorporate  PU and other CR opportunistic transmissions. The ability to reconstruct the signal from the collected sub-Nyquist samples enables the secondary user to both sense the spectrum and intercept/receive communications as in standard receivers. Advance in compressive sensing reconstruction algorithms  can be leveraged, possibly even for the randomised-nonuniform-sampling-based detection. \subsubsection{Implementation Complexity}
CS and DASP face similar implementation challenges where pseudorandom sampling sequences is commonly used as a compression strategy.  Certain CS approaches that do not utilise the aforementioned pseudorandom sampling were implemented and prototype systems were produced, e.g. MWC and RD. Nevertheless, they  require complex specialised analogue pre-conditioning modules prior to the low rate sampling (see Figures \ref{fig:RandomDemodulator} and \ref{fig:PNS_MWC}). Their subsequent processing can be also sensitive to sampler mismodeling. Thus such solutions are inflexible and are high Size, Weight, Power and Cost (SWAP-C). Designing flexible low SWAP-C sub-Nyquist samplers remains an open research question. It is noted that some nonuniform sampling schemes, e.g. random sampling on grid, are used in compressive sensing. The CS\ analogue to information converter in this case is a random time-domain sampler and the sensing matrix $\boldsymbol{\Upsilon}$ is random partial Fourier   matrix \cite{duarte2011structured}.   
\subsection{Numerical Examples}\label{sec:Experiments}
Consider a scenario that involves a SU monitoring $L=160$ subbands, each of width $B_{C}=7.5~ \text{MHz}$ and known central frequency, in search of a spectrum opportunity. The overseen frequency range  is $\mathfrak{B}=[0,1.2 ]~\text{GHz}$ and the processed total single-sided bandwidth is $1.2~\text{GHz}$. According to the Nyquist criterion, the sampling rate should be at least $f_{Nyq}=2.4~ \text{GHz}$. Let the the maximum expected  number of concurrently active subbands at any point in time or geographic location be $L_{A}=8$, i.e. the maximum occupancy is $5\%$. Here, an extensive set of Monte Carlo simulations  are conducted to evaluate  a selected number of the previously addressed sub-Nyquist wideband spectrum sensing algorithms. This carried out  in terms of the delivered probabilities of detection and false alarm. The objective is to gain an insight into their behavior for the available design resources and operation parameters. We are predominantly  interested in the impact of the  data acquisition rate, $\text{SNR}$ and sensing time $T_{ST}$ on the multiband detection outcome.  Whenever applicable, let $T_{W}=\left\vert \mathcal{T}_{j} \right\vert=0.2\mu s$ be the width of the  individual signal time analysis window and $J$  non-overlapping and equal time windows are used, i.e. $T_{ST}=JT_{W}$.  The maximum spectrum occupancy is considered in all the experiments to represent the extreme system conditions;  $L_{A}$ QPSK or 16QAM transmissions with randomly selected carrier frequencies are present in $\mathfrak{B}$. For simplicity, all active channels are assumed to have equal power levels. To maximise the opportunistic
use of  a given subband $\mathcal{B}_{l}$ and minimise the interference to the primary user, an adequate metric to
assess the sensing quality is given by
 \begin{equation} \label{eq:SC1}\small
\{\hat P_{D,l},\hat P_{FA,l}\}=\underset{P_{D,l}(i)\in \textbf{P}_{D,l},P_{FA,l}(i)\in \textbf{P}_{FA,l}}{\arg\max}P_{D,l}(i)+\left[1-P_{FA,l}(i)\right]
\end{equation}
where $\mathbf{P}_{D,l}=\left[P_{D,l}(1), P_{D,l}(2),...,P_{D,l}(p) \right]^{T}$ and $\mathbf{P}_{FA,l}=\left[P_{FA,l}(1),P_{FA,l}(2),...,P_{FA,l}(p) \right]^{T}$ are the attained probabilities for an extensive range of threshold values used to produce a complete ROC plot. Next, the basic CS-1 and CS-2 methods are examined along with the state-of-the-art MWC.  The compressed sampling matrix  $\mathbf{\Upsilon}$ in CS-1/CS-2 is a random partial Fourier matrix  and the greedy subspaces pursuit in \cite{dai2009subspace} is employed to recover the sparse vector. Random sampling on grid is the chosen RNUS scheme for the DASP approach.

\begin{figure}[t]
\centering
\begin{subfigure}[t]{1\linewidth}
\includegraphics[width=1\linewidth]{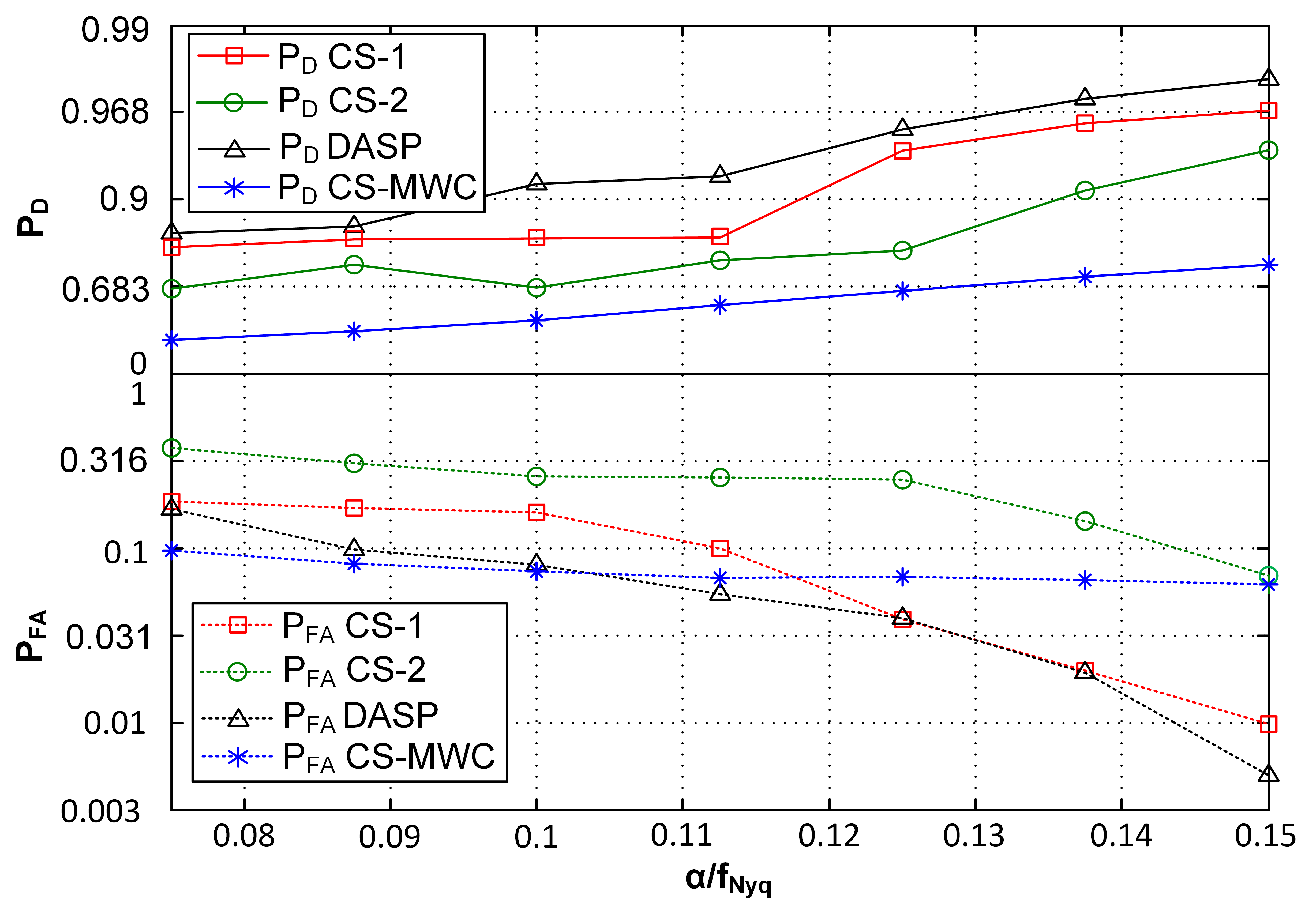}  
\caption{Total Sensing time $T_{ST}=15~\mu\text{s}$} 
\label{fig:AlphaVsT_ST1}
\end{subfigure}
\begin{subfigure}[t]{1\linewidth}
\includegraphics[width=1\linewidth]{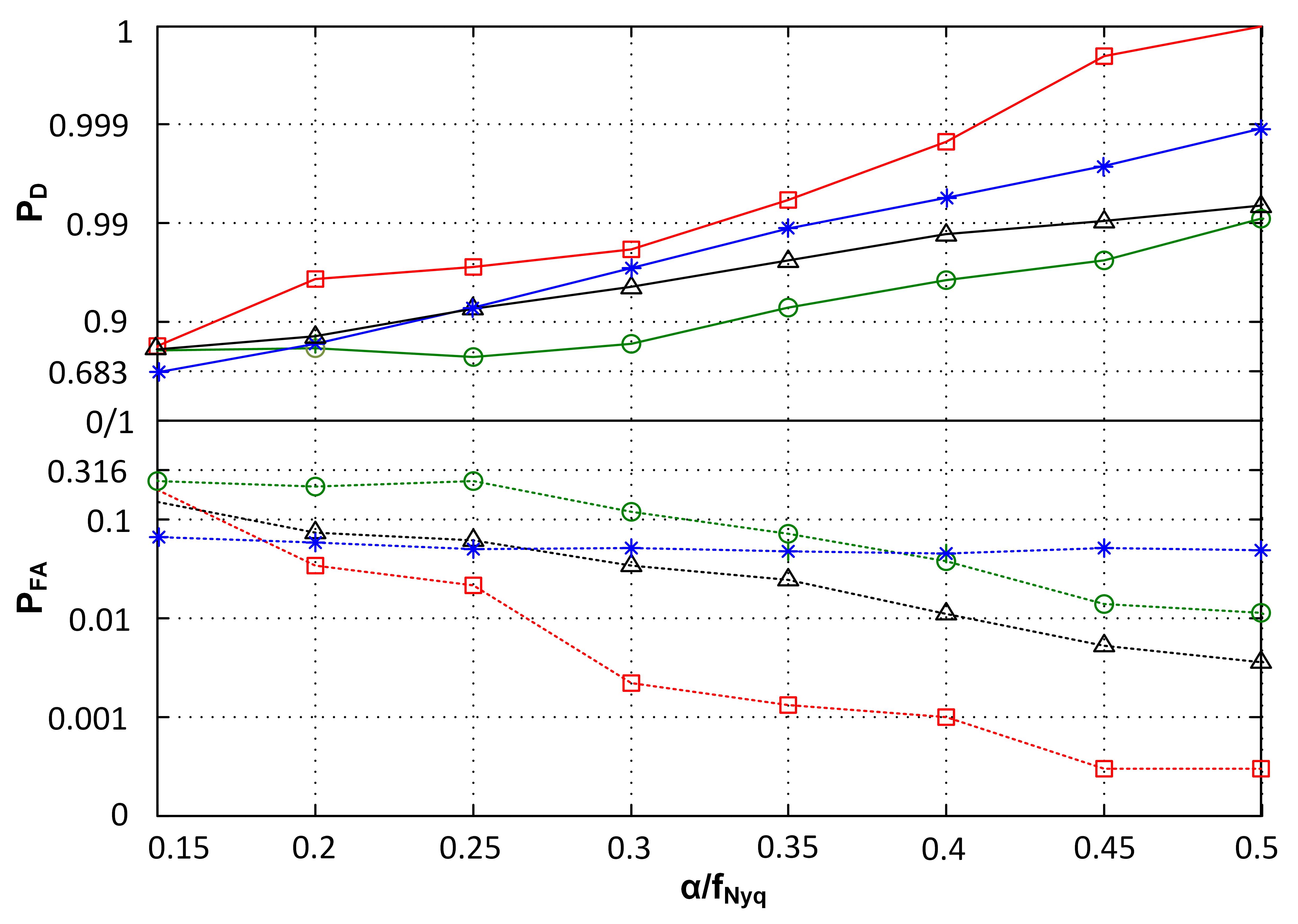} 
\caption{Total Sensing time is $T_{ST}=3~\mu\text{s}$.}  
\label{fig:AlphaVsT_ST2} 
\end{subfigure}
\caption{$\hat P_{D,l}$ and $\hat P_{FA,l}$ for selected  sub-Nyquist wideband spectrum sensing algorithms with varying compression ratios \cite{ahmad2013CSVsAliasfree_EUSIPCO2013}; $\text{SNR}=0~\text{dB}$.}
\label{fig:AlphaVsT_ST}
\end{figure}

Motivated by the goal of furnishing substantial savings on the data acquisition rates, Figure \ref{fig:AlphaVsT_ST1} depicts $\hat P_{D,l}$ and $ \hat P_{FA,l}$ in (\ref{eq:SC1}) for sub-Nyquist rates that achieve over 85\% reductions on $f_{Nyq}$ whilst $\text{SNR}=0~\text{dB}$ and a fixed total sensing time of $T_{ST}=15\mu s$. It is noted that MWC condition in (\ref{eq:MWC_Req1}) is satisfied when  $\alpha/f_{Nyq}\geq0.1$; the more recent lower rate limit in (\ref{eq:MWC_MCS_limit2_Sparse}) is exceeded when $\alpha/f_{Nyq}\geq0.05$. It can be noticed from the figure that DASP-based wideband spectrum sensing outperforms the compressive techniques, more noticeably the MWC for low sampling rates. In Figure \ref{fig:AlphaVsT_ST2}, $\alpha/f_{Nyq}\in[0.15,0.5]$ values are tested to assess  the sub-Nyquist sensing methods response to higher sampling rates. Due to the excessive high memory and computations requirements of the enormous sensing matrices associated with CS-1 for high $\alpha$ and/or $T_{ST}$ noting the large number of averaged Monte Carlo experiments, the sensing time is reduced to $T_{ST}=3\mu s$. In Figure \ref{fig:AlphaVsT_ST2}, it is evident that the CS-based algorithms deliver better detection quality compared with the DASP counterpart  for higher sampling rates, e.g. for $\alpha/f_{Nyq}\geq0.2$ in Figure \ref{fig:AlphaVsT_ST2}. Simulations also illustrate that MWC exhibits a high sensing quality only when the operation sampling rate significantly exceeds the lower theoretical bound in (\ref{eq:MWC_Req1}). Thus, Figure \ref{fig:AlphaVsT_ST} shows that DASP-based algorithms can achieve competitive, if not superior, spectrum sensing performance, compared with several compressive sensing approaches when tangible savings on the data acquisition rates are sought. This advantage degrades as $\alpha$   increases noting the substantial surge in the incurred computational cost for CS-based techniques.

\begin{figure}[t]
\centering
\includegraphics[width=1\linewidth]{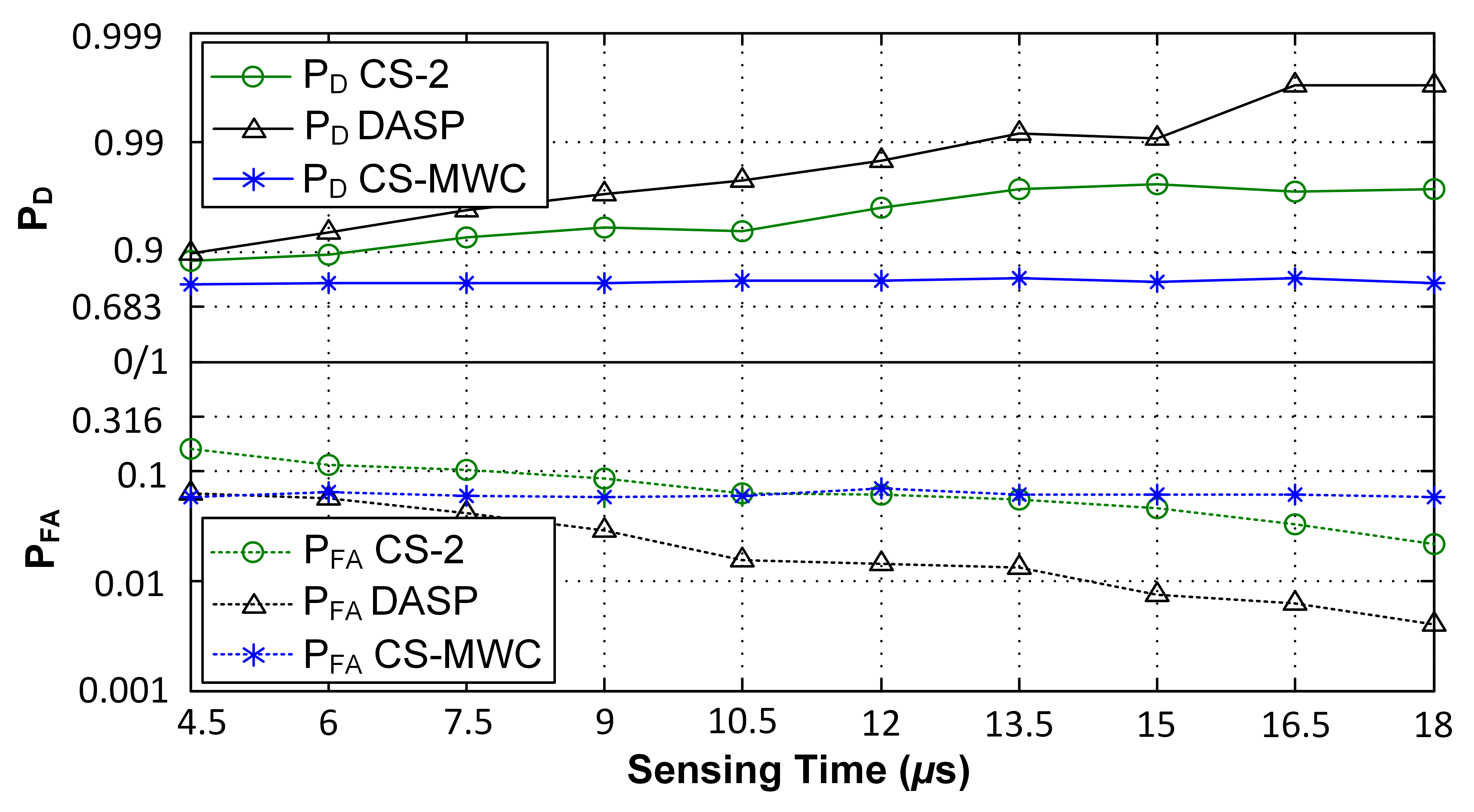}
\caption{Performance of sub-Nyquist wideband spectrum sensing algorithms for varying sensing time such that $\alpha/f_{Nyq}=0.15$ and $\text{SNR}=0~\text{dB \cite{ahmad2013CSVsAliasfree_EUSIPCO2013}.}$}
\label{fig:Exp_T_ST}
\end{figure}

In order to assess the effect of sensing time on the produced sensing results,  Figure \ref{fig:Exp_T_ST} displays the obtained probabilities of detection and false alarm for a changing $T_{ST}$ whilst $\alpha/f_{Nyq}=0.15$ and $\text{SNR=0}~\text{dB}$. CS-1 was omitted because of its prohibitive computational complexity. It is apparent from the figure that DASP-based and CS-2 exploit the  available sensing time to enhance their sensing capabilities. Whereas, MWC sensing probabilities remain nearly constant as the sensing time increases. This figure demonstrates that alias-free sampling and CS-2 approaches can effectively utilise or trade-off the sensing time for improved the spectrum sensing quality and vice versa.   

\begin{figure}[t]
\centering
\includegraphics[width=1\linewidth]{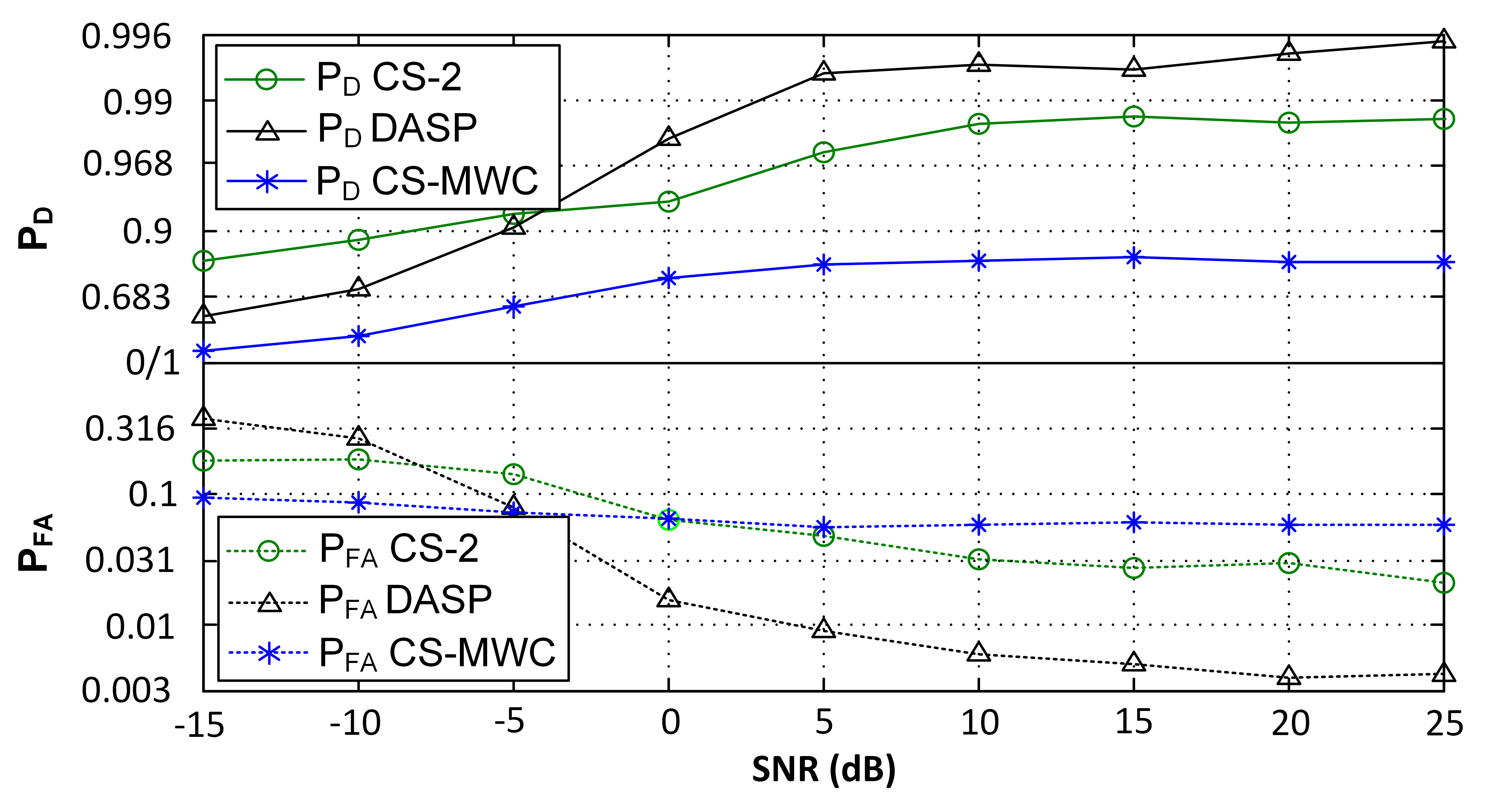}
\caption{Performance of sub-Nyquist wideband spectrum sensing algorithms for varying signal to noise ratio such that $\alpha/f_{Nyq}=0.15$ and $T_{ST}=15~\mu\text{s}$ \cite{ahmad2013CSVsAliasfree_EUSIPCO2013}.}
\label{fig:Exp_SNR}
\end{figure}

Finally, in Figure \ref{fig:Exp_SNR} the sub-Nyquist sensing algorithms are simulated for a varying SNR values whilst the sensing time and the inverse of the compression ratio  are fixed; $T_{ST}=15\mu s$  and  $\alpha/f_{Nyq}=0.15$. The figure illustrates that the DASP-based approach outperforms MWC and CS-2 as the SNR increases. The MWC distinctively continues to deliver poor results compared with the other methods despite  increasing signal power. Thus,  the sampling rate is a dominant  limiting factor in MWC. In Figure \ref{fig:Exp_SNR}, the sampling rate $\alpha$ is 1.5 times the MWC theoretical minimum admissible rate. The ability of MWC to achieve low $\hat P_{FA,l}$ for low $\alpha/f_{Nyq}$, $\text{SNR}$ and $T_{ST}$ in  Figures \ref{fig:AlphaVsT_ST}, \ref{fig:Exp_T_ST} and \ref{fig:Exp_SNR} can be the resultant of the considerably low attained  $\hat P_{D,l}$ in such ranges. 

Therefore, exploiting the aliasing suppression capabilities of random time-domain sampling can lead to a low-complexity and rather simple wideband spectrum sensing algorithms with competitive detection performance. They circumvent the need to undertake computationally intensive operations, e.g. solving complex optimisations. Nevertheless, with the ever expanding capability of DSP modules/cores and the emergence of new compressed sampling implementations, CS can still foster  effective and yet efficient  sub-Nyquist MSS solutions.  
\section{Conclusions and Open Research Challenges} \label{sec:Conclusions}
In this article,  we first introduced various aspects of the spectrum sensing functionality in a cognitive radio. The design and implementation challenges of  multiband detection were outlined and special attention was paid to the data acquisition limitation in the wideband regimes. Several wideband spectrum sensing algorithms were then discussed. Conventional parallel sensing methods that abide by the Nyquist sampling criterion commonly  employ complex analogue front-ends and a sweeping mechanism that can result in severe intolerable delays. As an alternative, the sub-Nyquist techniques were addressed and categorised as being either compressed-sampling-based or nonuniform-sampling-based. They offer new opportunities and mitigate the data acquisition bottleneck of digitally accomplishing the parallel multiband detection task. Both CS-based and NUS-based approaches have their own merits. Generally, DASP\ main advantage is simplicity and low-computational complexity compared to CS. However, CS offers a more concrete framework that can be used not only for spectrum sensing but also for subsequent CR\ functionalities such as PU characterisation and transmission interception/decoding. Simulations demonstrate that for substantially low sub-Nyquist sampling rates, DASP-based sensing can produce a higher quality detections.  

In most sub-Nyquist wideband sensing systems,
the required sampling rate is proportional to the spectrum utilisation (i.e. sparsity level). Assuming maximum spectrum occupancy can lead to pessimistically over-conservative measures to ensure the sensing reliability, i.e. cater for the worst sense scenario. This approach can waste the portable device valued resources such as power, space and memory. In practice, the
sparsity level of the wideband signal is 
time-varying due to the the dynamic nature of  PUs transmissions. Future cognitive radio networks
should be capable of performing efficient wideband
spectrum sensing for  unknown or time-varying spectrum occupancies. This calls for adaptive wideband detection techniques
that can swiftly and efficiently select the appropriate resources, e.g. sensing time sub-Nyquist sampling rate and even data acquisition scheme, without prior knowledge of the signal sparsity level. This can be a very challenging task, especially with the time-varying fading channels between the PU(s) and the CR. 

Whilst the majority of CS-based multiband detectors  assume knowledge of the sparsifying basis/frame (e.g. DFT/IDFT), a future research direction can focus on robust compressed sampling with unknown basis/frame. This becomes more pressing as the emerging cognitive radio networks are expected to alleviate the spectrum under-utilisation by facilitating dynamic opportunistic spectrum access. Hence, the radio spectrum will be no longer sparse in the frequency domain and adopting alternative sparsifying basis/frames will be mandatory to use the CS methodology. With regards to DASP, similar challenge is faced where estimators other that those targeting the frequency representation will be required. 

Due\ the hidden terminal problem and channel fading effects, a practical dependable wideband spectrum sensing will necessitate collaborative detection routines. Hence, sub-Nyquist multiband detection algorithms that promote collaborative sensing are highly desirable and are expected to be the focus in the future. For example,  how to appropriately combine information from several CRs in real time. Finally, realising dynamic low SWAP-C sub-Nyquist samplers with their subsequent processing tasks is  an open research question since portable devices supporting multimedia communications are expected to have limited power, space and memory resources. 
\bibliographystyle{IEEEtran}
\bibliography{WidebandSpectrumSensing}
\end{document}